\let\accentvec\vec
\documentclass[epj,nopacs,final]{svjour}
\usepackage{graphics}
\usepackage{latexsym,overpic,rotating}
\usepackage{subfigure,color,showkeys}
\usepackage{epsfig,color,rotating,amsmath,delarray,array,multirow}
\usepackage{makeidx,pifont,float,amssymb}
 
 \let\vec\accentvec
\newcommand{\oo}{$^o$}
\definecolor{gray01}{gray}{0.9}
\definecolor{gray02}{gray}{0.8}
\definecolor{gray03}{gray}{0.7}
\definecolor{gray04}{gray}{0.6}
\definecolor{gray05}{gray}{0.5}
\definecolor{gray06}{gray}{0.4}
\definecolor{gray07}{gray}{0.3}
\definecolor{gray08}{gray}{0.2}
\definecolor{gray09}{gray}{0.1}

\newcommand{\vecl}{{\vec\lambda}}
\newcommand{\vecr}{{\vec\rho}}

\newcommand{\gptpiz}{\ensuremath{\gamma p \to p\;\pi^0 \pi^0}}

\newcommand{\er}{$\pm$}
\newcommand{\be}{\begin{eqnarray}}
\newcommand{\ee}{\end{eqnarray}}
\newcommand{\bc}{\begin{center}}
\newcommand{\ec}{\end{center}}

\graphicspath{{pic}}

\title{High statistics study of the reaction \boldmath$\gamma p\to p\;2\pi^0$
}
\titlerunning{High statistics study of the reaction $\gamma p\to p\;2\pi^0$}
\authorrunning{The CBELSA/TAPS Collaboration}
\author{The CBELSA/TAPS Collaboration \medskip \\
 V.~Sokhoyan\inst{1}$^,$\thanks{{\em Present address:} Institut f\"{u}r Kernphysik, Universit\"{a}t Mainz, Germany},
 E.~Gutz\inst{1,2},
 V.~Crede\inst{3},
 H.~van Pee\inst{1},
 A.V.~Anisovich\inst{1,4},
 J.C.S.~Bacelar\inst{5},
  B.~Bantes\inst{6},
 O.~Bartholomy\inst{1},
 D.~Bayadilov\inst{1,4},
  R.~Beck\inst{1},
 Y.A.~Beloglazov\inst{4},
 R.~Castelijns\inst{5},
 H.~Dutz\inst{6},
 D.~Elsner\inst{6},
 R.~Ewald\inst{6},
 F.~Frommberger\inst{6},
  M.~Fuchs\inst{1},
  Ch.~Funke\inst{1},
 R.~Gregor\inst{2},
 A.B.~Gridnev\inst{4},
 W.~Hillert\inst{6},
 Ph.~Hoffmeister\inst{1},
 I.~Horn\inst{1},
 I.~Jaegle\inst{7},
 J.~Junkersfeld\inst{1},
 H.~Kalinowsky\inst{1},
 S.~Kammer\inst{6},
 V.~Kleber\inst{6}$^,$\thanks{{\em Present address:} German Research School for Simulation Sciences, J\"{u}lich, Germany},
 Frank~Klein\inst{6},
 Friedrich~Klein\inst{6},
 E.~Klempt\inst{1},
 M.~Kotulla\inst{2,7},
 B.~Krusche\inst{7},
 M.~Lang\inst{1},
 H.~L\"ohner\inst{5},
 I.V.~Lopatin\inst{4},
 S.~Lugert\inst{2},
 T.~Mertens\inst{7},
  J.G.~Messchendorp\inst{5},
  V.~Metag\inst{2},
  B.~Metsch\inst{1},
    M.~Nanova\inst{2},
  V.A.~Nikonov\inst{1,4},
  D.~Novinsky\inst{1,4},
  R.~Novotny\inst{2},
  M.~Ostrick\inst{6}$^,$\footnotemark[1],
  L.~Pant\inst{2}$^,$\thanks{{\em On leave from:} Nucl. Phys. Div., BARC, Mumbai, India},
  M.~Pfeiffer\inst{2},
  D.~Piontek\inst{1},
  A.~Roy\inst{2}$^,$\thanks{{\em On leave from:} Department of Physics, IIT, Mumbai, India},
  A.V.~Sarantsev\inst{1,4},
  Ch.~Schmidt\inst{1},
  H.~Schmieden\inst{6},
  T.~Seifen\inst{1},
	S.~Shende\inst{5},
  A.~S\"ule\inst{6},
  V.V.~Sumachev\inst{4},
  T.~Szczepanek\inst{1},
  A.~Thiel\inst{1},
  U.~Thoma\inst{1},
  D.~Trnka\inst{2},
  R.~Varma\inst{2}$^,$\footnotemark[4],
  D.~Walther\inst{1,6},
  Ch.~Wendel\inst{1}, and
  A.~Wilson\inst{1,3}
	\\
}                     
\institute{\inst{1}Helmholtz-Institut f\"ur Strahlen- und Kernphysik, Universit\"at Bonn, Germany\\
\inst{2}II. Physikalisches Institut, Universit\"at Gie{\ss}en, Germany\\
\inst{3}Department of Physics, Florida State University, Tallahassee, USA\\
\inst{4}Petersburg Nuclear Physics Institute, Gatchina, Russia\\
\inst{5}Kernfysisch Versneller Instituut, Groningen, The Netherlands\\
\inst{6}Physikalisches Institut, Universit\"at Bonn, Germany\\
\inst{7}Institut f\"ur Physik, Universit\"at Basel, Switzerland}

\date{Received: \today / Revised version: }

\abstract{The photoproduction of 2$\pi^0$ mesons off protons was studied
with the Crystal Barrel/TAPS experiment at
the electron accelerator ELSA in Bonn. The energy of photons
produced in a radiator was tagged in the energy range from 600\,MeV
to 2.5\,GeV. Differential and total cross sections and $p\pi^0\pi^0$ Dalitz plots are
presented. Part of the data was taken with a diamond radiator
producing linearly polarized photons, and beam asymmetries were derived. Properties of nucleon and
$\Delta$ resonances contributing to the $p\pi^0\pi^0$ final state
were determined within the BnGa partial wave analysis. The
data presented here allow us to determine branching ratios
of nucleon and $\Delta$ resonances for their decays into
$p\pi^0\pi^0$ via several intermediate states. Most prominent are
decays proceeding via $\Delta(1232)\pi$, $N(1440)1/2^+\pi$, $N(1520)3/2^-\pi$,
$N(1680)5/2^+\pi$, but also $pf_0(500)$, $pf_0(980)$,
and $pf_2(1270)$ contribute to the reaction.
 }
\begin{document}
\maketitle
%
\section{Introduction}

Multi-meson decays of baryon resonances are supposed to become
significantly more important with increasing baryon masses but so
far, little is known about the dynamics of the decay process. It is
known that the $N\pi$ decay fractions become small for high-mass
nucleon and $\Delta$ resonances. For some resonances, large branching
ratios for sequential decays via $\rho$ formation in the intermediate
state were reported, others have significant branching ratios for
decays into $\Delta(1232)\pi$ \cite{Manley:1984jz}. But many questions
remain open. Do high-mass resonances decay into ground state
nucleons plus higher-mass mesons like $f_0(980)$ or $f_2(1270)$, or
do they prefer to decay via excited baryons? Is there a preference
for decays with small momenta via high-mass intermediate resonances as observed in $\bar pp$ annihilation
\cite{Vandermeulen:1992eh}? What is the role of the angular momentum barrier in the decay of baryon resonances? Photoproduction of multi-body final
states cannot only shed light on these questions but there is also
the hope that {\it missing resonances} may be discovered which have
been predicted by quark models \cite{Capstick:1986bm,Loring:2001kx}
and in QCD calculations on the lattice (even though with pion masses of 400\,MeV)
\cite{Edwards:2011jj} but which were not (yet) found in
experiments. Indeed, quark model predictions suggest that many of
the so far unobserved states should have a significant
$\Delta(1232)\pi$-coupling \cite{Capstick:2000qj} and their helicity
amplitudes should not be anomalously low \cite{Capstick:1992uc}.

The study of the reaction
\begin{equation}
\gptpiz
\label{2pi0}\end{equation}
opens a good chance to search for the missing resonances and to
study sequential decays of high-mass resonances. With\-in the different
$\gamma p \to N 2\pi$-channels, the $\gamma p \to p
\pi^0\pi^0$-channel is the one best suited to investigate the
$\Delta(1232)\pi$ decay of baryon resonances. Compared to other
isospin-channels, many non-resonant-``background'' amplitudes do not
contribute, like diffractive $\rho$-production or the direct
dissociation of the proton into $\Delta^{++}\pi^-$, the so-called
Kroll-Rudermann term. In addition, the number of possible Born terms
and t-channel processes is strongly reduced; e.g. $\pi$-exchange is
not possible. This leads to a high sensitivity of the $\gamma p \to
p \pi^0\pi^0$-channel to baryon resonances decaying into $\Delta(1232)\pi$
or into higher mass baryon resonances and a pion. However, other
isospin channels will be important to separate contributions from
$N$ and $\Delta$ isobars.

High-quality $2\pi^0$ photoproduction data covering the region above
1.8~GeV where the {\it missing resonances} are expected do not exist
so far. Good angular coverage is needed to extract the contributing
resonances in a partial wave analysis. The {\bf EL}ectron {\bf S}tretcher {\bf
A}ccelerator
ELSA \cite{Hillert:2006yb} in combination with the Crystal
Barrel/TAPS experiment offers a powerful tool for studying these
nucleon resonances at large masses also in multi-particle final
states.

In this paper, the reaction $\gamma p \to p \pi^0\pi^0$ is studied for photon energies up to 2.5\,GeV. A full description of the acceptance correction and of the method of how
total and differential cross sections are determined is given in \cite{Gutz:2014wit}.
The acceptance correction requires a partial wave analysis which allows us to integrate differential cross sections into regions where the acceptance vanishes. The partial wave analysis includes
the data presented here, the data on $\gamma p\to p\pi^0\eta$ \cite{Gutz:2014wit}, and a large number of other photo- and pion-induced reactions \cite{Anisovich:2011fc,Anisovich:2013vpa}. A part of the results presented here were already communicated in two letter publications \cite{Thiel:2015xba,Sokhoyan:2015eja}.

The paper is organized as follows: In Section \ref{SectionOldData}, we give a survey on the $\rm\gamma p\rightarrow p\pi^0\pi^0$-data already published before the CBELSA/TAPS experiment was performed. The selection of the data is described in Section~\ref{SectionExperimentalSetup}. Results are presented in Section~\ref{SectionCrossSections}: total and differential cross sections are  discussed in subsections~\ref{Tot}-\ref{Mass}; part of the data were taken with a linearly polarized photon beam, polarization observables are defined and results presented  in Section~\ref{SectionPol}. Section~\ref{SectionPWA} summarizes  the results of a partial wave analysis. We give our interpretation of the results in Section~\ref{SectionInterpretation}. The paper ends with a short summary (Section~\ref{SectionSummary}). The properties of $N^*$ and $\Delta^*$ resonances as derived from the data presented here are listed in an Appendix.

\section{Previous results on {\it 2\boldmath$\pi$\unboldmath} -photoproduction}
\label{SectionOldData}
\subsection{\boldmath Data on $\gamma {p}\to {N}\pi\pi$}

\paragraph{Early experiments:}
Early bubble chamber experiments benefited from the large solid angle
coverage and the good reconstruction efficiency but suffered from the limited statistics. Photoproduction of mesons off
protons was pioneered in the sixties using bubble chambers at Cambridge
\cite{CambridgeBubbleChamberGroup:1968zz}, DESY \cite{Erbe:1970cq},
and at SLAC by different collaborations
\cite{Ballam:1971wq,Ballam:1971yd,Davier:1973fy} (only references to
the latest collaboration papers are given). The experiments used
wide band photon beams where the photon energy was reconstructed
from the event kinematics or, alternatively, positron annihilation
on a thin foil was exploited to generate photon beams in a narrow
energy band. The two-body intermediate states $\gamma p\to \pi^-
\Delta^{++}$, $\rho^- \Delta^{++}$, and $a_2(1320) n$ were studied
and the energy dependence of their cross section was determined, partly also by
using a polarized photon beam. Later experiments
\cite{Gialanella:1969ng,Carbonara:1976tg} studied the role of the
$\Delta(1232)$ in two-pion photoproduction for $E_{\gamma}$ below
1\,GeV.\vspace{-3mm}

\paragraph{Experiments at MAMI:} DAPHNE at the Mainz Microtron
MAMI gave total cross sections for $\gamma p \to p\pi^0\pi^0$,
$\gamma p \to n\pi^+\pi^0$ and $\gamma p \to p\pi^+\pi^-$
\cite{Braghieri:1994rf} at photon energies from 400 to 800\,MeV.
Double neutral pion photoproduction off the proton was
measured with the TAPS photon spectrometer from threshold
\cite{Kotulla:2003cx} to 792 MeV~\cite{Harter:1997jq} and to
820~MeV~\cite{Wolf:2000qt}, respectively. The reaction was
identified by reconstructing two neutral pions from the four photons
and by exploiting a missing mass analysis. In~\cite{Harter:1997jq},
data with one detected neutral pion and a detected photon were
included in the analysis. Below the $\eta$ threshold, this was
sufficient to identify the reaction. In \cite{Wolf:2000qt}, total
and the differential cross sections $d\sigma/dM(p\pi^0)$,
$d\sigma/dM(\pi^0\pi^0)$ and Dalitz plots were presented. The total
cross section reached a maximum of about 10\,$\mu$b at $E_{\gamma}$
= 740~MeV. Above 610~MeV a strong contribution from the
$\Delta$(1232) as intermediate state was observed. The analysis of
the reaction $\gamma p \to n\pi^+\pi^0$~\cite{Langgartner:2001sg}
revealed $N\rho$ as important decay mode of $N(1520)3/2^-$.
The helicity difference $\sigma_{3/2}-\sigma_{1/2}$ was
measured by the GDH/A2 collaboration, again with the DAPHNE
detector, for incident photon energies from 400 to 800\,MeV
\cite{Ahrens:2005ia,Ahrens:2007zzj}. The largest contribution to
the $p\pi^0\pi^0$ final state was provided when photon and proton
had a parallel spin orientation. Yet, the configuration with anti-parallel spins
provided a non-negligible contribution.

Beam-helicity asymmetries were measured in the three isospin
channels ($\overrightarrow{\gamma}p\to n\pi^{+}\pi^0$, $\overrightarrow{\gamma}p\to
p\pi^{0}\pi^0$ and $\overrightarrow{\gamma}p\to \pi^{+}\pi^{-}p$) with
circularly polarized photons in a detector configuration which
combined the Crystal Ball calori\-meter with the TAPS detector
\cite{Krambrich:2009te}.

Recently, MAMI was upgraded with a further acceleration stage to a
maximal electron energy of 1604\,MeV. Using the Crystal Ball and
TAPS photon spectrometers together with the photon tagging facility,
the reaction $\gamma p \to p\pi^0 \pi^0$ was studied from
threshold to $E_\gamma=1.4$\,GeV. Total and differential cross
sections and angular distributions were reported. A partial-wave
analysis revealed strong contributions of the $3/2^+$ wave even in
the threshold region \cite{Kashevarov:2012wy}. Beam helicity
asymmetries for the reactions $\overrightarrow\gamma p\to p\pi^0\pi^0$ and
$n\pi^+\pi^0$ were measured in the second resonance region
(550\,$<E_\gamma <$\,820\,MeV) as well as total cross-sections and
invariant-mass distributions \cite{Zehr:2012tj}. The energy range
was extended in \cite{Oberle:2013kvb} to a maximal photon energy of
1450\,MeV. The experiment also reported beam helicity asymmetries and
cross sections for the photoproduction of pion pairs off quasi-free
protons and neutrons bound in deuterons and found that the
asymmetries using protons or neutrons are very similar.\vspace{-3mm}

\paragraph{The GRAAL experiment:}
The GRAAL collaboration measured the cross
section for reaction (\ref{2pi0})~\cite{Assafiri:2003mv}
and for $\gamma d\to n\pi^0\pi^0 (p)$ \cite{Ajaka:2007zz} in the beam energy
range from $0.65 - 1.5$\,GeV. The total and differential cross
sections $d\sigma/dM(p\pi^0)$, $d\sigma/$ $dM(\pi^0\pi^0)$ and the
beam asymmetry $\Sigma$ with respect to the nucleon were extracted. Prominent peaks in the second
and third resonance region were observed.\vspace{-3mm}

\paragraph{Experiments at ELSA:}
The SAPHIR collaboration presented total and differential
cross-sections for $\gamma p\to p\pi^+\pi^-$ and determined the contributions from
$\rho^0$-mesons and $\Delta$-baryons for photon
energies up to 2.6\,GeV. At high photon energies,  $s$-channel
helicity is conserved but not near threshold. The energy
dependencies of the decay angular distributions suggest that $s$- or
$u$-channel resonance contributions are small \cite{Wu:2005wf}.

Published data from CB-ELSA, taken with the Crystal Barrel detector
at ELSA, covered the photon energy range from 0.4 to 1.3\,GeV
\cite{Thoma:2007bm,Sarantsev:2007aa}. The total cross section
exhibits the second and third resonance region at $W=1.5$ and
$1.7$\,GeV, respectively. Dalitz plots and decay angular
distributions were shown. The total cross section was decomposed into
partial wave contributions derived from an event based partial wave
analysis \cite{Thoma:2007bm,Sarantsev:2007aa}. \vspace{-3mm}

\paragraph{Experiments at Tohoku:}
The photoproduction of $\pi^+\pi^-$ pairs off protons and deuterons has
been studied in a photon energy range of 0.8 - 1.1\,GeV at the
Laboratory of Nuclear Science, Tohoku University
\cite{Hirose:2009zz} and the cross section for the
$\Delta^{++}\Delta^-$ production was deduced. \vspace{-3mm}

\paragraph{Experiments at JLab:}
The CLAS collaboration has concentrated the efforts on
electro-production of pion pairs which is beyond the scope of this
paper. The most recent result can be found in \cite{Mokeev:2012vsa},
reviews can be found in \cite{Tiator:2011pw,Aznauryan:2011qj}.
The beam-helicity asymmetry $I^\odot$ for $\gamma p\to p\pi^+\pi^-$ was
studied \cite{Strauch:2005cs} over a wide range of energies and
angles.

\subsection{Interpretations}

After early attempts to understand two-pion photoproduction, L\"uke
and S\"oding presented a very successful description of the reaction
$\gamma p\to p\pi^+\pi^-$ \cite{Luke:book}. (References to earlier
work can be found therein.) A contact (Kroll-Rudermann) term
dominated the leading $\Delta^{++}\pi^-$ production at low energies. A decisive
step forward was the inclusion of interference between the
diffractive resonant $\rho$ amplitude and background amplitudes in
which the photon splits into a $\pi^+\pi^-$ pair and where one pion
is rescattered off the proton field. In total five Feynman diagrams were used with surprisingly few
parameters.

The model was extended to include $N(938)$, $\Delta(1232)$, $N(1440){1/2^+}$, and $N(1520){3/2^-}$ as intermediate baryonic states and the $\rho$-meson as an intermediate $2\pi$-resonance. Later extensions included more baryon resonances. This type of model is based on the coupling of photons and pions to nucleons and resonances, and exploits effective Lagrangians thus leading to a set of Feynman diagrams at the tree level. The models reproduce fairly well the experimental cross sections and the invariant mass distributions. Examples of this approach can be found in \cite{GomezTejedor:1995kj,GomezTejedor:1995pe,Hirata:1997xva,Nacher:2000eq,Penner:2002md,Hirata:2002tp,Fix:2005if}.

In MAID, photoproduction amplitudes of two pseudoscalars on a nucleon were
presented in \cite{Fix:2012ds}. Expansion coefficients were defined
which correspond to the multipole amplitudes for single meson
photoproduction. Within a given set of contributing resonances,
moments of the inclusive angular distribution of an incident photon
beam with respect to the c.m.\ coordinate system for $\pi^0\pi^0$
and $\pi^0\eta$ were calculated. In the BnGa approach \cite{Anisovich:2004zz},
most channels are not treated as missing channels with unknown
couplings but rather, a large variety of channels is included in a
real ``multi-channel'' analysis. A list of presently included
reactions can be found in \cite{Anisovich:2011fc,Anisovich:2013vpa}.
SAID has not extended their ``territory'' to multi-particle final states.

\section{New ELSA Data}
\label{SectionExperimentalSetup}

The data presented here on $\gamma p \to p \pi^0\pi^0$ cover the
photon energy range from 600\,MeV to 2500\,MeV and give access to the fourth resonance region.
Results on decays of positive-parity $N^*$ and $\Delta^*$ resonance in the fourth resonance region
have been reported in two letters \cite{Thiel:2015xba,Sokhoyan:2015eja}. The data were taken in
parallel with those on $\gamma p \to p \pi^0\eta$ \cite{Gutz:2014wit}. The experimental setup is hence identical and
the event selection is similar. For experimental details, including an
account of calibration procedures and of the Monte Carlo simulation used to
determine the acceptance, we refer the reader to \cite{Gutz:2014wit}. The data selection
uses slightly different cuts than used in \cite{Gutz:2014wit} and are documented here in some detail.

\subsection{Data and data selection}

The data were obtained using the tagged photon beam of ELSA \cite{Hillert:2006yb} at the University of Bonn,
and the Crystal Barrel (CB) detector \cite{Aker:1992ny} to which the TAPS
detector \cite{Novotny:1991ht,Gabler:1994ay} was added in a forward-wall
configuration. The $\pi^0$ mesons were reconstructed from their $2\gamma$
decay by a measurement of the energies and the directions of the two
photons in CsI(Tl) (CB) and BaF$_2$ (TAPS) crystals. The proton direction was determined from its hit in a three-layer scintillation fiber detector (the inner detector)
surrounding the target \cite{Suft:2005cq} and its hit in the CsI(Tl)
(CB) or BaF$_2$ (TAPS) crystals, assuming that it originated from the center of the target.  The reconstruction efficiency for
reaction (\ref{2pi0}) is about 30\% in the energy range considered here.

\begin{figure}[pt]
\begin{center}
\includegraphics[width=0.46\textwidth]{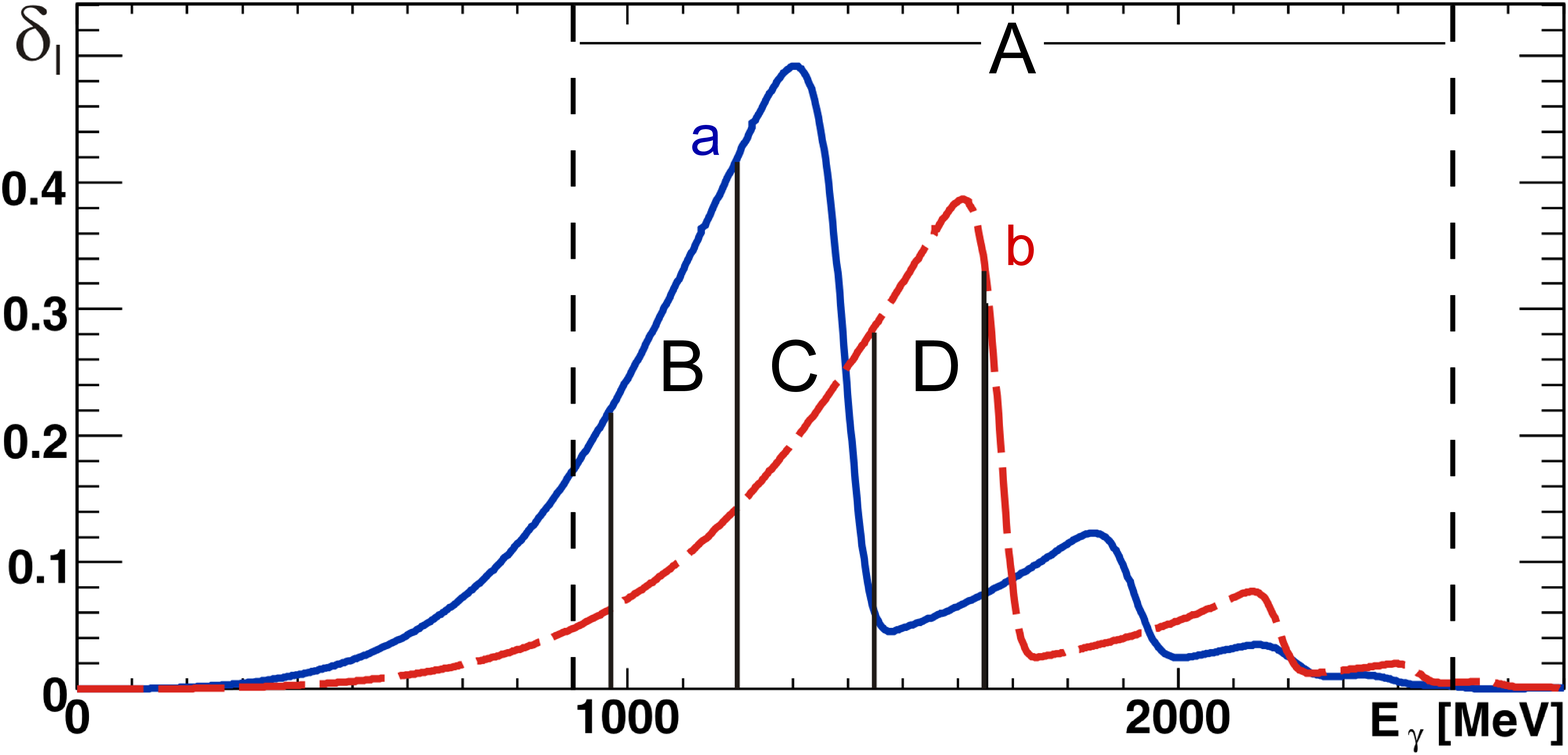}
\end{center}
\caption{\label{fig:pol}The degree of linear polarization for the
two beam times, (a) and (b). The calculation relies on an Analytic
Bremsstrahlung Calculation \cite{Natter:2003} and on measured photon intensity
distributions. The highest polarizations were 49.2\% at $E_\gamma =
1305$\,MeV (a) and 38.7\% at 1610\,MeV (b), respectively
(see \cite{Elsner:2008sn} for details). Vertical lines indicate the
energy range chosen for the extraction of cross sections (A) and
polarization observables (B, C, D), respectively.\vspace{-3mm}}
\end{figure}

The data have been acquired in different run periods in 2002/2003, CBELSA/TAPS2
with unpolarized (985,000 events) and CBELSA/TAPS1 with polarized photons
(620,000 events). Photons with linear
polarization were created by Brems\-strahlung of the 3.175\,GeV electron beam
off a diamond crystal \cite{Elsner:2008sn}. For the extraction of
total and differential cross sections, both data sets were used.
These data span the energy range (A) in Fig. \ref{fig:pol}. Data
with linear photon polarization were taken with two different
settings (a) and (b) of the diamond crystal. In the analysis, the
data were divided into three subsets, (B) from period (a), (C) from
period (a) and (b), and (D) from period (b). Figure~\ref{fig:pol}
shows the degree of polarization as a function of photon energy. The
polarization reached its maximum of 49.2\% at 1305\,MeV in period
(a), and 38.7\% at 1610\,MeV in period (b).

In the event selection of the CBELSA/TAPS1 data, events with four or five hits in the CB and TAPS calorimeters were selected. It was required that not more than one charged particle was detected based on the information from the scintillating fiber detector or the veto counters in front of the TAPS modules. The corresponding ``charged flag'' assignment to a final-state particle was however not used in the analysis to avoid possible azimuthal-dependent systematic effects on the polarization observables. Instead, a full combinatorial analysis was performed to identify the proton among the five detector hits. For each combination, it was tested if the invariant mass of both photon pairs agreed
with the $\pi^{0}$ mass within $\pm 35$~MeV and if the missing mass of the proton was compatible with the proton mass within $\pm 100$~MeV. The missing-proton direction also had to agree with the direction of the detected charged hit within $\pm 10^{\circ}$ in the azimuthal angle, $\varphi$, and within $\pm 10^{\circ}$ in the polar angle, $\theta$, in the CB or within $\pm 5^{\circ}$ in $\theta$ in TAPS. If more than one combination passed these kinematic cuts, a kinematic fit was performed and the combination with the greatest Confidence Level (CL) for the $\gamma p\rightarrow p\pi^0\pi^0$ hypothesis was selected. In a further step, all events were subjected to kinematic fitting and the CL for
the hypothesis $\gamma p\rightarrow p\pi^0 \pi^0$ was required to be greater than 10\,\% and had to exceed the CL for the competing $\gamma p\rightarrow p\pi^0\eta$ hypothesis. The direction of the proton before and after kinematic fit had to agree within $\pm 8^{\circ}$ in $\varphi$ and $\pm 10^{\circ}$ in $\theta$ and within $\pm 4^{\circ}$ in $\theta$ for the protons detected in the CB and in TAPS, respectively. Additionally, it was required that the number of crystals in a proton cluster had to be less than five for both calorimeters. The maximal energy deposited by protons was restricted to 450~MeV in the CB and to 600~MeV in TAPS. The polar angle of the proton was restricted to the kinematic limit of $70^{\circ}$.

In contrast to the reaction $\gamma p\rightarrow p\pi^0\eta$ \cite{Gutz:2014wit}, the contribution of events with only four calorimeter hits is not negligible for the reaction $\gamma p\rightarrow p\pi^0 \pi^0$. To suppress the background contribution to this class of events, the same cuts (as above) were imposed on the invariant mass of the photon pairs and the missing mass of the proton. Events were selected if either the direction of the missing proton was consistent with the forward opening of TAPS ($0^{\circ} < \theta_{\rm proton} < 5^{\circ}$) or the energy of the proton was too low to be detected in any of the calorimeters. To further suppress the remaining background, additional
cuts on the polar angle, $\theta_{\rm proton}$, and on the momentum of the missing protons were applied. In the forward direction covered by TAPS ($5^{\circ} < \theta < 30^{\circ}$), events with momenta below 350~MeV/$c$ were selected. For $30^{\circ} < \theta_{\rm proton} < 60^{\circ}$ and proton momenta below 250~MeV/$c$, no detected hit in the inner detector was required since the protons with such low momenta cannot reach this detector. If the missing-proton momentum was in the range $250 < P_{\rm proton} < 450$~MeV/$c$, the direction had to be consistent with a hit in the inner detector within $\pm 15^{\circ}$ in $\varphi$ and $\pm 20^{\circ}$ in $\theta$. Furthermore, all selected four-hit events were subject to kinematic fitting and the same 10\,\% cut as for the five-hit events was applied
for the $\gamma p\rightarrow p\pi^0 \pi^0$ hypothesis. The CL also had to exceed the confidence level for the $\gamma p \to p \pi^0 \eta$ hypothesis. The background contamination of the final event sample was determined to be below 1\,\%.

\begin{figure*}[pt]
\begin{center}
\includegraphics[height=0.3\textheight,width=0.66\textwidth]{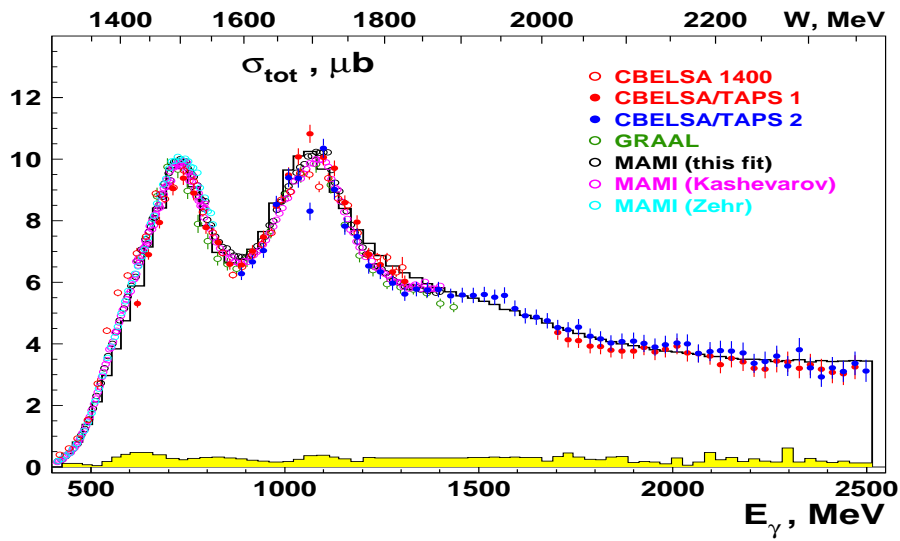}
\vspace{-3mm}
\end{center}
\caption{\label{FigureTotal}The total cross section for
$\gamma p\to p\pi^0\pi^0$. Red and blue dots represent our results derived from two different run periods. Results from other experiments are represented by open circles: CB-ELSA 1400 \cite{Thoma:2007bm,Sarantsev:2007aa} in red, those from GRAAL \cite{Assafiri:2003mv} in green, and those from TAPS/A2 at MAMI  in pink \cite{Kashevarov:2012wy} and cyan \cite{Zehr:2012tj}. The black open circles represent the data from \cite{Kashevarov:2012wy} integrated within the BnGa partial wave analysis. The result of the partial wave analysis including the data from \cite{Kashevarov:2012wy,Assafiri:2003mv} is shown as a histogram. Our systematic uncertainty is shown as yellow band.
In addition, there is a normalization uncertainty of 10\%, see Table~\ref{systerr}. }
\end{figure*}

The selection of the unpolarized CBELSA/TAPS2 data follows a similar strategy but differs in the basic particle identification. In the forward direction, no (or at most one) signal in a TAPS photon-veto belonging to a cluster defines a photon (or a charged particle). For polar angles above $30^\circ$, a CB cluster is assigned to a charged particle if the trajectory from the target center to the barrel hit forms an angle of less than $20^\circ$ with a trajectory from the target center to a hit in the scintillating fiber detector. Events with four photons and at most one charged particle in the final state were selected. In five-cluster events, a coplanarity cut required  $|180^\circ - \Delta\phi| < 20^\circ$ between the total momentum of the four photons and the detected charged particle. In a following step, the proton momentum was then reconstructed from event kinematics in “missing-proton” kinematic fitting. It was required that CL $> 10$\,\% for the hypothesis $\gamma p\to p\pi^0\pi^0$ and CL $<  1$\,\% for the hypothesis $\gamma p\to p\pi^0\eta$. In this way, the four-photon (no charged particle) and the five-cluster events (with no more than one charged particle) were treated alike. For five-particle events, the direction of the reconstructed and the kinematically fitted proton had to agree within $15^\circ$ and $5^\circ$ in CB and TAPS, respectively.

\section{Results}
\label{SectionCrossSections}

\subsection{\label{Tot}The total cross section}
Figure~\ref{FigureTotal} shows the total  cross section for $\gamma p\to p\;\pi^0\pi^0$. The cross section is determined from a partial wave analysis to
the data described below. The partial wave analysis allows us to generate a Monte Carlo event sample representing the ``true'' physics. For any distribution, the efficiency can then be calculated as fraction of the reconstructed to the generated events. Our data points in Fig.~\ref{FigureTotal} were determined from the efficiency corrected number of events. The red and blue full dots represent the two run periods, CBELSA/TAPS1 and CBELSA/TAPS2, open circles represent earlier data \cite{Kashevarov:2012wy,Zehr:2012tj,Assafiri:2003mv,Thoma:2007bm,Sarantsev:2007aa}.
Due to the sharp edges of the coherent peaks in the photon energy spectrum, the determination of the photon flux
suffers from additional uncertainties. Therefore, this energy range is omitted in the determination of the cross section but used to derive polarization observables.
At about 1100 MeV, two tagger fibers had a large noise rate. This may be responsible
for the additional fluctuations observed here.

\begin{table}[pb]
\caption{\label{systerr}The normalization uncertainty in the determination of the cross sections.}
\renewcommand{\arraystretch}{1.2}
\begin{center}
\begin{tabular}{lr}
\hline\hline
flux normalization        &  8\%\\
reconstruction efficiency &  5\%\\
target density            &  2\% \\
\hline
total normalization uncertainty &  10\% \\
\hline\hline
\end{tabular}
\end{center}
\end{table}

The systematic uncertainty in the determination of the total (and the differential) cross sections contains several contributions. The first uncertainty depends on the extrapolation of the partial wave analysis into the region where no data exist. Since the region is very different for the data taken at ELSA and MAMI, we take the difference between the fits to our data including or excluding the MAMI data \cite{Kashevarov:2012wy} as estimate of the systematic uncertainty due to the PWA. This uncertainty is a function of energy and the scattering angle and is plotted as yellow band in the histograms. For energies above 1.7\,GeV we use the difference between our two data sets to estimate the uncertainty. In the energy range where no red data exist, the systematic errors are
interpolated between the low and the high-energy region. In addition there is an overall systematic error, see Table~\ref{systerr}.

Our cross section agrees reasonably well with our earlier data and with those from TAPS/A2~\cite{Kashevarov:2012wy,Zehr:2012tj} at MAMI and those from GRAAL~\cite{Assafiri:2003mv}. In the low energy region, the CB-ELSA data show a slightly larger cross section than those from MAMI. The partial wave analysis described below assigns the shoulder to production of the Roper resonance $N(1440)1/2^+$ decaying into $\Delta(1232)\pi$. In the fit we use both data sets which reproduces the mean.

The data exhibit clear evidence for the second and third resonance regions, the peak cross section reaches about 10\,$\mu$b. We hence expect contributions from $N(1440)$ ${1/2^+}$, $N(1535){1/2^-}$, and $N(1520){3/2^-}$ ($2^{\rm nd}$ resonance region) and from $N(1710){1/2^+}$, $N(1650){1/2^-}$, $N(1680){5/2^+}$, $N(1700){3/2^-}$, $\Delta(1600){3/2^+}$, $\Delta$$(1620){1/2^-}$, and $\Delta(1700){3/2^-}$ ($3^{\rm rd}$ resonance region). Above the third resonance region, the cross section falls off smoothly from 6\,$\mu$b and reaches 4\,$\mu$b at 2.5\,GeV photon energy. There is some indication for a small enhancement at about $E_\gamma = 1.5$\,GeV due to the fourth resonance region.

\begin{figure*}[pt]
\begin{center}
\begin{tabular}{cc}
\hspace{-3mm}\includegraphics[height=0.31\textheight]{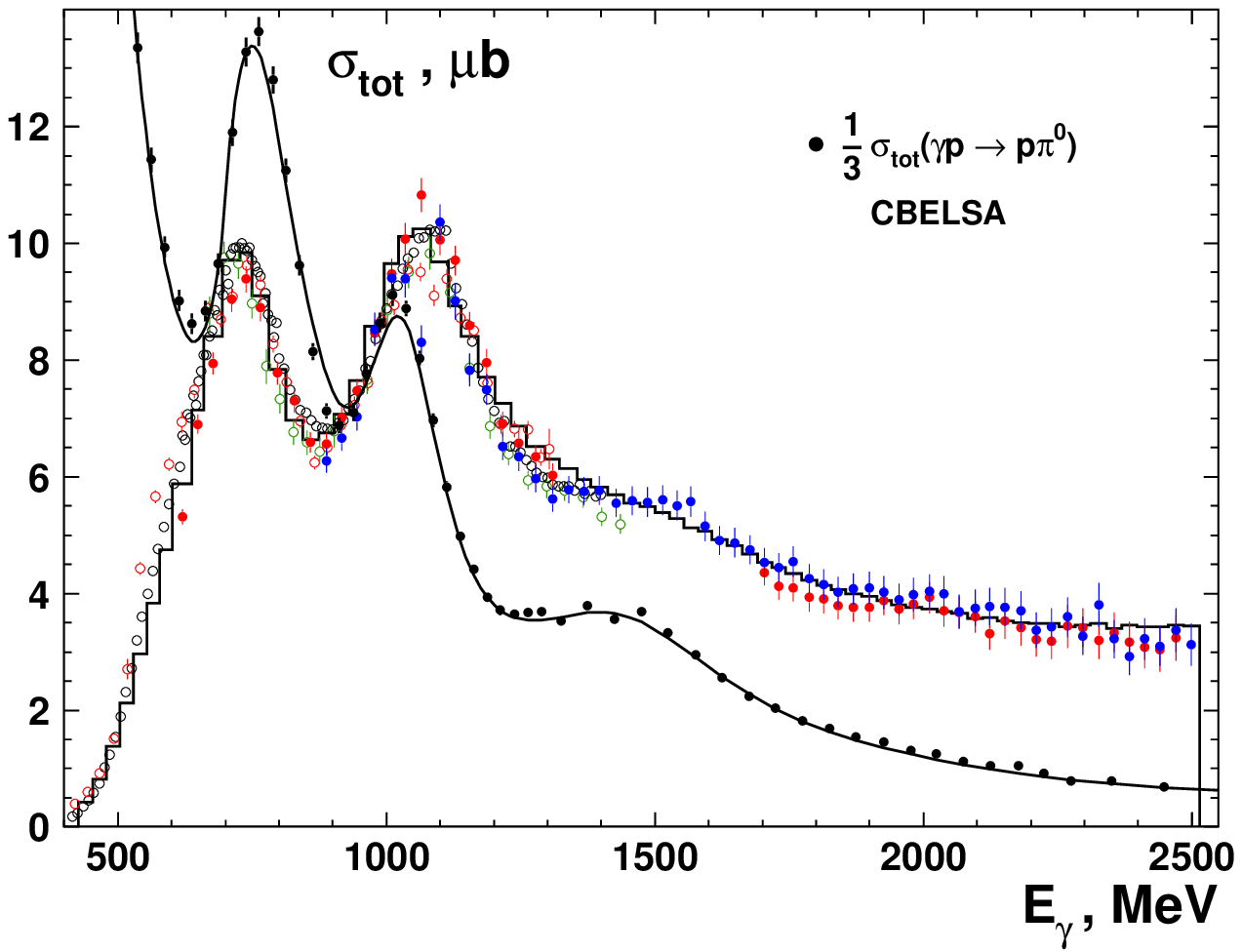}&
\hspace{-6mm}\includegraphics[height=0.31\textheight]{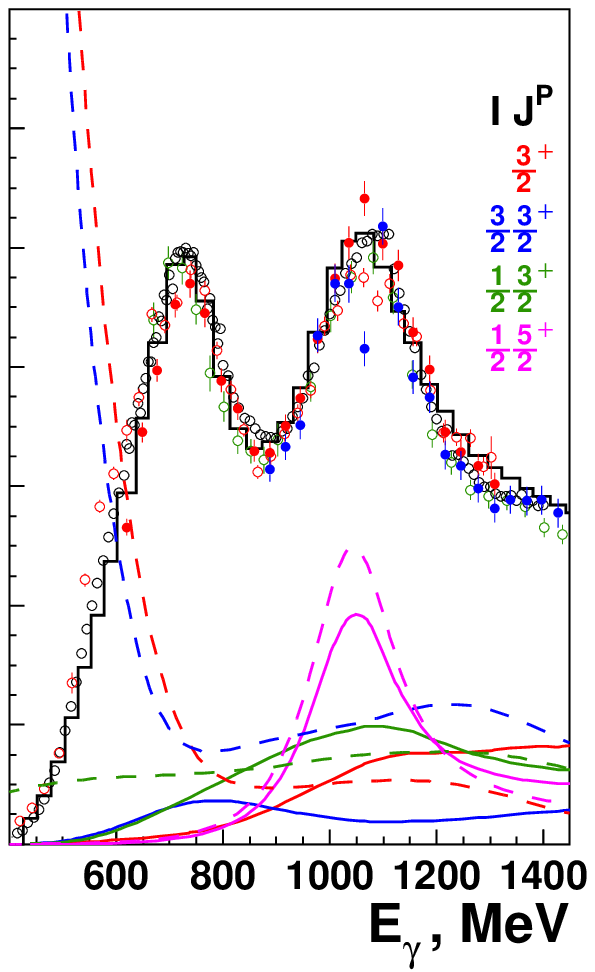}
\hspace{-4mm}\includegraphics[height=0.31\textheight]{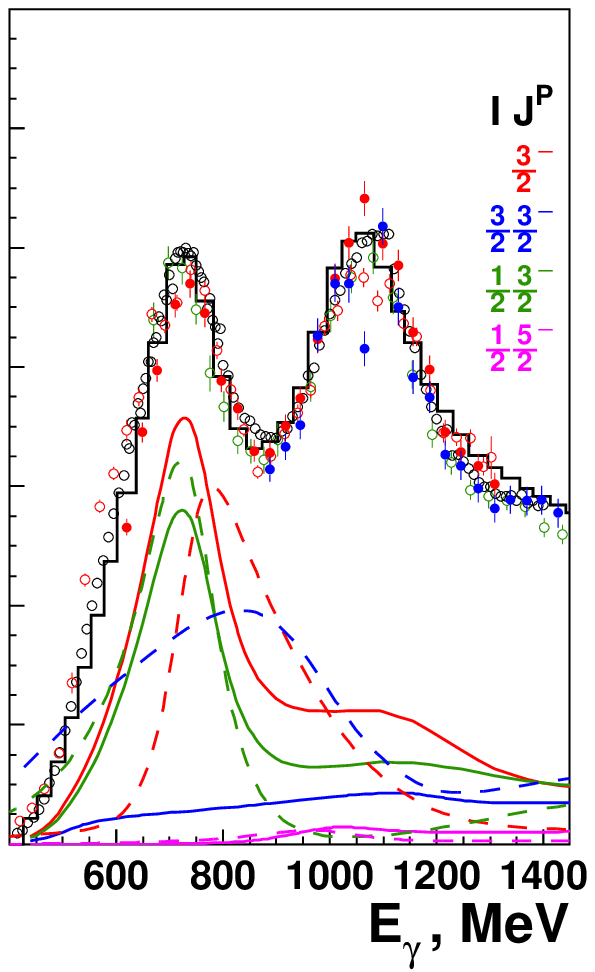}
\end{tabular}
\end{center}
\caption{\label{FigureTerms}Left: Comparison of the total cross sections for $\gamma p\to p\pi^0$ \cite{vanPee:2007tw} (black dots) and $\gamma p\to p\pi^0\pi^0$ (for symbols, see Fig. \ref{FigureTotal}). The results of the partial wave analysis are shown as solid curve or histogram. Right: The most important partial wave contributions to the cross sections (solid curve $\gamma p\to p\pi^0\pi^0$, dashed curves: $\gamma p\to p\pi^0$; the color code for different $I J^P$ partial waves is given in the figures).
For spin parity $J^P=3/2^\pm$, the total contributions are shown as well as the contributions to isospin $I=3/2, J^P=3/2^\pm$ and $I=1/2, J^P=3/2^\pm$. The $I=1/2$ contributions to $5/2^\pm$ are also shown.}
\end{figure*}

In Fig.~\ref{FigureTerms} (left) the cross section for photoproduction of two neutral pions is compared to the cross section for single $\pi^0$ photoproduction (adapted from \cite{Thiel:2015xba}). The latter cross section is scaled down by a factor 3 to allow for a direct comparison of the two cross sections. The positions and shapes of the structures in the $N\pi$ and $N\pi\pi$ mass distributions are strikingly different: The low-mass peak in the $N\pi$ mass distribution is found at a higher mass than in the $N\pi\pi$ mass distribution, and for the higher mass peak, the reverse is true.

For $\gamma p\to p\pi^0$, the tail of $\Delta(1232)$ is seen followed by the second and third resonance regions. The visible height in $\frac13\sigma(\gamma p\to p\pi^0)$ in the second (third) resonance region is a bit larger (smaller) than in $\sigma(\gamma p\to p\pi^0\pi^0)$; however, the third resonance peak appears to be considerably wider. Thus we expect the ratio of $N\pi$ to $N\pi\pi$ branching ratios to be of the order of 4:1 larger for the resonances in the second and of the order of 2:1 in the third resonance region. In Table~\ref{tab:br-2-3} we list the $N\pi$ and $N\pi\pi$ branching ratios.

\begin{table}
\renewcommand{\arraystretch}{1.2}
\caption{\label{tab:br-2-3}Width and branching ratios of resonances in the $2^{\rm nd}$ and  $3^{\rm rd}$ resonance region for decays into $N\pi$ and $N\pi\pi$ \cite{Agashe:2014kda}. The $N\rho$ contribution to $N\pi\pi$ -- not contributing to $N\pi^0\pi^0$ -- is mostly small.}
\begin{center}
\begin{tabular}{cccc}
\hline\hline
                   &    $\Gamma$\,(MeV)   & $N\pi$         & $N\pi\pi$ \\
                                    \hline
$N(1440)1/2^+$     &325\er125& 0.65\er 0.10  & 0.35\er 0.05\\
$N(1520)3/2^-$     &113\er 13& 0.60\er 0.10  & 0.25\er 0.10\\
$N(1535)1/2^-$     &150\er 25& 0.45\er 0.10  & 0.05\er 0.05\\
\hline
$N(1650)1/2^-$     &155\er 30& 0.70\er 0.20  & 0.15\er 0.05\\
$N(1675)5/2^-$     &150\er 20& 0.40\er 0.05  & 0.55\er 0.05\\
$N(1680)5/2^+$     &130\er 10& 0.68\er 0.03  & 0.35\er 0.05\\
$N(1700)3/2^-$     &175\er 75& 0.65\er 0.10  & 0.35\er 0.05\\
$N(1710)1/2^+$     &150\er100& 0.13\er 0.08  & 0.65\er 0.25\\
$N(1720)3/2^+$     &275\er125& 0.11\er 0.03  & 0.80\er 0.10\\
$\Delta(1620)1/2^-$&140\er 10& 0.25\er 0.05  & 0.15\er 0.05\\
$\Delta(1700)3/2^-$&300\er100& 0.15\er 0.05  & 0.75\er 0.05\\
\hline\hline
\end{tabular}\vspace{-2mm}
\end{center}
\renewcommand{\arraystretch}{1.0}
\end{table}

There are several reasons which may be responsible for the shifts in the peak position observed in Fig.~\ref{FigureTerms} when the cross sections for $\gamma p\to p\pi^0$ and $\gamma p\to p\pi^0\pi^0$ are compared. First, the peak positions are not directly related to the pole positions. Interference with a `background' amplitude can lead to a shift of the observed mass distribution; only the pole position does not depend on the particular reaction dynamics. Second, the observed distribution depends on the interference of resonances with different isospin. Thus, $N(1520)3/2^-$ can interfere with the tails of $N(1700)3/2^-$ and $\Delta(1700)3/2^-$; $N(1535)1/2^-$ can interfere with $\Delta(1620)1/2^-$,  and the interference may be different in the $p\pi^0$ and $p\pi^0\pi^0$ final states. (Of course, interference between, e.g., $N(1535)1/2^-$ and  $N(1520)3/2^-$ is possible as well but this leads to changes in angular distributions but not to changes of the total cross section.) A third possibility to explain the mass shifts between the peaks in the $N\pi$ and $N\pi\pi$ cross sections are differences in the relative importance of individual resonances contributing to the second or third resonance region. Obviously, the interpretation of the mass shifts requires a full partial wave analysis.

\begin{figure*}[pt]
\begin{center}
\begin{tabular}{cc}
\includegraphics[width=0.45\textwidth]{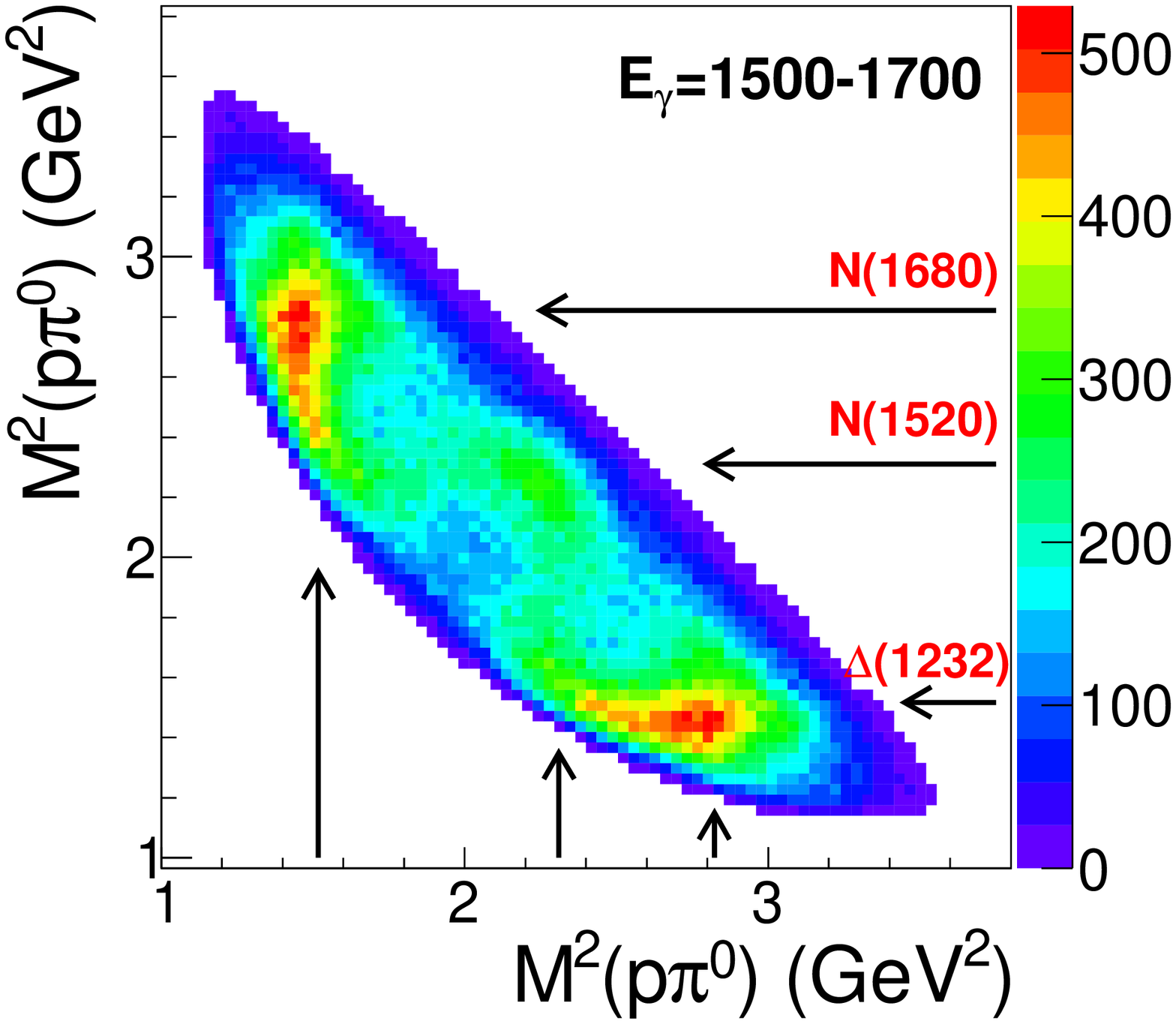}&
\includegraphics[width=0.45\textwidth]{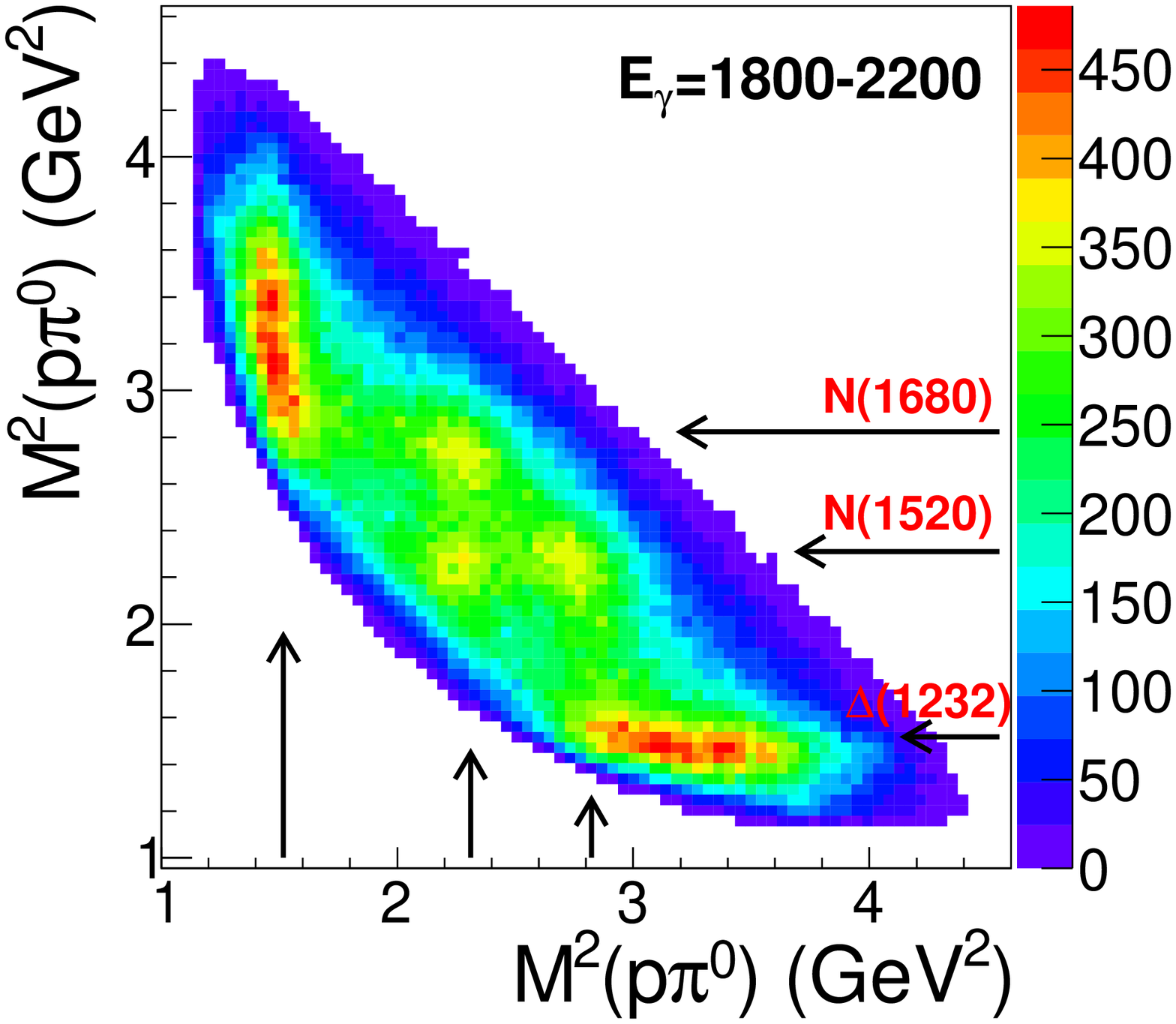}
\end{tabular}
\end{center}
\caption{\label{FigureDalitz-large}Dalitz plots for 1500\,MeV $<$ E$_\gamma$ $<$ 1700\,MeV (left)
and 1800\,MeV $<$ E$_\gamma$ $<$ 2200\,MeV. The arrows indicate the positions of $\Delta(1232)$, $N(1520)3/2^-$, and $N(1680)5/2^+$.}
\end{figure*}

Some main results for the leading partial waves are shown in
Fig.~\ref{FigureTerms}, right. We first notice that the two
prominent peaks in the total cross sections are dominantly due to
$N(1520)3/2^-$ and $N(1680)5/2^+$. The $N(1520)3/2^-$ state has a
branching fraction $0.61\pm0.02$ into the $\pi N$ channel and
$0.28\pm0.06$ into the $\Delta\pi$ channel. Therefore the expected
contribution of this state in the photoproduction calculated from
the $\pi^0 p$ channel is 96 $\mu$b. This number yields an expected
contribution to $\sigma(\gamma p\to p\pi^0\pi^0)$ of 6 $\mu b$ which
perfectly corresponds to the observed value (see
Fig.~\ref{FigureTerms}). However even in the case of the
$N(1680)5/2^+$ state the situation is a more complicated one. The
contribution to $p\pi^0$ reaching a peak value of 14\,$\mu$b, which
after correcting for the $N\pi$ branching fraction of $0.62\pm0.04$
(our value, see Table~\ref{nucleon}) corresponds to the total
contribution 68\,$\mu$b. The dominant intermediate states are
$\Delta(1232)\pi^0$ and $p\sigma$ where $\sigma$ stands for the full
$(\pi\pi)_{\rm S-wave}$. The $\Delta\pi$ channel is later determined
to have a $17$\% branching ratio yielding an expected contribution
of $68 \cdot 17\% \cdot 2/9 = 2.5\,\mu$b. The $N\sigma$ isobar is
observed with a branching ratio of 14\% and a Clebsch-Gordan
coefficient 1/3 yielding an expected contribution to $\sigma(\gamma
p\to p\pi^0\pi^0)$ of 3.2\,$\mu$b in $p\pi^0\pi^0$. These numbers are perfectly
reproduced for the $N(1680)5/2^+$ state when all non-resonant
contributions (including high mass states) are switched off. However,
due to interference with these non-resonant contributions, the full
partial wave reaches only 2.2 $\mu b$ in the $\Delta\pi^0$ and
2.4\,$\mu$b in the $p\sigma$ channels. Further interference between
$\Delta\pi$ and $p\sigma$ reduces the maximum of the total
contribution to 4\,$\mu$b instead of 4.6\,$\mu$b which is expected
from a simple addition of the branching ratios.

The $J^P=3/2^+$ contribution to $p\pi^0$ is of course dominated by
the $\Delta(1232)$ resonance; only its tail is shown here. Above
800\,MeV in $E_\gamma$, the $(I)J^P=(3/2)3/2^+$ contribution exceeds
the total contribution from both isospins: there is destructive
interference between the two isospin components. The
$(I)J^P=(3/2)3/2^+$ contribution is larger than the \,corresponding
\,contribution \,from \,isospin \,$1/2$: $\Delta(1920)3/2^+$ has
larger $p\gamma$ and $N\pi$ coupling constants than $N(1900)3/2^+$.
In the reaction $\gamma p\to p\pi^0\pi^0$, the isodoublet
contribution exceeds the isoquartet contribution in the $3/2^+$
partial wave: nucleons with $J^P=3/2^+$ couple more strongly to
$N\pi\pi$ than the corresponding $\Delta$ states. The interference
of the two isospin components changes from destructive to
constructive with increasing photon energy. The two isospin
amplitudes are identified by taking SAPHIR data on
$\Delta^{++}(1232)\pi^-$ \cite{Wu:2005wf} into the global fit.

Remarkable are the contributions from the two $3/2^-$ waves. We first discuss the contributions to single $\pi^0$ production. The narrow $N(1520)3/2^-$ resonance (green, dashed) interferes with the broad $\Delta(1700)3/2^-$ (blue, dashed) leading to a very significant mass shift of the peak in the total $3/2^-$ contribution (red, dashed). In contrast, there is no significant mass shift in the $\gamma p\to p\pi^0\pi^0$ reaction. The high-mass tail of the $3/2^-$ contribution to the $\gamma p\to p\pi^0\pi^0$ cross section (red, solid) shows a rapid fall and is similar in shape to the $N(1520)3/2^-$ contribution, the isospin $I=3/2$ contribution to the $J^P=3/2^-$ wave (blue, solid) is small. Clearly, the distributions would lead to different masses and different widths if the data were fitted with simple Breit-Wigner amplitudes. Both distributions are fitted well with one pole position if background amplitudes and interference with resonances at higher masses are properly taken into account.

\begin{figure*}
\centerline{\includegraphics[width=0.25\textwidth]{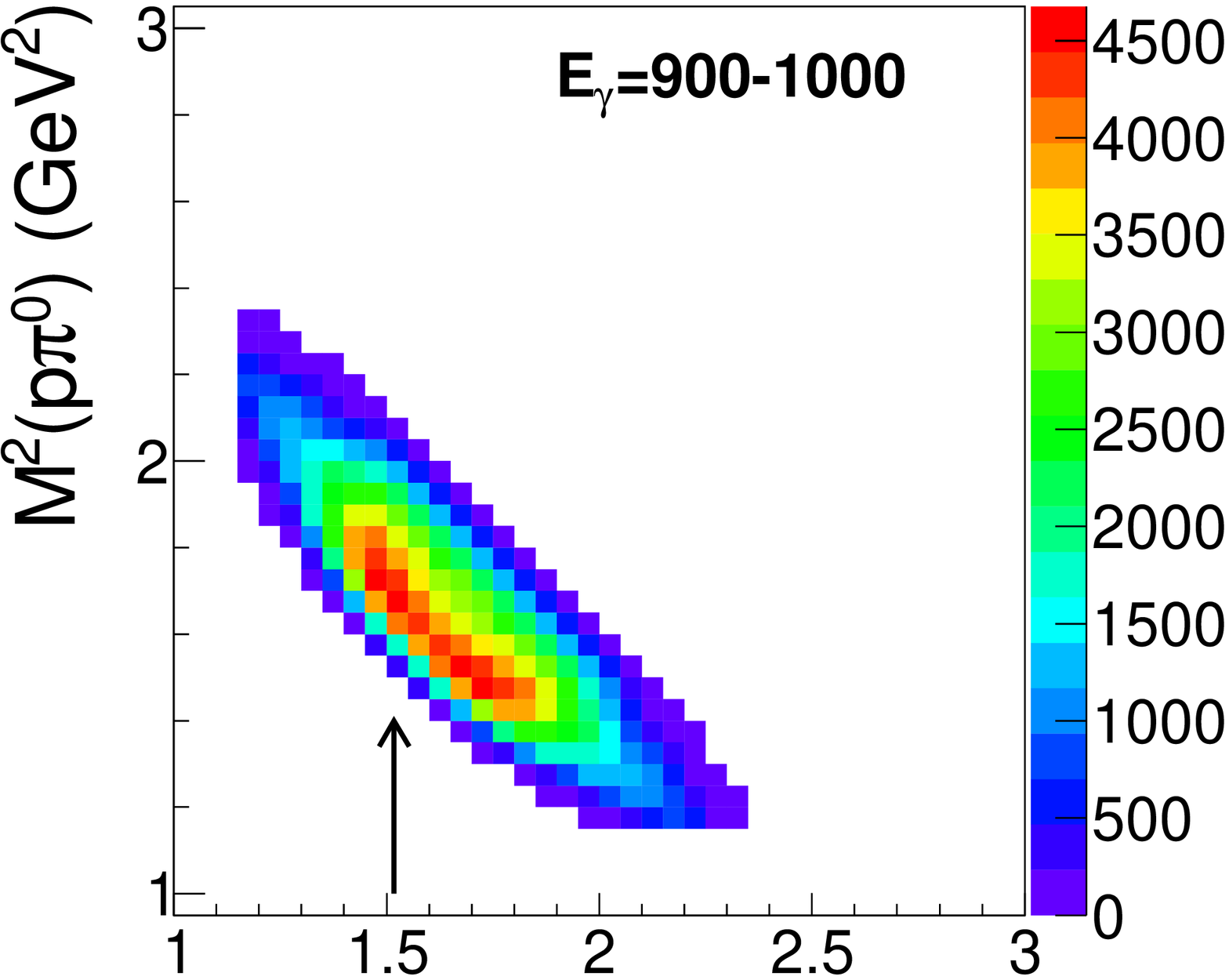}\hspace{-1mm}
            \includegraphics[width=0.25\textwidth]{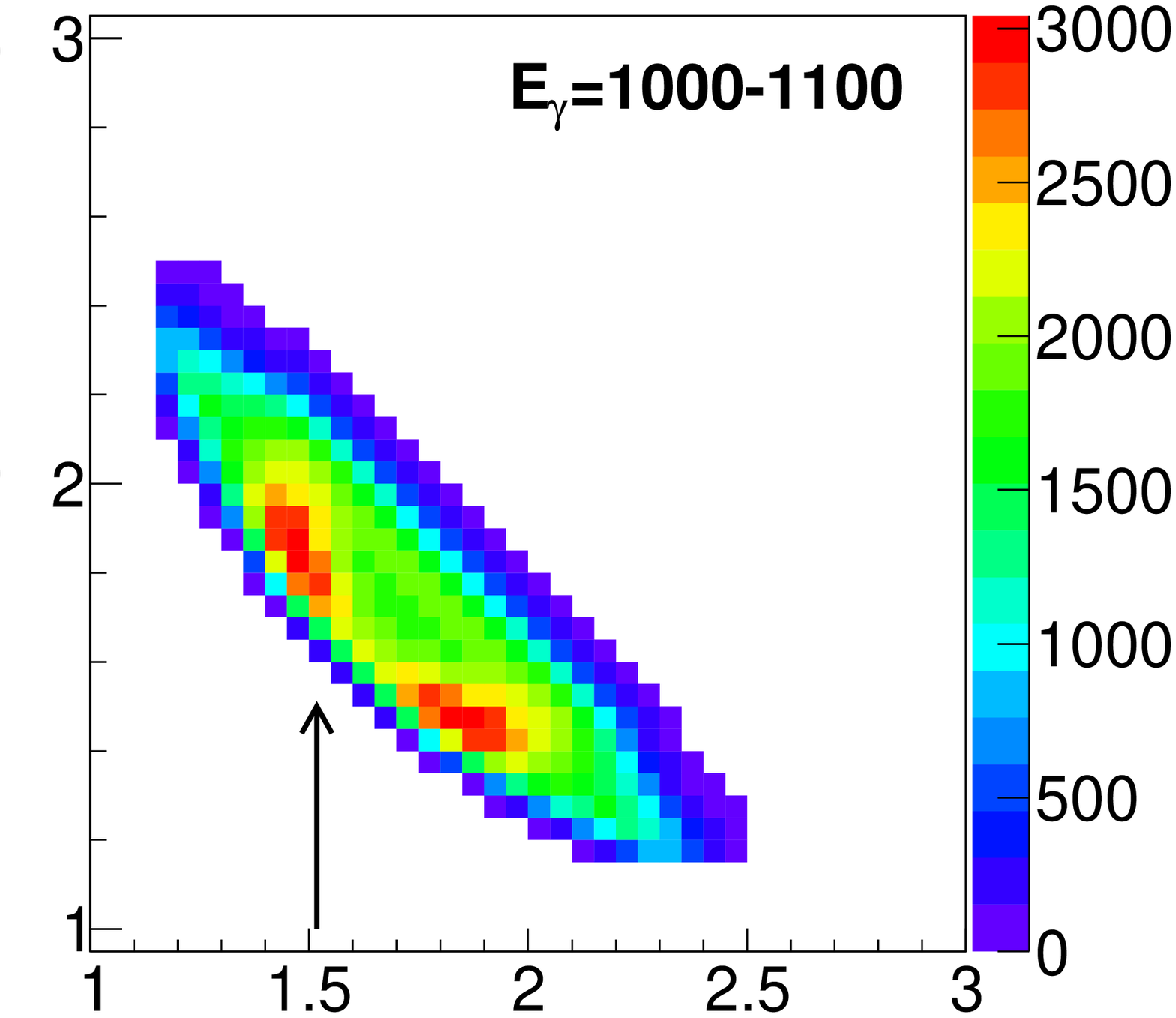}\hspace{-3mm}
            \includegraphics[width=0.25\textwidth]{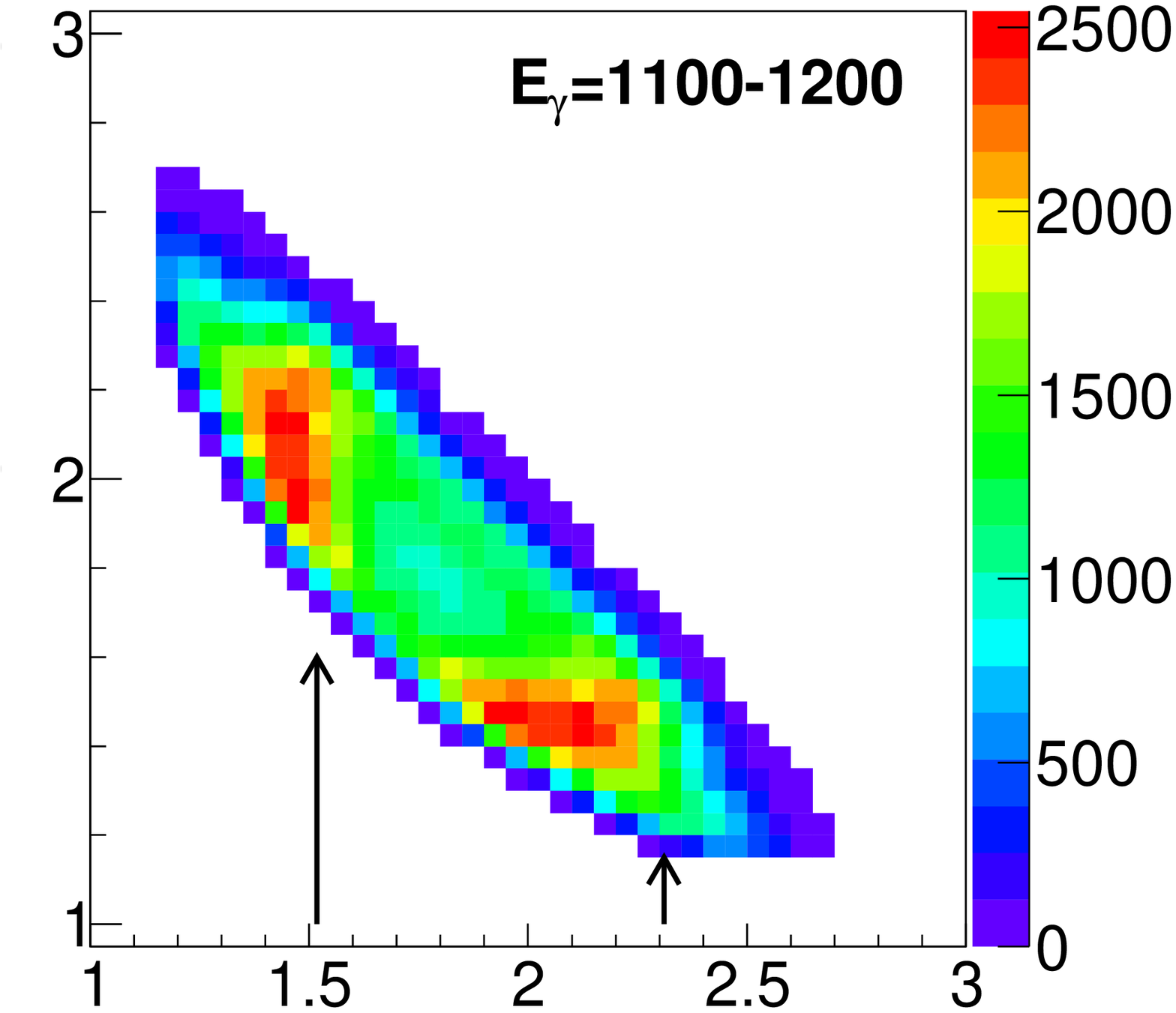}\hspace{-3mm}
            \includegraphics[width=0.25\textwidth]{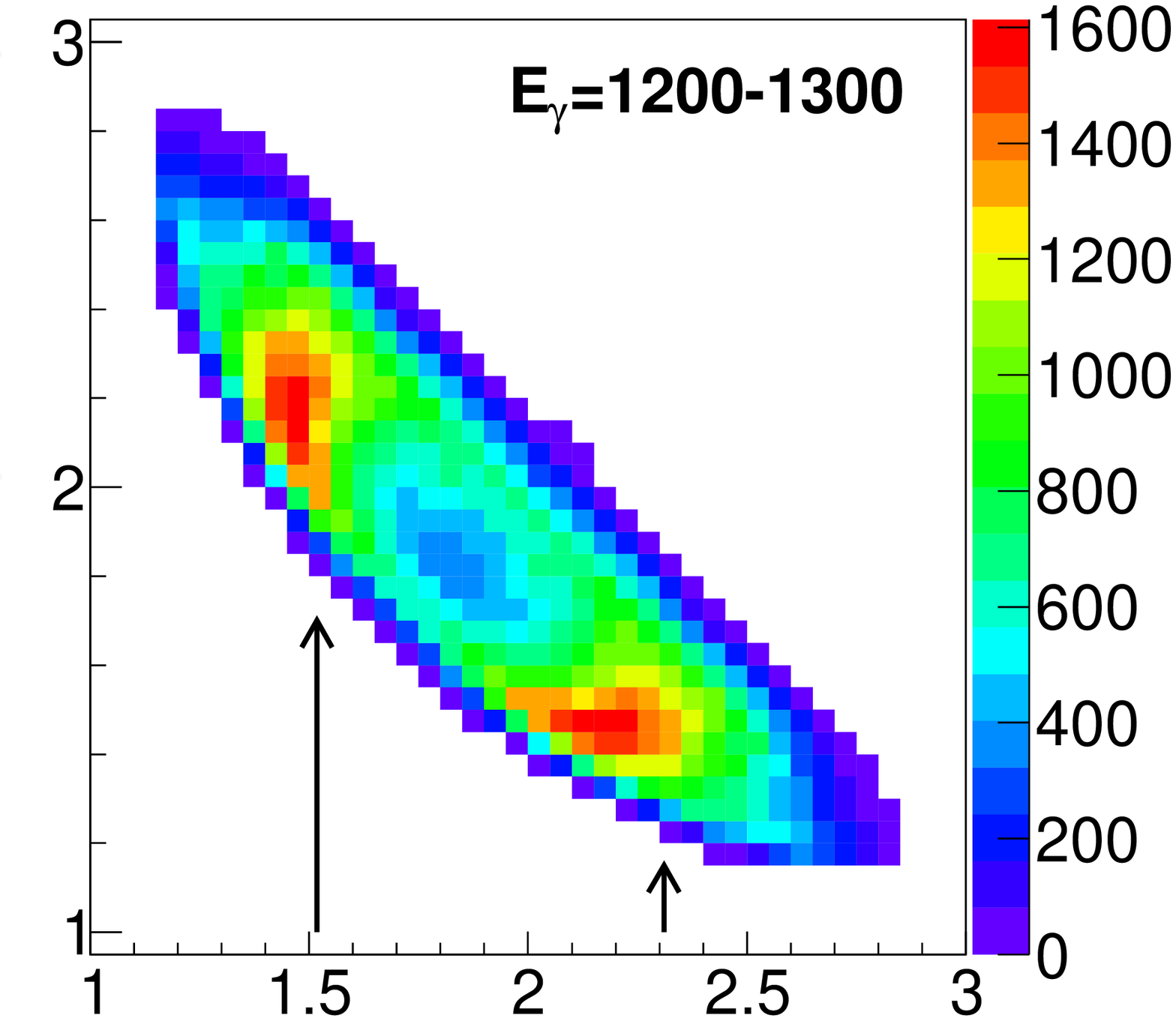}}
\vspace{-3mm}
\centerline{\includegraphics[width=0.25\textwidth]{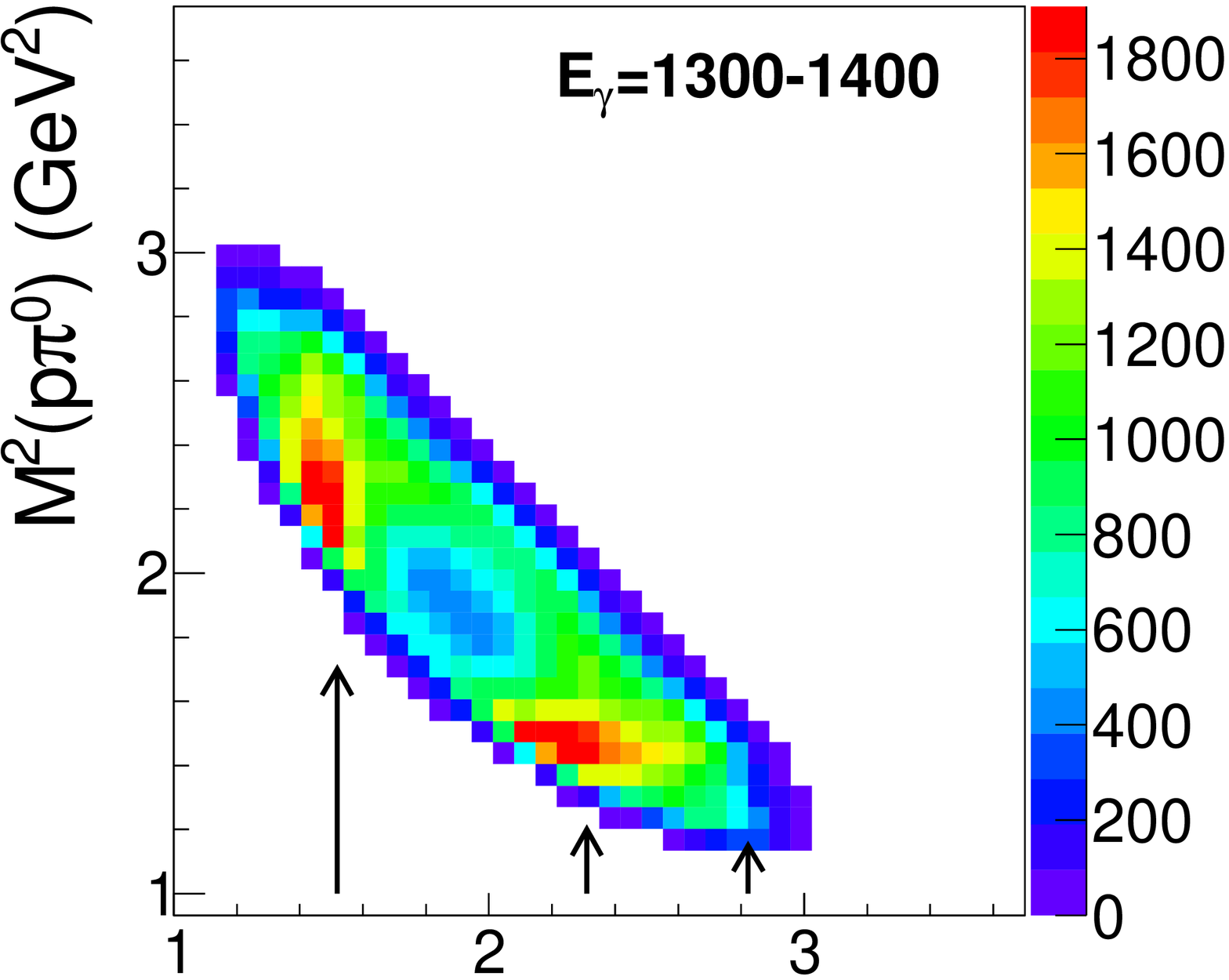}\hspace{-1mm}
            \includegraphics[width=0.25\textwidth]{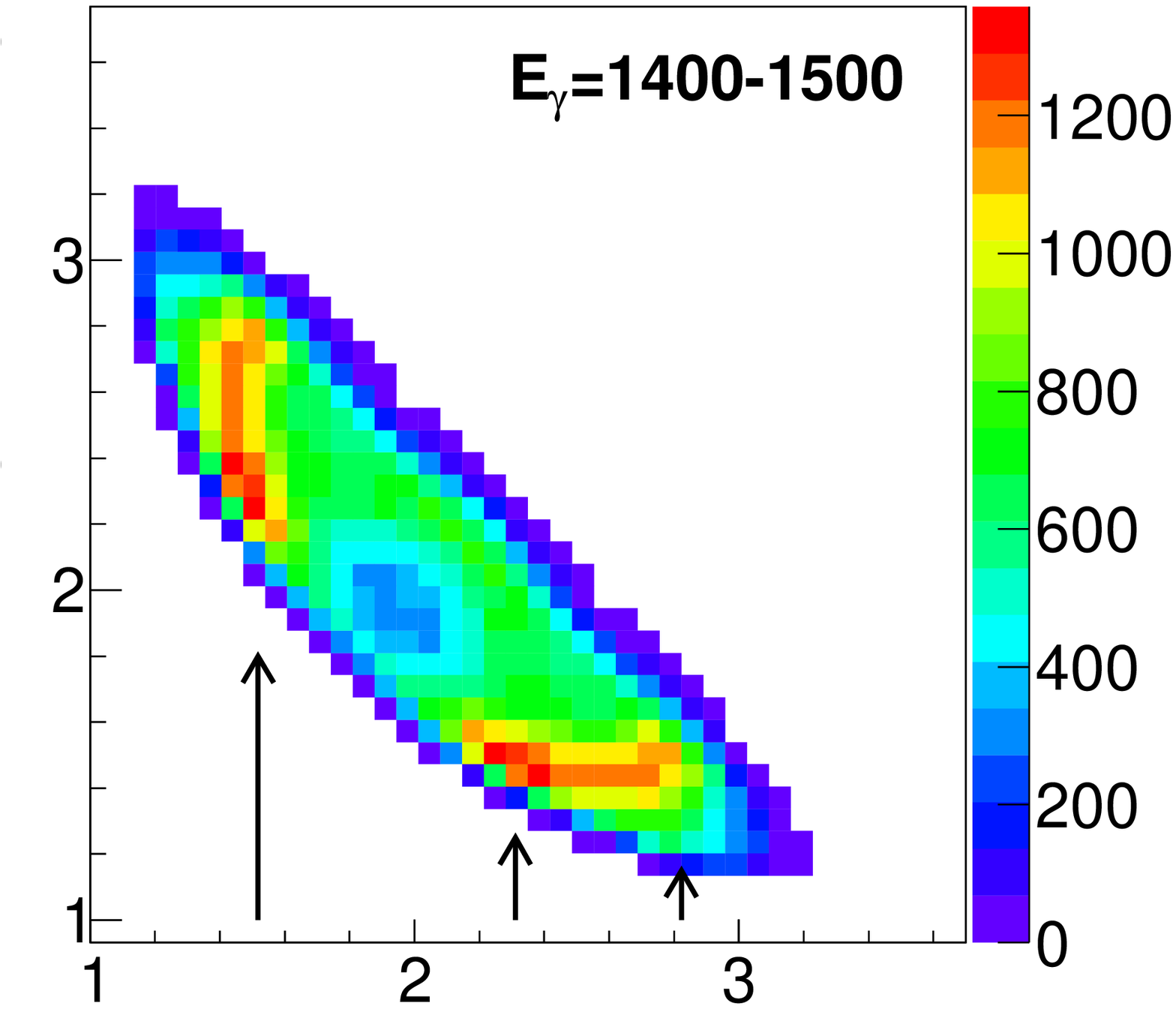}\hspace{-3mm}
            \includegraphics[width=0.25\textwidth]{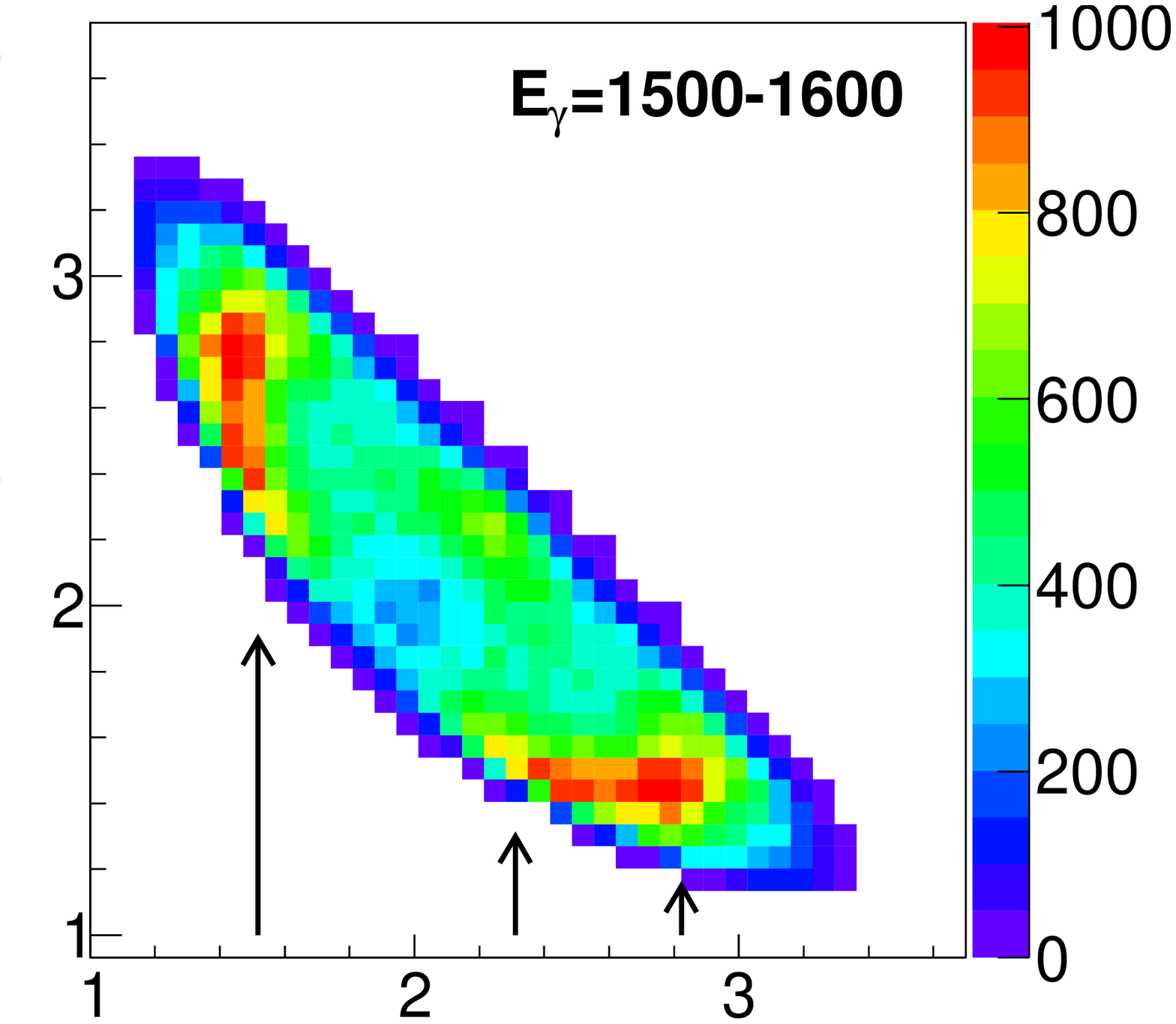}\hspace{-3mm}
            \includegraphics[width=0.25\textwidth]{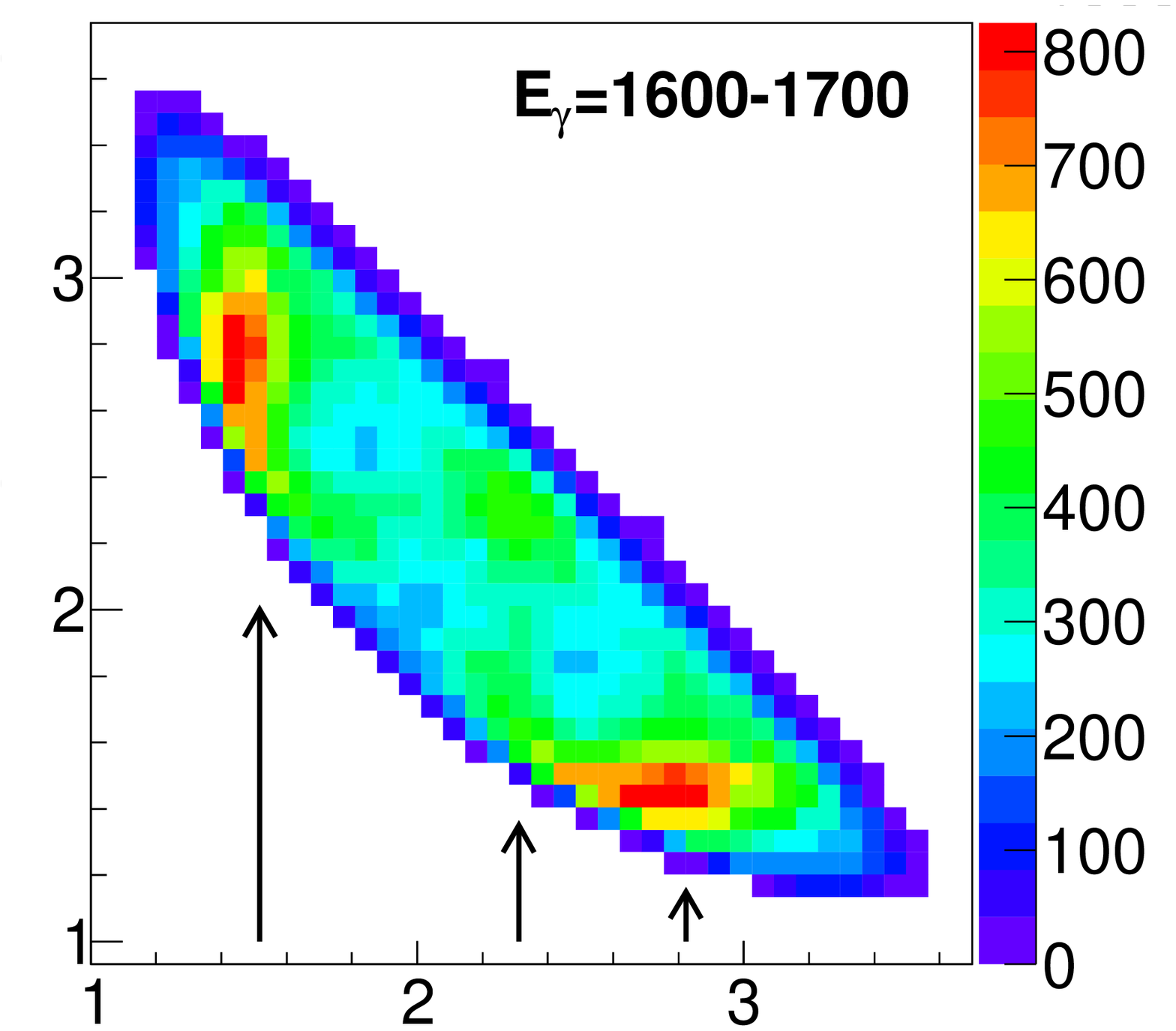}}
\vspace{-3mm}
\centerline{\includegraphics[width=0.25\textwidth]{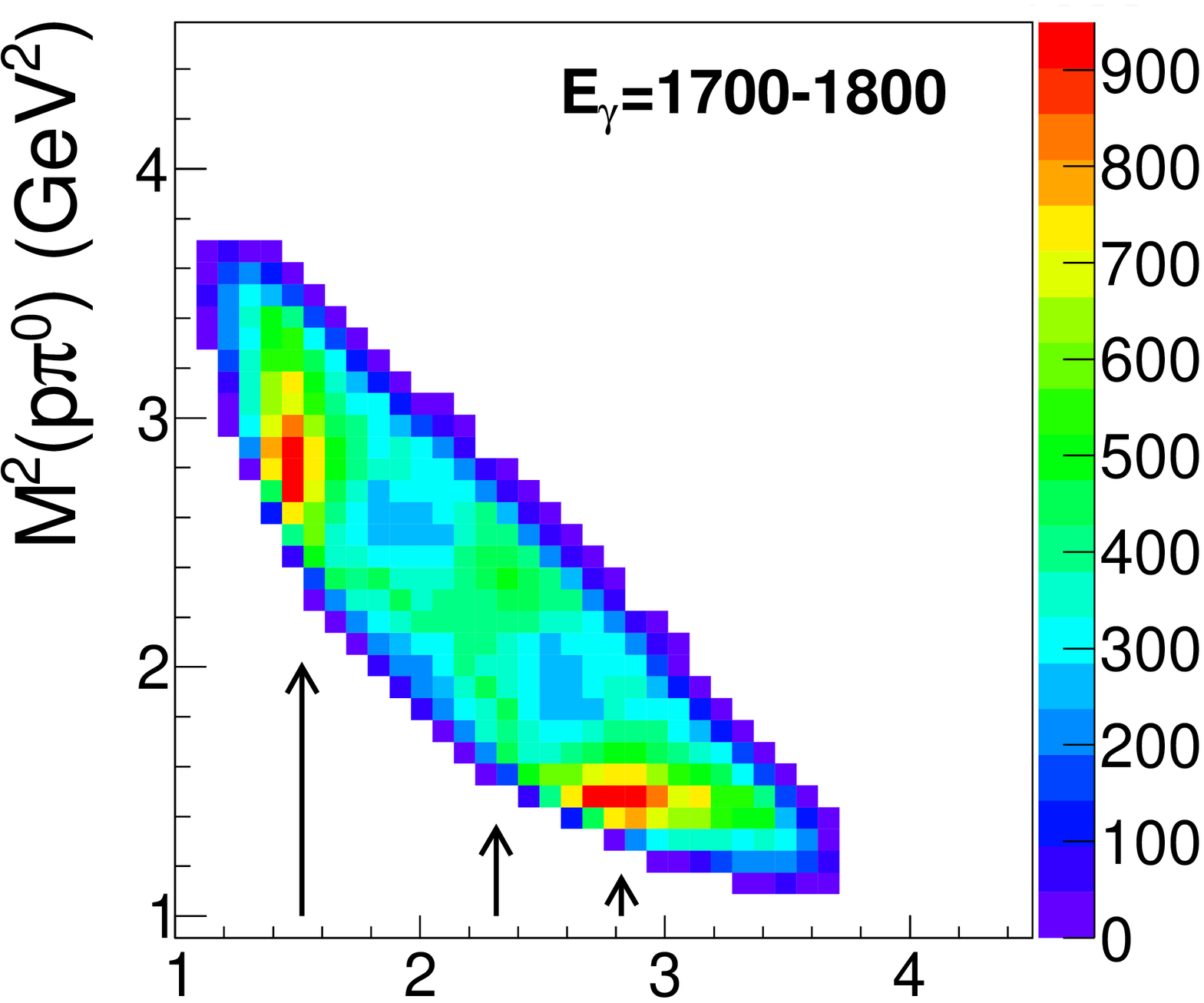}\hspace{-1mm}
            \includegraphics[width=0.25\textwidth]{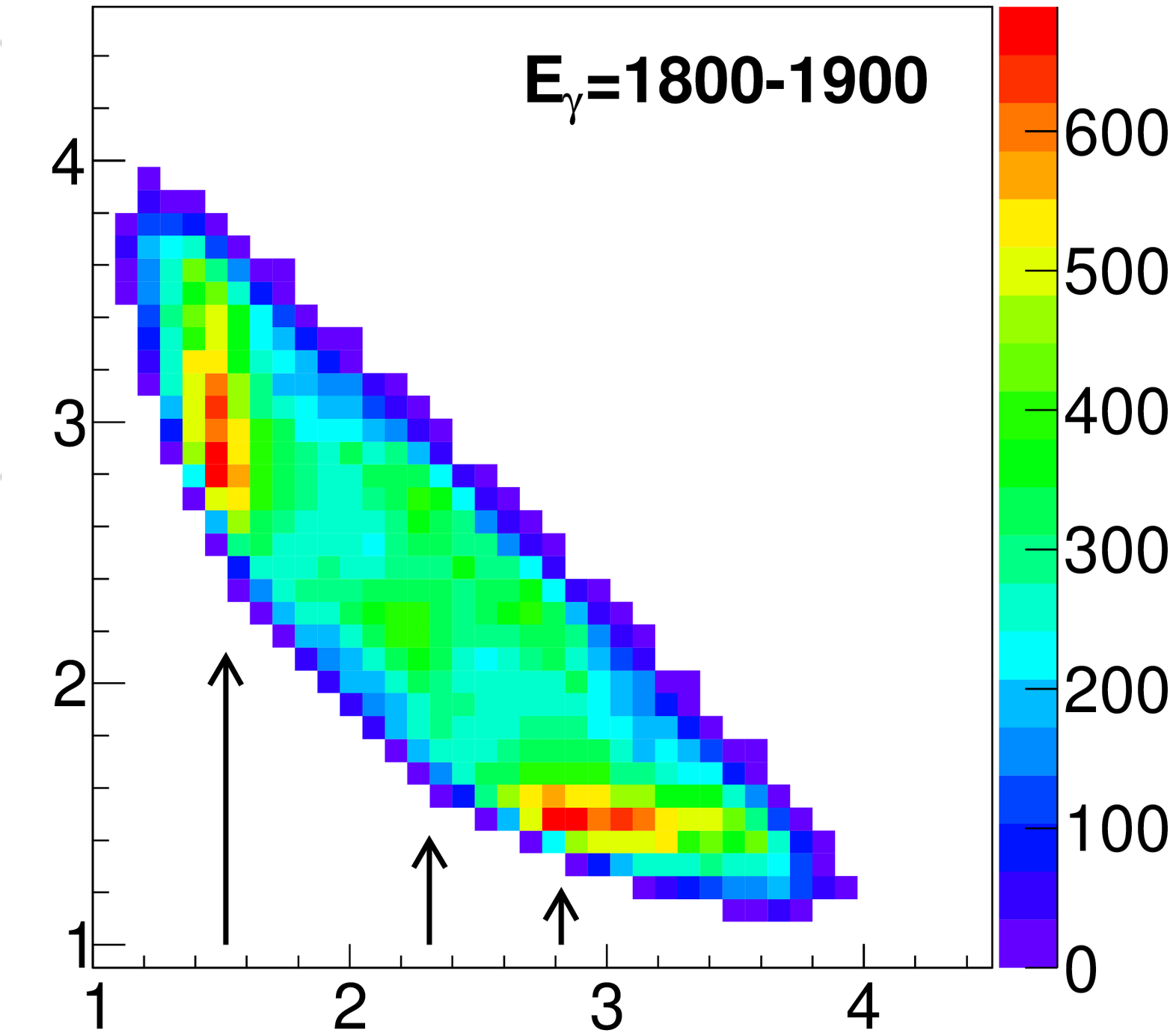}\hspace{-3mm}
            \includegraphics[width=0.25\textwidth]{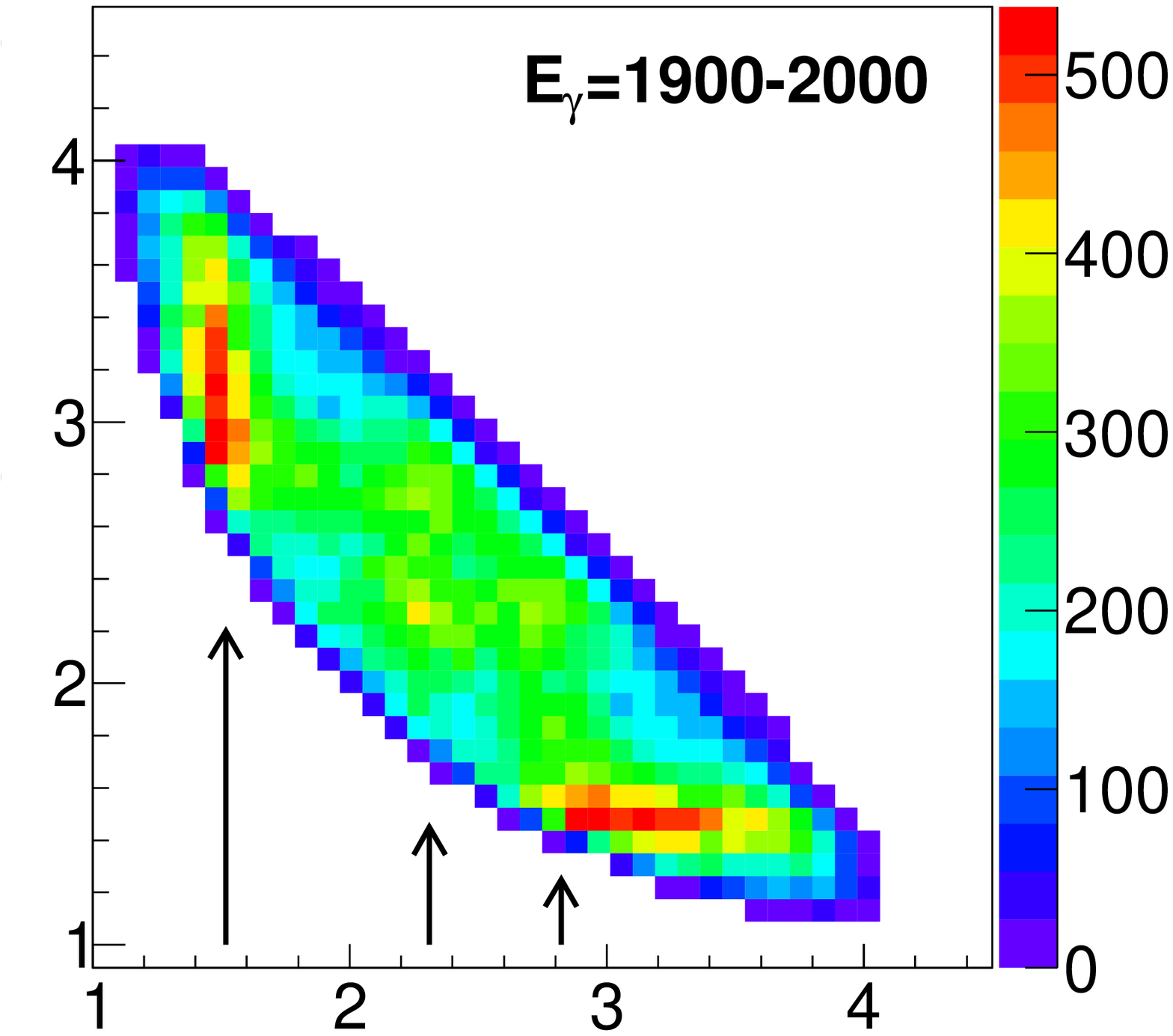}\hspace{-3mm}
            \includegraphics[width=0.25\textwidth]{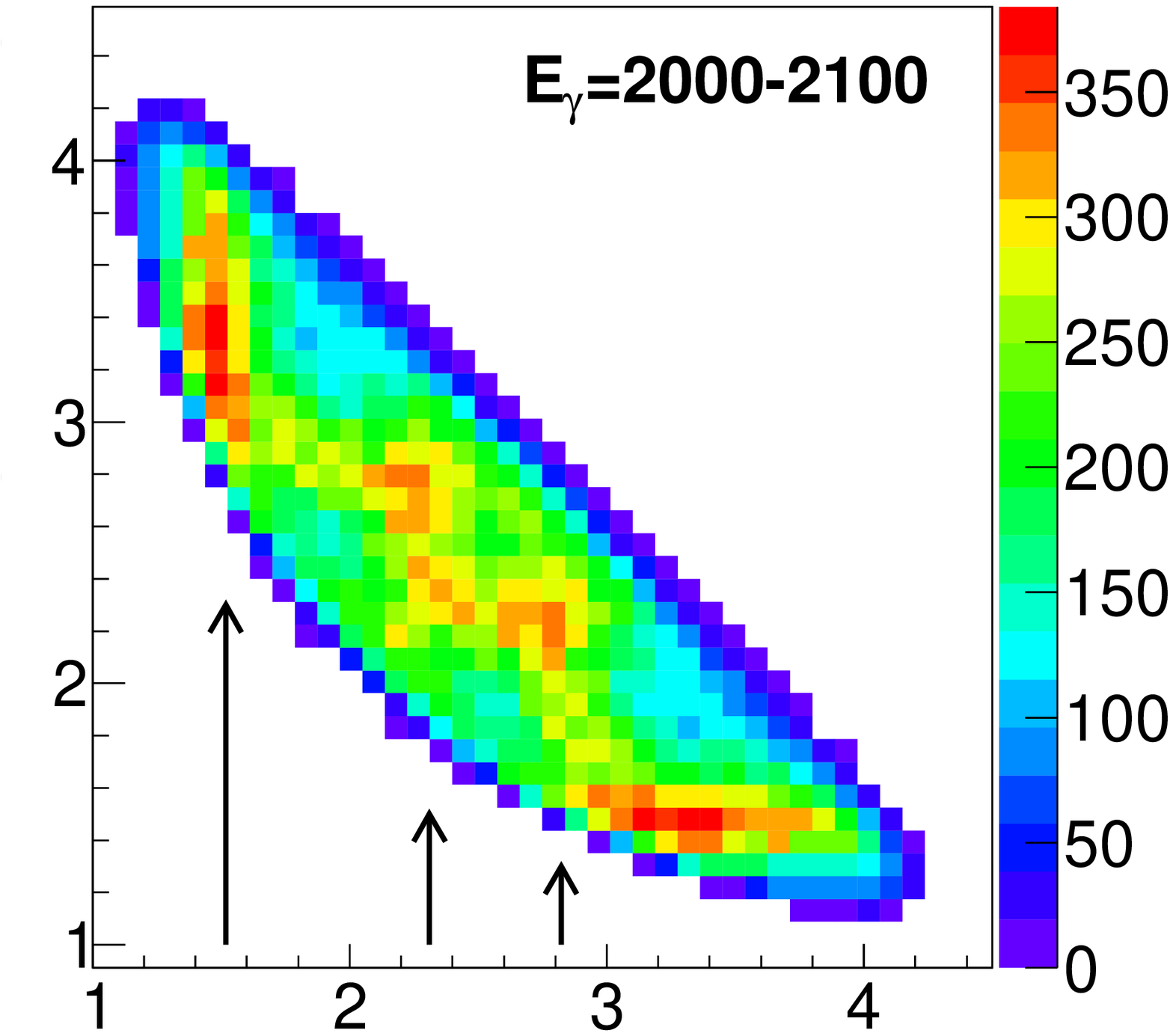}}
\vspace{-3mm}
\centerline{\includegraphics[width=0.25\textwidth]{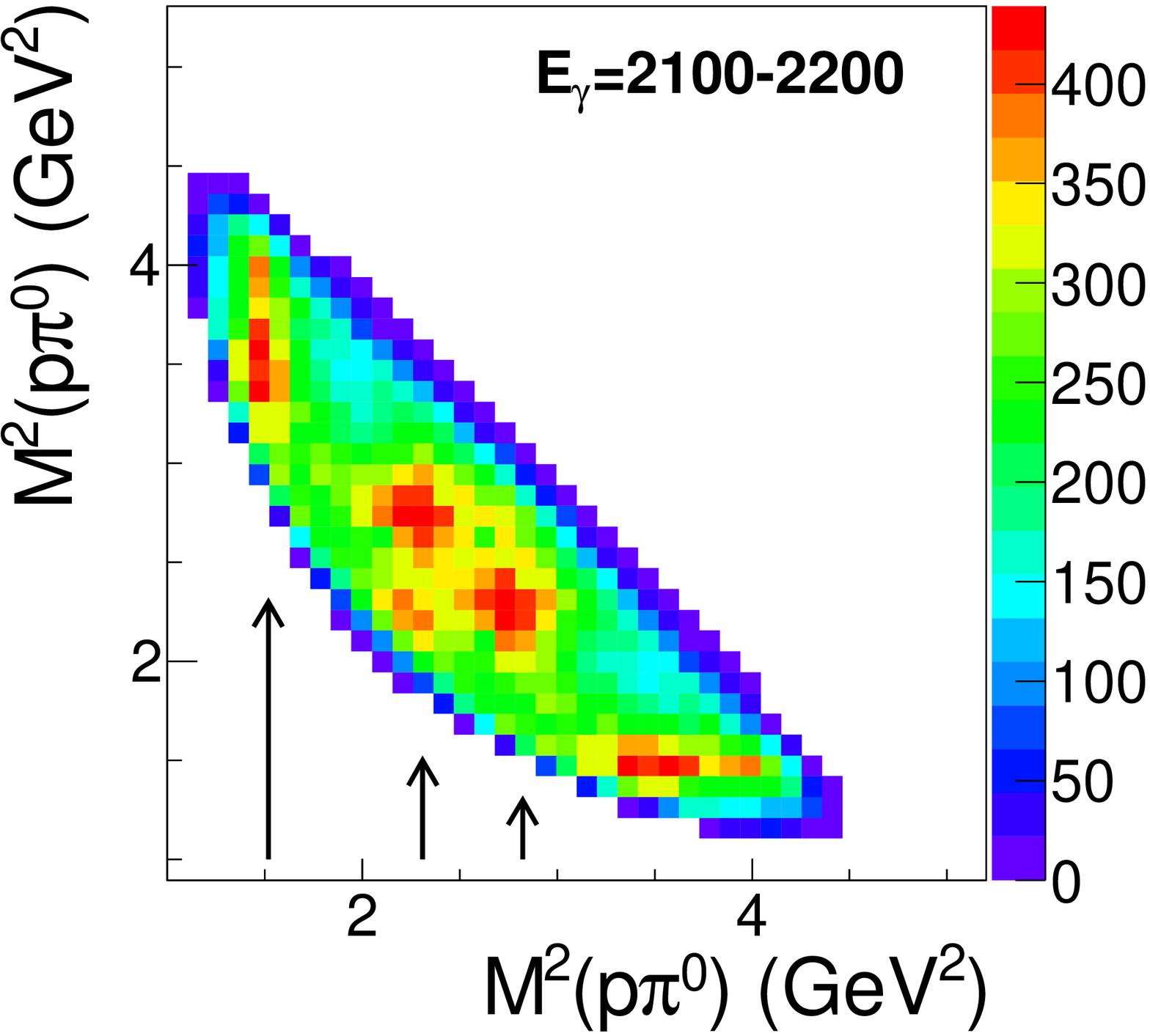}\hspace{-1mm}
            \includegraphics[width=0.25\textwidth]{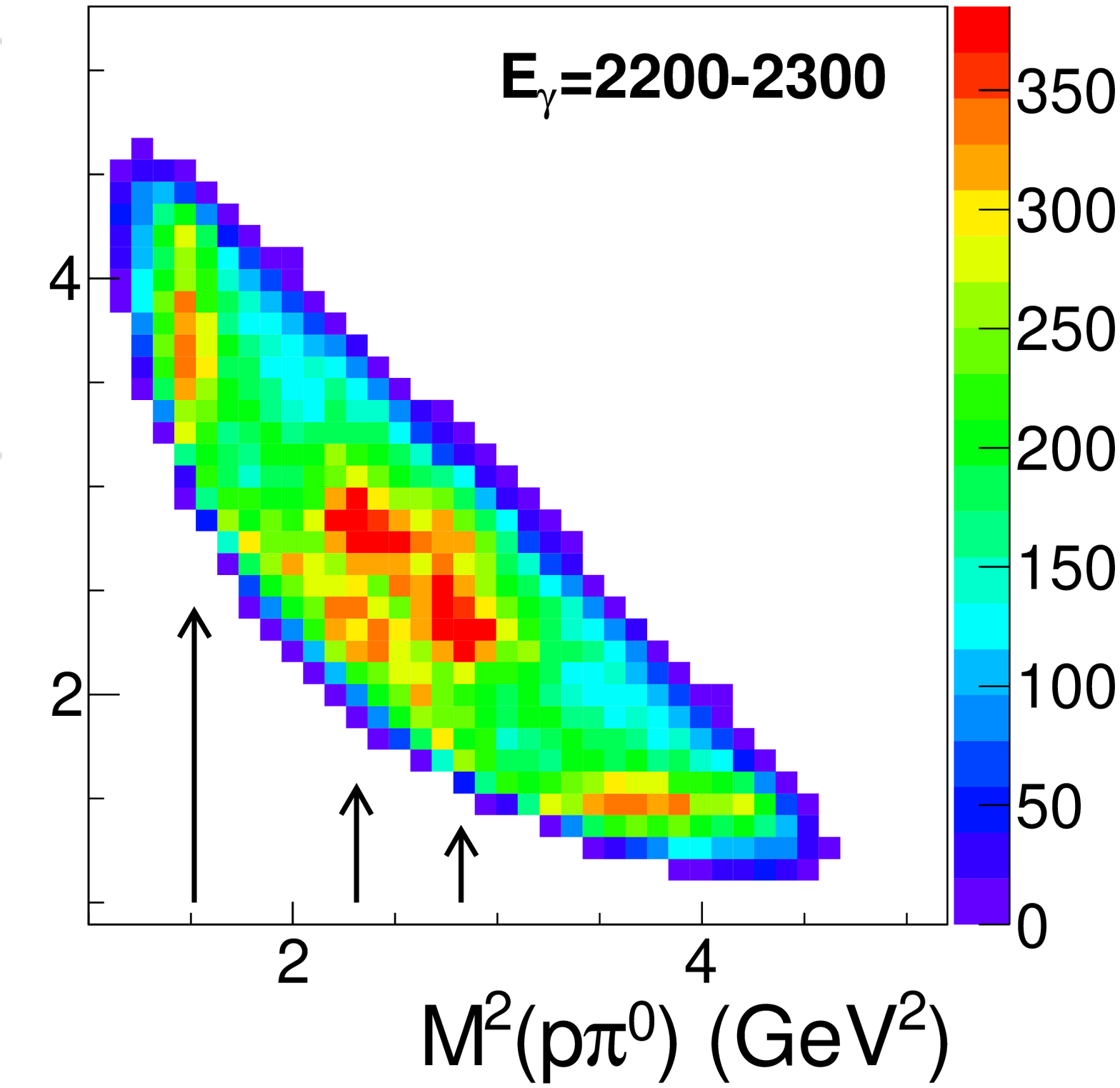}\hspace{-3mm}
            \includegraphics[width=0.25\textwidth]{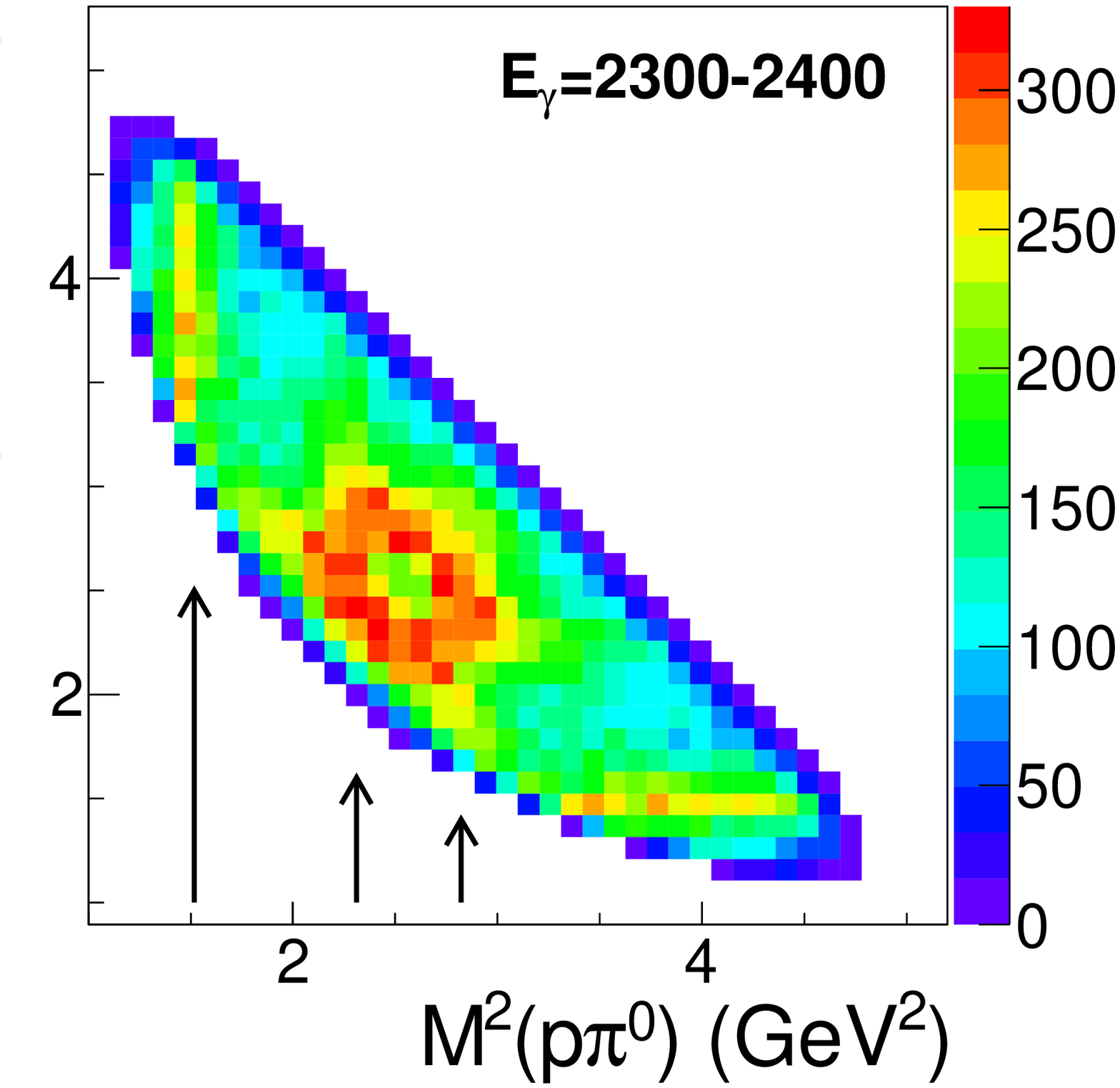}\hspace{-3mm}
            \includegraphics[width=0.25\textwidth]{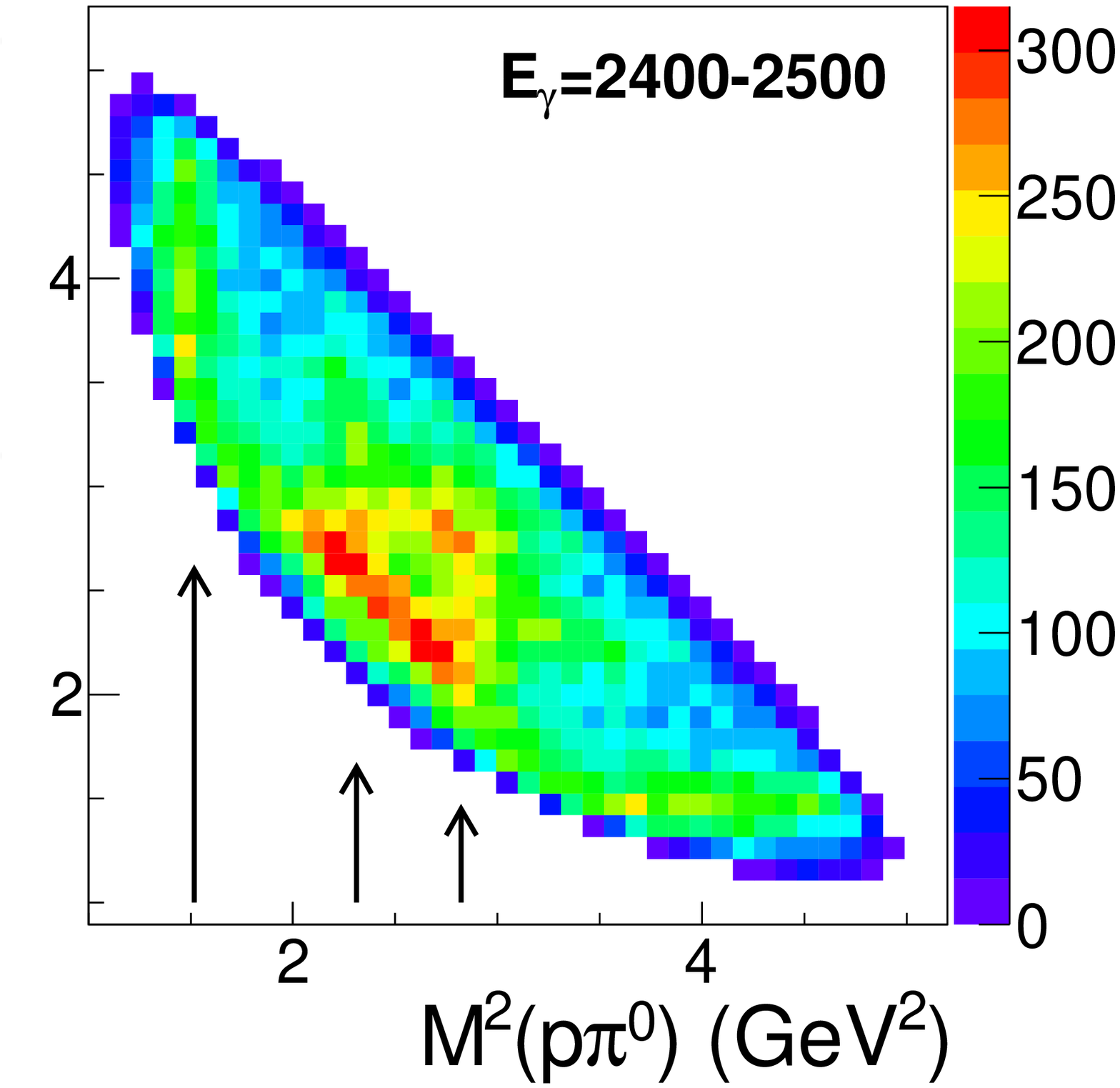}}
\caption{\label{FigureDalitz}Dalitz plots in E$_\gamma$  bins from 900\,MeV to 2500\,MeV. The data are not corrected for acceptance and photon flux. The arrows indicate the positions of $\Delta(1232)$, $N(1520)3/2^-$, and $N(1680)5/2^+$. \vspace{-3mm}}
\end{figure*}
\subsection{\label{Dalitz}Dalitz plots}

Above 1.25\,GeV photon energy, or above 1.8\,GeV invariant mass, the total cross section does not exhibit any significant features. This does not imply that the reaction has no internal energy-dependent dynamics.
Figure~\ref{FigureDalitz-large} shows two Dalitz plots for photons in the range 1500\,MeV $<$ E$_\gamma$ $<$ 1700\,MeV (or for the invariant-mass range from 1.92 - 2.02\,GeV) and for 1800\,MeV $<$ E$_\gamma$ $<$ 2200\,MeV (or 2.07 - 2.24\,GeV in mass). In the Dalitz plots, the squared $p\pi^0_1$ invariant mass formed by choosing one $\pi^0$ is plotted against the squared $p\pi^0_2$ invariant mass with the second pion. The two $\pi^0$ are identical and hence the Dalitz plots are filled with two entries per event. This leads to a symmetry of the Dalitz plots with respect to the diagonal. There is a low-intensity region at the Dalitz plot border. This is an artifact due to the finite energy bin: the outer border of the Dalitz plots is given by the upper limit of $E_\gamma$, the low-intensity region extends into the inner region of the Dalitz plot and is defined by the lower limit of $E_\gamma$.

The highest intensity in both subfigures of Fig.~\ref{FigureDalitz-large} is observed along vertical and horizontal bands at $M^2\approx1.5$\,GeV$^2$ corresponding to the (squared) $\Delta(1232)$ mass. The partial wave analysis assigns the bands to sequential decays of high-lying $N^*$ and $\Delta^*$ resonances decaying via
\begin{eqnarray}
N^{*+}, \Delta^{*+} \to & \Delta^+(1232)\frac32^+ \pi^0 & \to p\pi^0\pi^0\label{1232}
\end{eqnarray}
The intensity along the $\Delta(1232)$ bands in Fig.~\ref{FigureDalitz-large} (left) is not uniform; it increases with increasing mass of the second $p\pi^0$ system. This is due to two reasons: there is additional intensity due to $N(1680)5/2^+$ (to be discussed below). The partial wave analysis also identifies $pf_0(500)$ (or $p\sigma$) as intermediate isobar. The scalar $f_0(500)$ is rather broad; the
$pf_0(500)$ isobar creates additional intensity along the secondary diagonal. In addition, there is a weak pair of en\-hanced-intensity bands at $M^2\approx2.3$\,GeV$^2$, i.e. at the mass of the $N(1520)3/2^-$ resonance. It is assigned to the decay sequence~(\ref{1520}). The two $N(1520)3/2^-$ bands interfere constructively.

\begin{figure*}[pt]
\begin{center}\begin{tabular}{cc}
\hspace{-2mm}\includegraphics[width=0.50\textwidth]{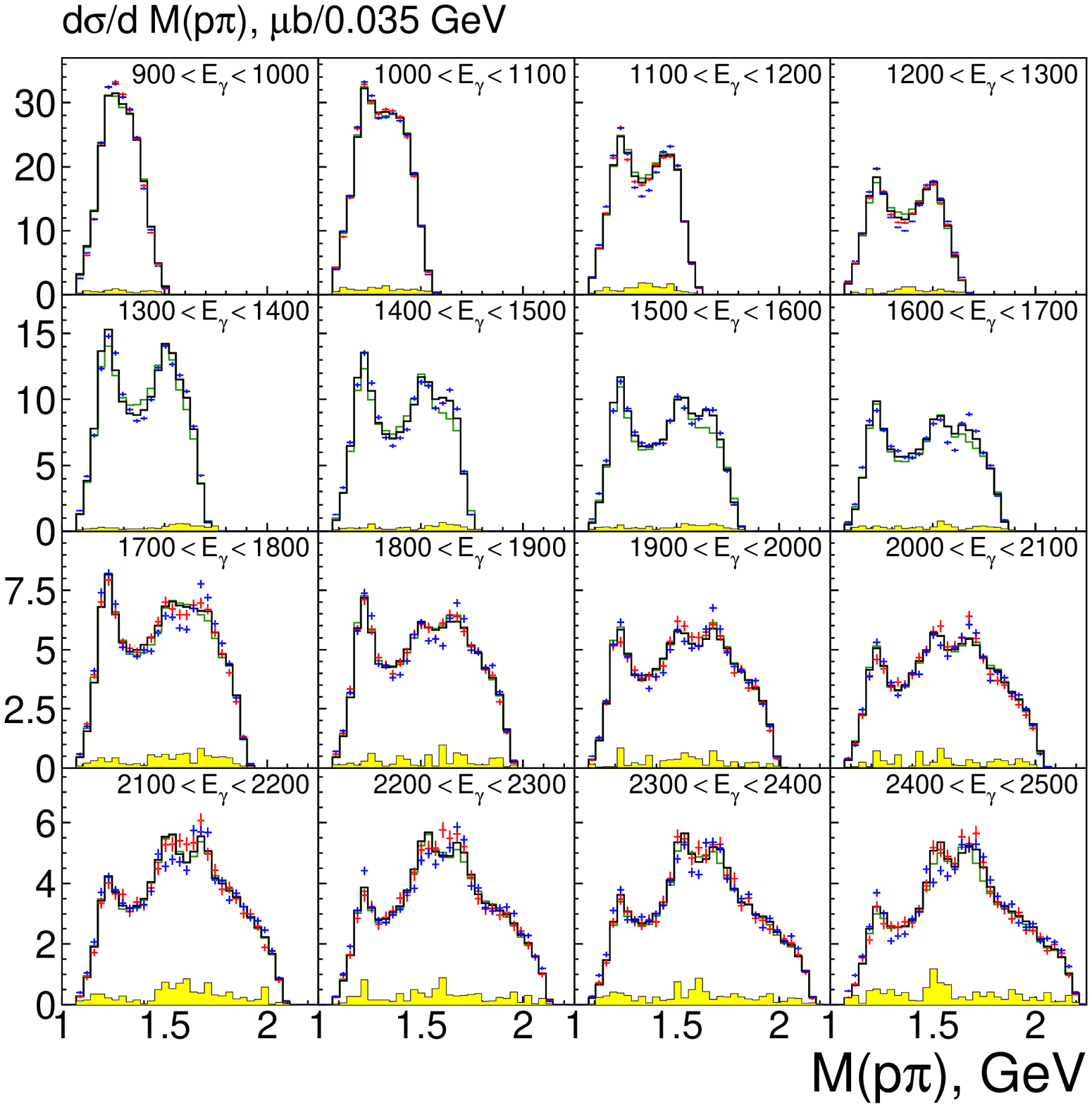}
&
\hspace{-2mm}\includegraphics[width=0.50\textwidth]{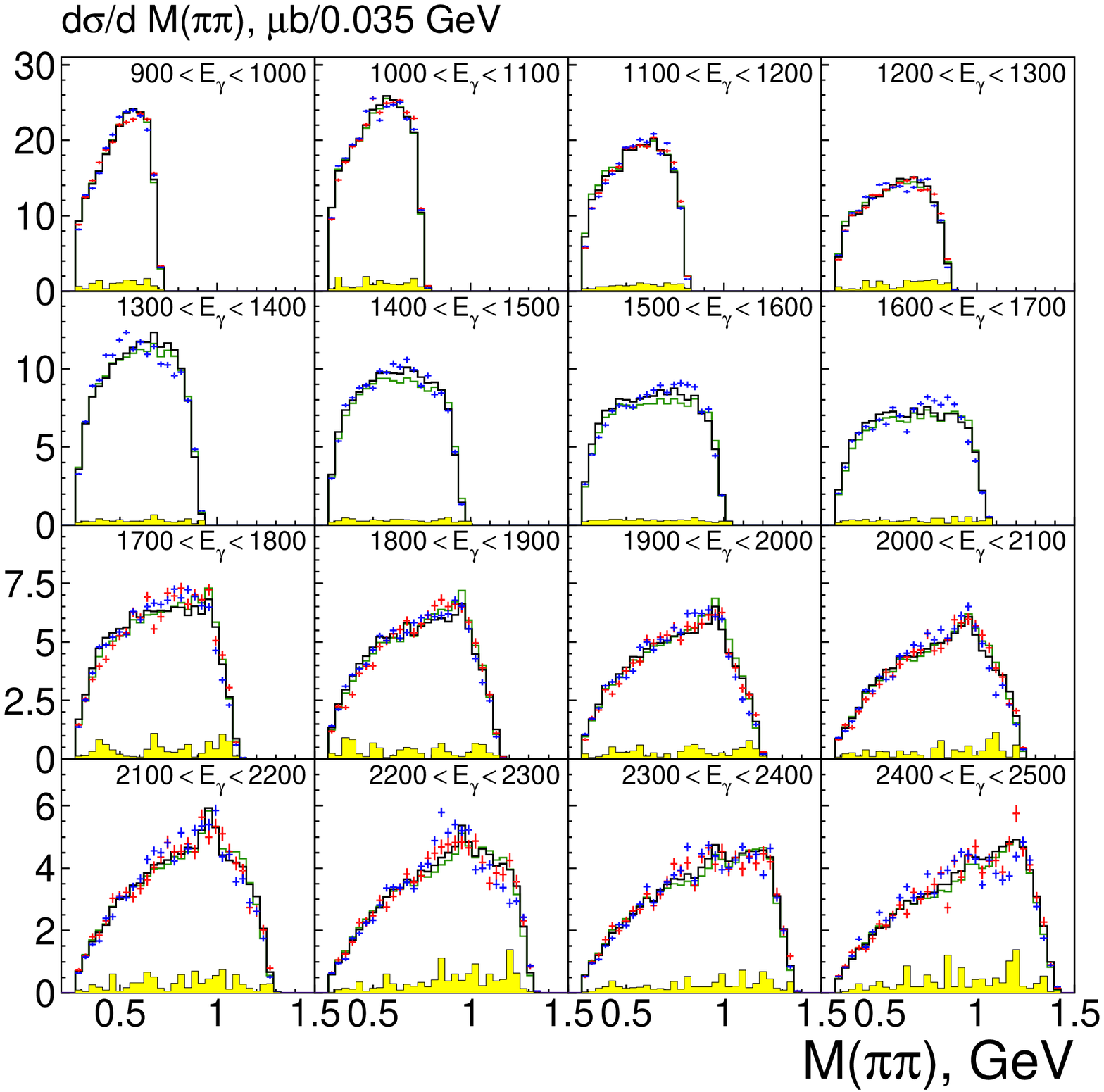}
\end{tabular}\vspace{-3mm}
\end{center}
\caption{\label{mppi-proj}The $p\pi^0$ (left) and
$\pi^0\pi^0$ (right) invariant mass distributions for increasing photon energies, for $E_\gamma= 0.9 $ to $2.5$\,GeV. In the left figure, there are two entries per event. Red and blue crosses: data from CBELSA/TAPS1 and CBELSA/TAPS2. Histograms: BnGa full fit (black), BnGa fit without $N(1900)3/2^+$ (green). The systematic uncertainty is shown as yellow band. In addition, there is a normalization uncertainty of 10\%. \vspace{-3mm}}
\end{figure*}

In Fig.~\ref{FigureDalitz-large} (right) the $\Delta(1232)$ band extends over a wider range, and three additional enhancements turn up. The enhancement on the diagonal is due to the interference of the two $N(1520)3/2^-$ bands; the other two enhancements can be traced to the interference between the amplitudes for $N(1520)3/2^-$ and $N(1680)5/2^+$ production. Thus we have evidence for several sequential decay chains:
 \begin{eqnarray}
N^{*+} \to & N^+(1520)\frac32^- \pi^0 & \to p\pi^0\pi^0\label{1520}\,,\\
N^{*+} \to & N^+(1680)\frac52^+ \pi^0 & \to p\pi^0\pi^0\label{1680}\,,\\
N^{*+} \to & p f_0(500) & \to p\pi^0\pi^0\label{500}\,.
\end{eqnarray}
Surprisingly, the partial wave analysis (Section~\ref{SectionPWA}) assigns these decay chains to $ N^{*}$ resonances and (nearly) not to $\Delta^*$'s. An interpretation of this observation will be given in Section~\ref{Cascades}.

\begin{figure*}[pt]
\begin{center}
\begin{tabular}{cc}
\hspace{-2mm}\includegraphics[width=0.45\textwidth,height=0.42\textwidth]{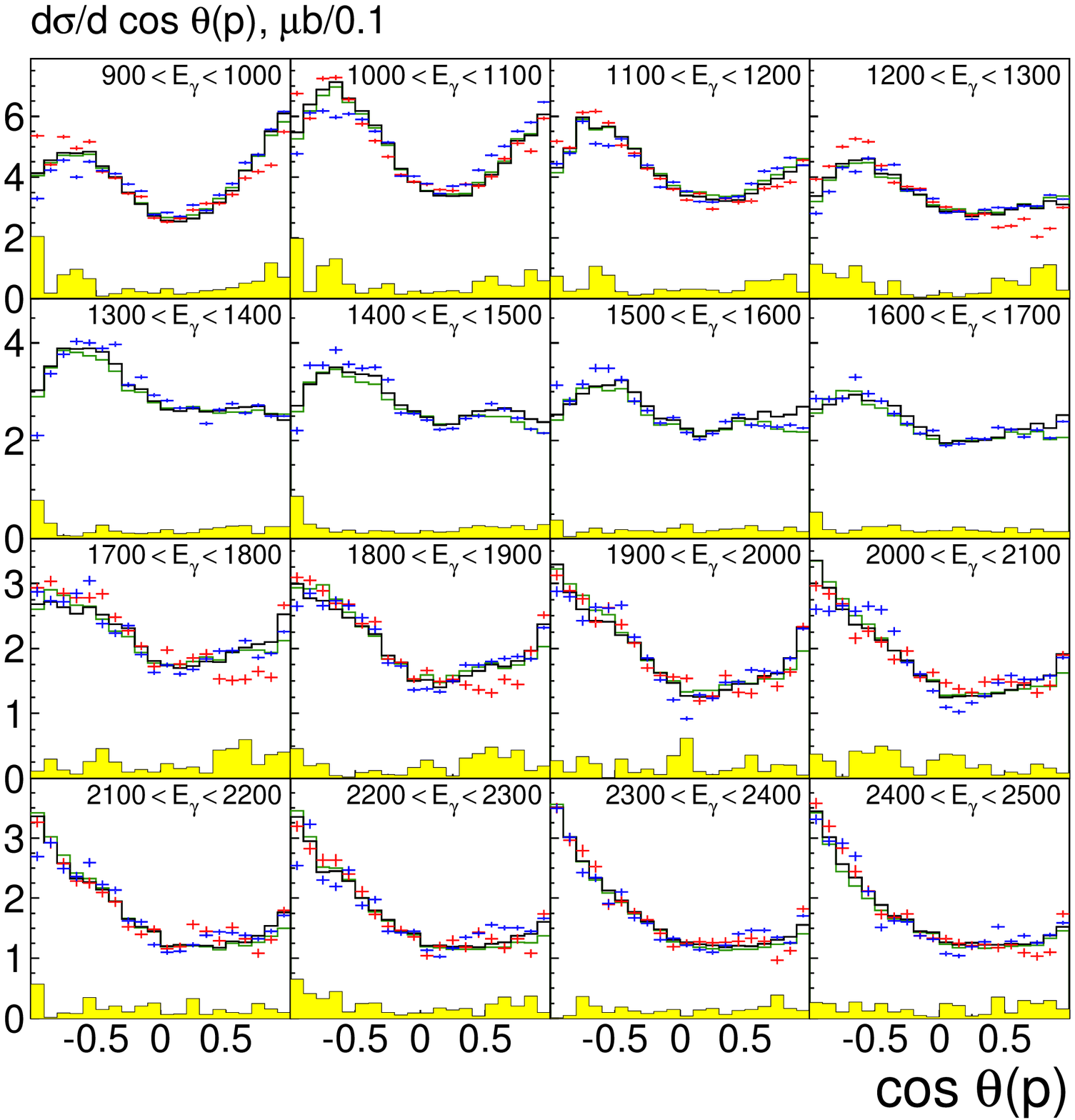}
&
\hspace{-2mm}\includegraphics[width=0.45\textwidth,height=0.42\textwidth]{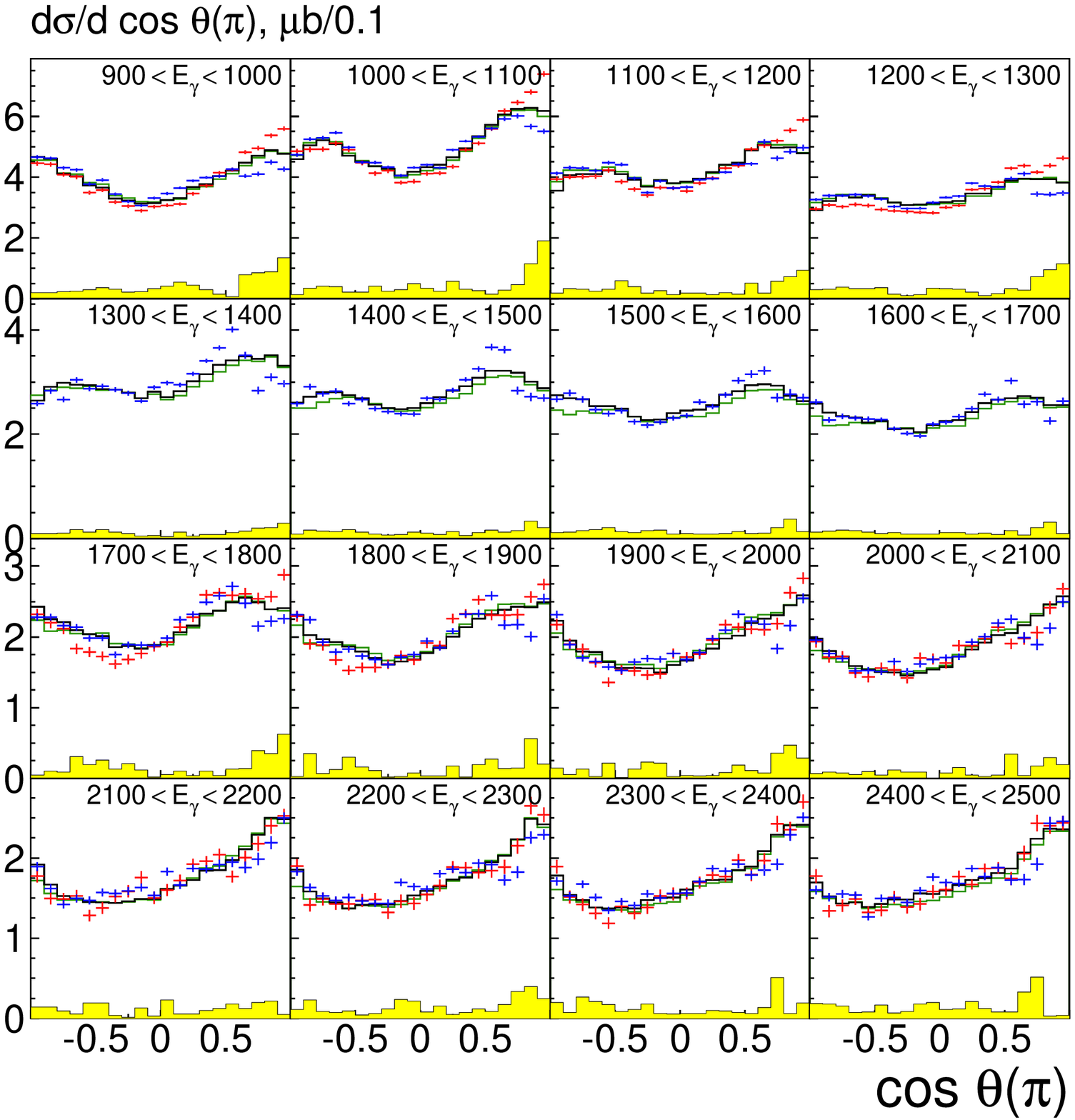}
\end{tabular}
\begin{tabular}{cc}
\hspace{-2mm}\includegraphics[width=0.45\textwidth,height=0.42\textwidth]{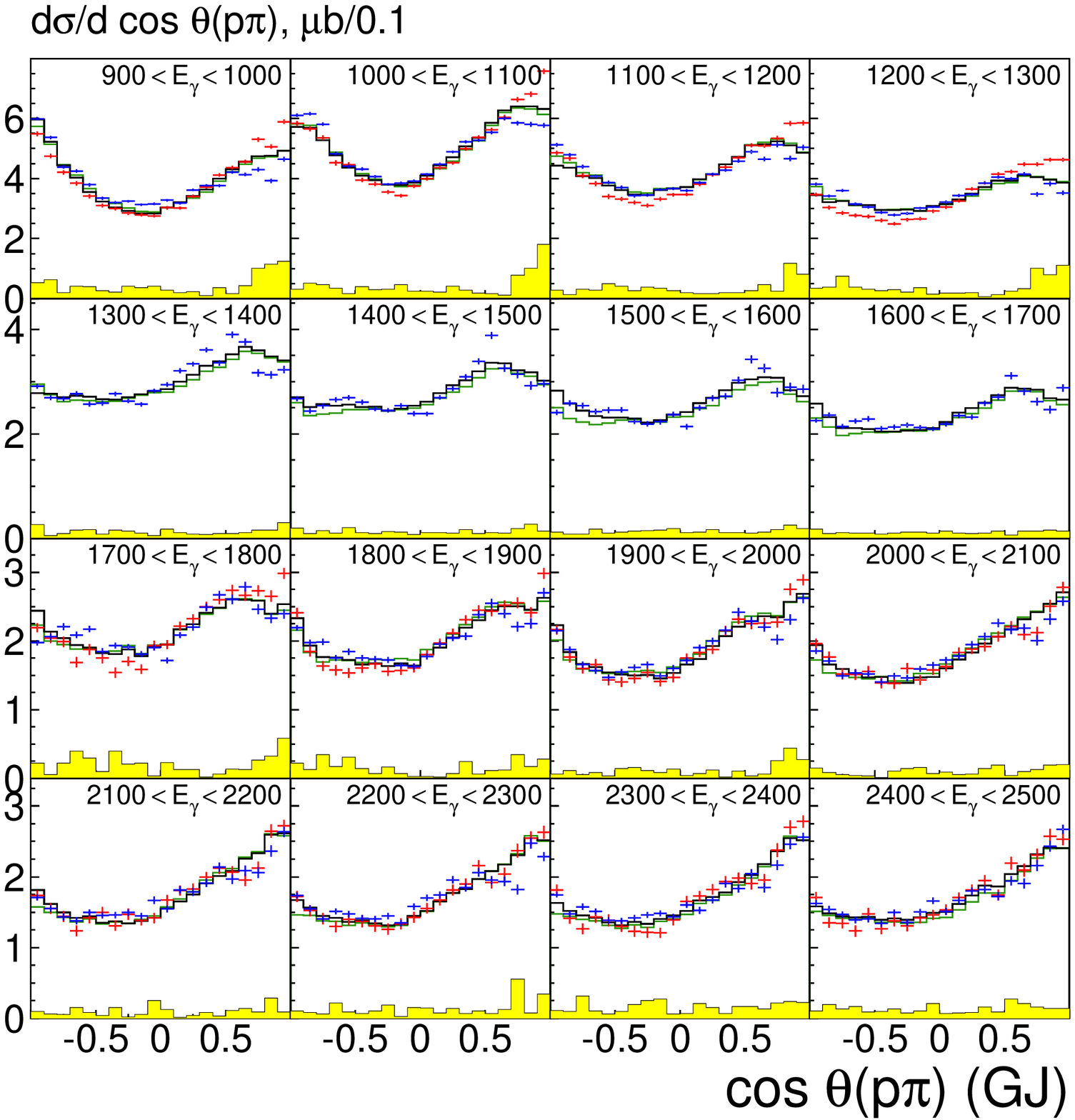}
&
\hspace{-2mm}\includegraphics[width=0.45\textwidth,height=0.42\textwidth]{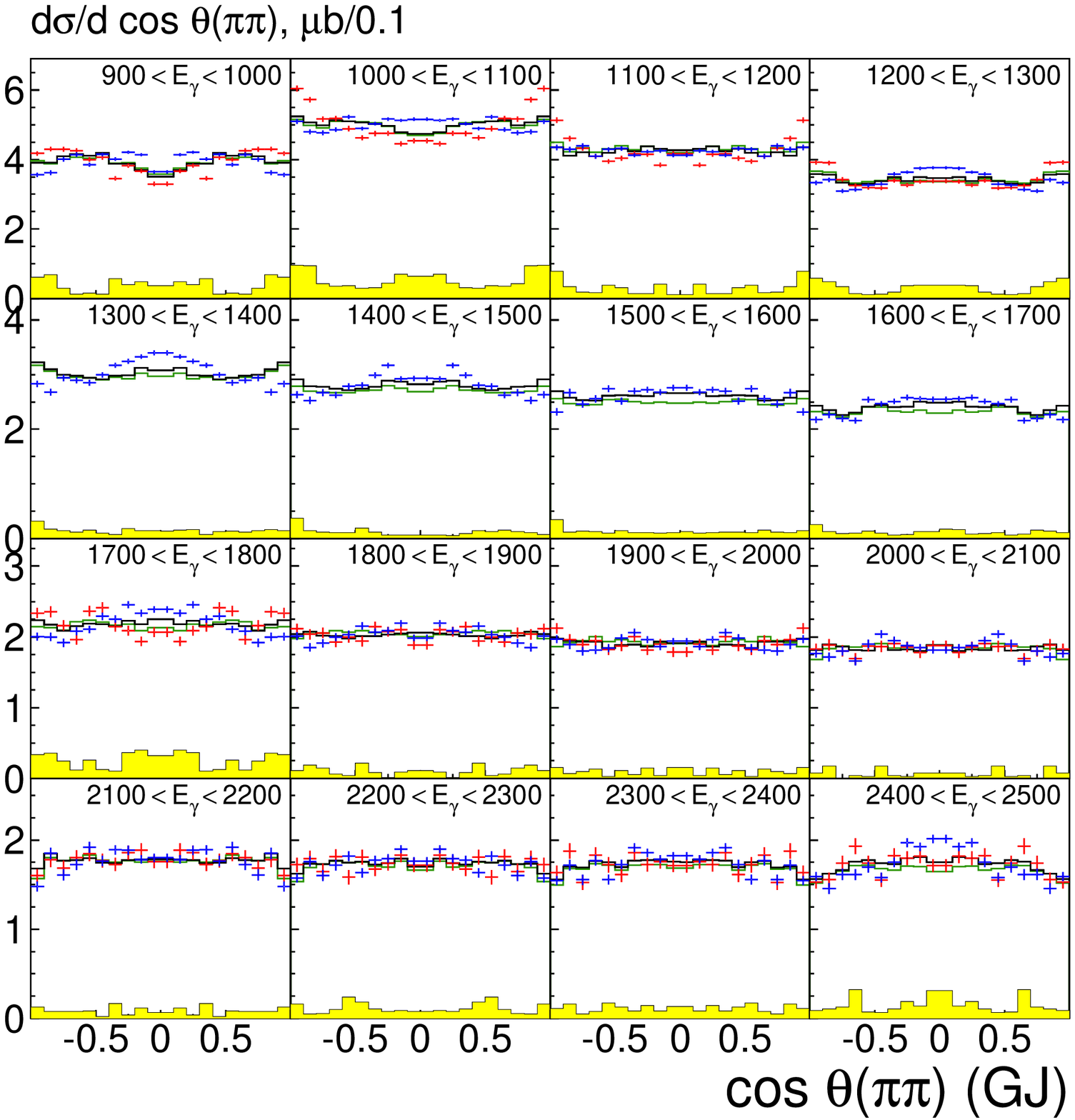}
\end{tabular}
\begin{tabular}{cc}
\hspace{-2mm}\includegraphics[width=0.45\textwidth,height=0.42\textwidth]{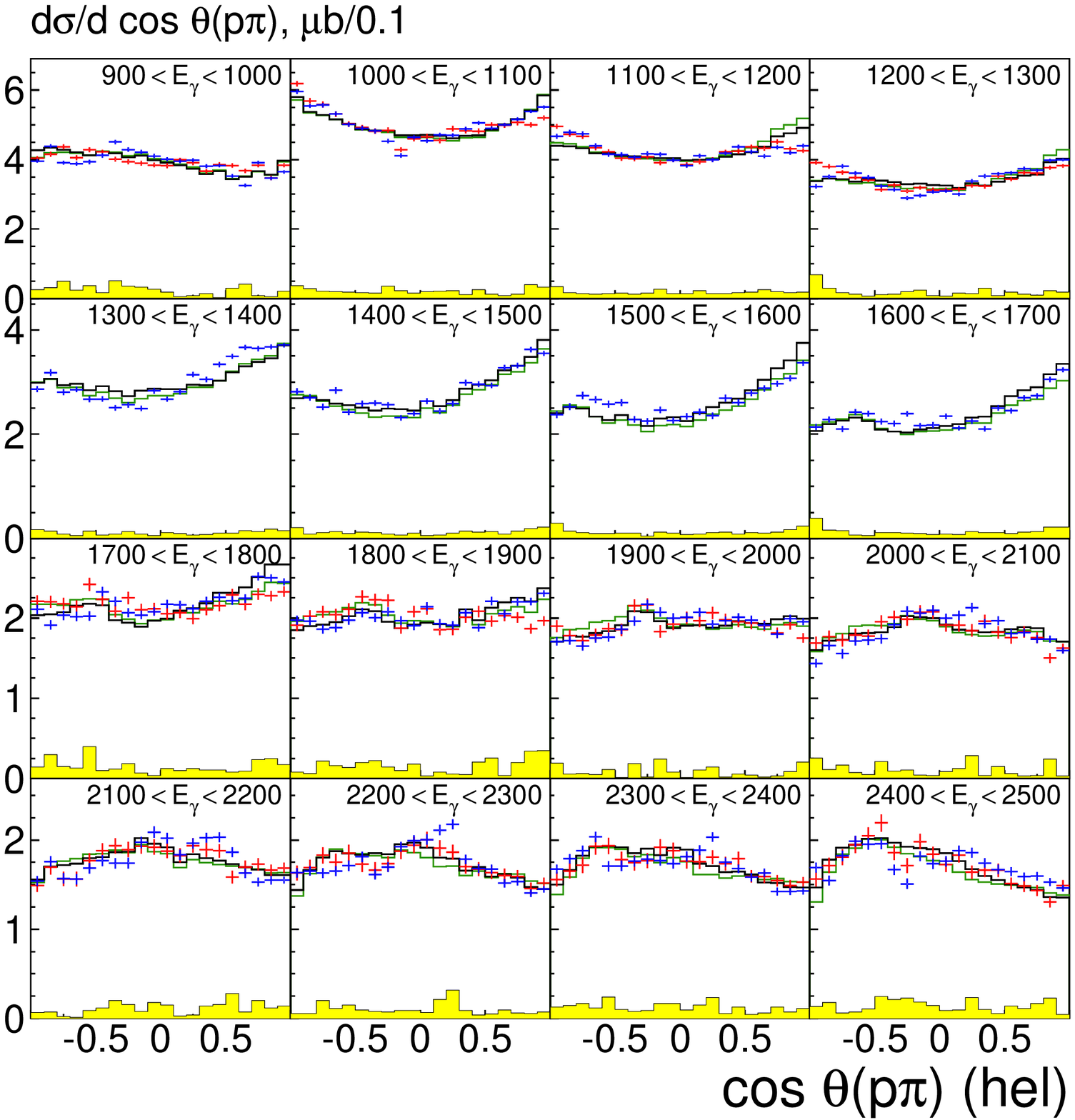}
&
\hspace{-2mm}\includegraphics[width=0.45\textwidth,height=0.42\textwidth]{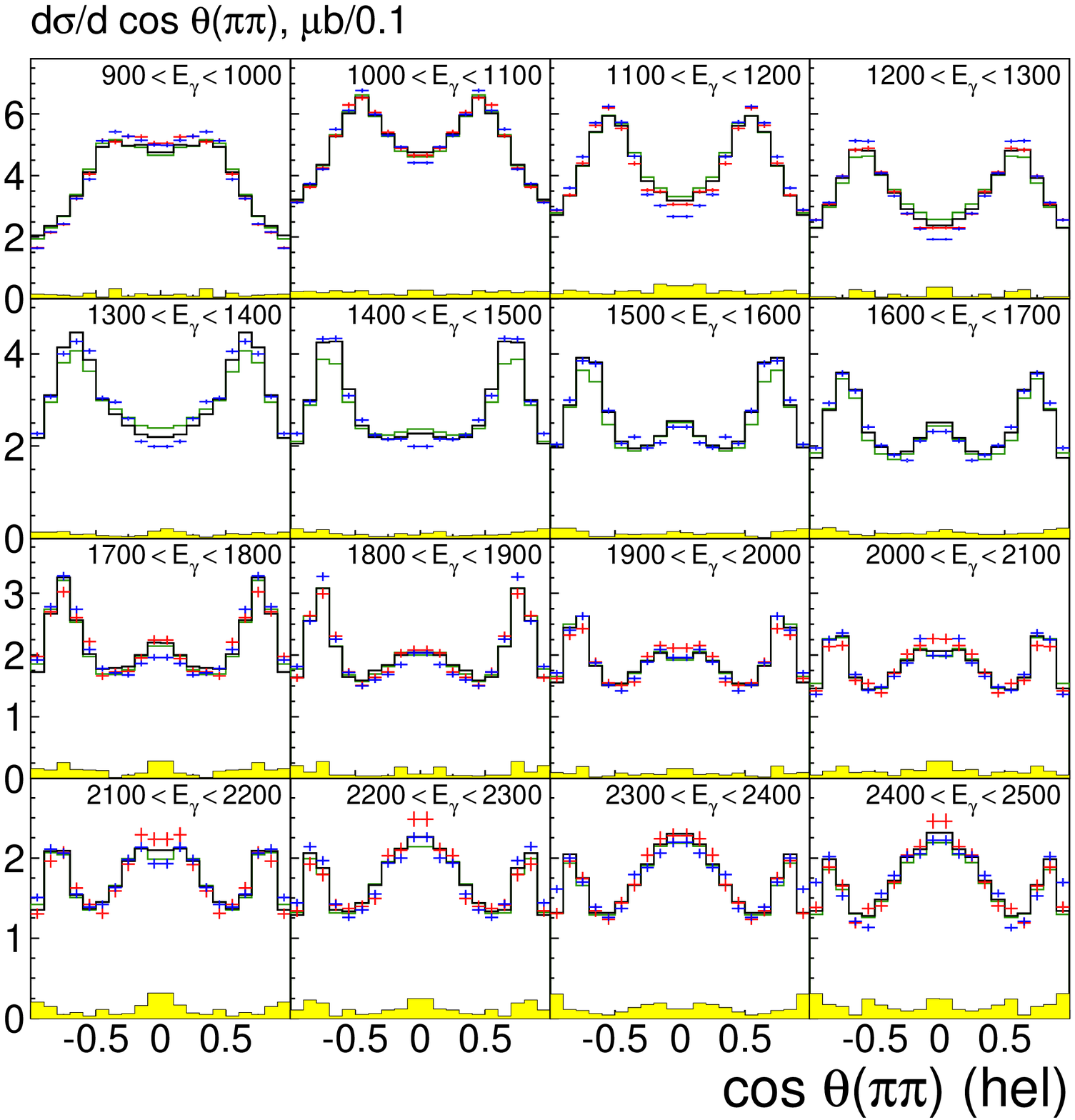}
\end{tabular}\vspace{-3mm}
\end{center}
\caption{\label{cosp}Angular distributions of the proton (left) and pions (right) with respect to the photon beam (top), in the Gottfried-Jackson frame (middle) and in the helicity frame (bootom). The systematic uncertainty is shown as yellow band.  See caption of Fig.~\ref{mppi-proj} for symbols.\vspace{-3mm}}
\end{figure*}

Figure~\ref{FigureDalitz} shows a series of Dalitz plots for increasing
values of the photon energy. (The Dalitz plots for the lowest energy bins are featureless and not shown here.) In the first Dalitz plot, an enhancement
is observed which stretches along the secondary diagonal. It is, however,
not a $\pi^0\pi^0$ effect (even though its mass corresponds to the ABC
effect, we quote the original finding and two recent results
\cite{Abashian:1960zz,Adlarson:2011bh,Adlarson:2012fe}); instead it can
be assigned to the coherent production of $\Delta(1232)$ as
intermediate isobar in both $p\pi^0$ pairs. In the next Dalitz plot
covering the $1000<E_\gamma<1100$\,MeV photon energy range, the two
$\Delta(1232)$ isobars (formed by the proton and one of the two pions)
separate, and one horizontal and one vertical line due to
$\Delta(1232)$ production can been seen. The Dalitz plots at $E_\gamma
= 1150\pm50$ and $1250\pm50$\,MeV confirm that the latter enhancement
is associated with a defined $M^2$. These two lines continue to be seen
up to the highest photon energies $E_\gamma=(2.4 - 2.5)$\,GeV. In the
Dalitz plots covering the photon energy range from 1400 to 1800\,MeV, a
small enhancement is observed at a squared mass $M^2\approx
2.2$\,GeV$^2$. This small enhancement is interpreted as $N(1520)3/2^-$.
At  $E_\gamma = 1450\pm50$\,MeV and  $E_\gamma = 1550\pm50$\,MeV it
leaves its trace as small enhancement on the diagonal at the position
where the two $N(1520)3/2^-$ - formed with the two $\pi^0$ - overlap.
At $E_\gamma = 1950\pm50$ and $2050\pm50$\,MeV, this enhancement
separates into two faint bands at 2.2 and 2.7\,GeV$^2$ indicating the
presence of two resonances, $N(1520)3/2^-$ and $N(1680)5/2^+$. In the
next two bins with $E_\gamma = 2150\pm50$\,MeV and $E_\gamma =
2250\pm50$\,MeV, the $N(1680)5/2^+$ becomes the most prominent feature
while both,  $\Delta(1232)$ and $N(1520)3/2^-$ can still be recognized.
At the highest photon energies, a faint band close to the lower Dalitz
plot boundary develops (corresponding to the highest $\pi^0\pi^0$
mass). In the partial wave analysis described below, the latter is
identified as $f_2(1270)$. \vspace{-2mm}

\subsection{\label{Mass}Mass and angular distributions}

In Fig.~\ref{mppi-proj} the mass distributions, in Fig.~\ref{cosp} three angular
distributions are shown. The mass distributions are given as functions of $M_{p\pi^0}$  and $M_{\pi^0\pi^0}$, the angular distributions depend on the cosine of the angle between proton
momentum and the incoming photon direction in the center-of-mass
system or, alternatively, of the scattering angle in the Gottfried-Jackson or in the
helicity frame \cite{Schilling:1969um}. The cms frame is useful to characterize the decay
angular distribution of a resonance, the Gottfried-Jackson and the
helicity frames are useful to discuss production mechanisms of vector
mesons. Here, in the case of $2\pi^0$ production, they do not provide
immediate insight into the production mechanism. We show these
distributions for completeness and to facilitate a judgement of the fit
quality. The systematic uncertainties are calculated from the difference of the distributions resulting from the two data sets and from the variation of the PWA results when the data from MAMI \cite{Kashevarov:2012wy} are taken into account or not. The calculated systematic uncertainties are thus much smaller where only one data set exists, and may be underestimated.
The solid lines represent the results of the partial wave analysis (see section~\ref{SectionPWA}). The decay pattern
of resonances via the cascade $N^*\to N(1520)3/2^-\pi^0\to p\pi^0\pi^0$ evidences significant contributions
from $N(1900)3/2^+$. For this reason, we display the PWA fit results also when this resonance is omitted from the fit.

The $M_{p\pi^0}$ shows the features discussed above: clear peaks are
seen due to production of $\Delta(1232)$, $N(1520)3/2^-$, and
$N(1680)5/2^+$ as intermediate isobars. The  $M_{\pi^0\pi^0}$ mass
distributions have little structure. In the mass interval from 1800 to
1900\,MeV, a small peak evolves at 1\,GeV which becomes more
significant in the next mass intervals. It is identified with $f_0(980)$.
At the highest photon energy (2400 - 2500\,MeV), a
comparatively narrow structure seems to be present. In the fit it is
assigned to $f_2(1270)$ production. Due to the large $f_2(1270)$ width, the narrow
structure is not well described by the partial wave analysis. The difference
between the two data sets (red and blue) is significant, indicating that small discrepancies should not be overinterpreted.\vspace{-2mm}

\subsection{Polarization}
\label{SectionPol}

\begin{figure}[pb]
\vspace{-10mm}\begin{center}
\includegraphics[width=.5\textwidth]{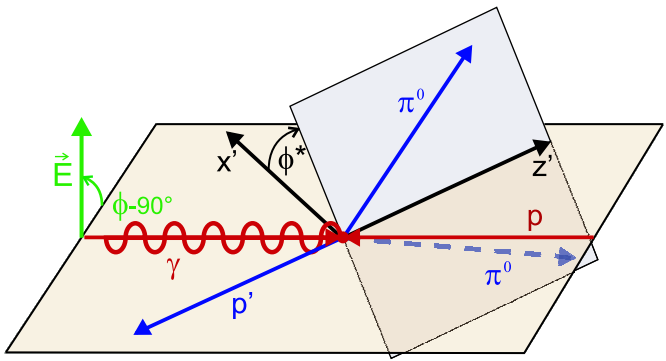}
\vspace{-10mm}\end{center}
\caption{\label{fig:angles}Kinematics of the three-body final state
in the cms frame: incoming photon and outgoing proton $p'$ define
one of three production planes. Its normal vector (not shown) and the photon polarization span
an angle $\phi-90^\circ$. The decay plane (above the production plane in light blue) is defined by the three particles in the final state. $\phi^{*}$ is defined as the angle between production
and decay planes.}
\end{figure}

In photoproduction of single pseudoscalar mesons with a linearly polarized beam, the beam asymmetry $\Sigma$ can be determined from the distribution of the angle $\phi$ between two planes
(shown with an offset of $90^\circ$ in Fig.~\ref{fig:angles}). One plane is defined by the photon polarization vector and the direction of the photon (polarization plane), the second plane is called production plane and defined by the direction of the photon and the outgoing proton or one of the outgoing mesons. In photoproduction of a single meson, the outgoing proton or the outgoing meson define, of course, the same plane. In two-meson production, the production plane can be chosen, and three beam asymmetries can be defined, $\Sigma_p$, $\Sigma_{m_1}$, $\Sigma_{m_2}$. Here, the two mesons are both neutral pions, and two beam asymmetries, $\Sigma_p$ and $\Sigma_\pi$
exist. These two asymmetries are, however, not the only ones which can be determined.

With three particles in the final state, the beam asymmetries depend on the kinematical variables defining the recoiling two-particle system. E.g., a third plane can be defined by the
\begin{figure}[pb]
\begin{center}
\includegraphics[width=.4\textwidth]{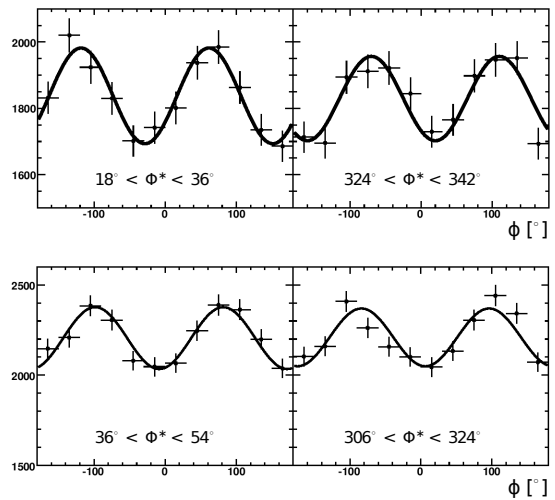}\vspace{-3mm}
\end{center} \caption{\label{fig:phi}Examples of
distributions of azimuthal angle $\phi$, binned in limited ranges
of $\phi^{*}$. The two upper figures: recoiling pion, $E_{\gamma} =
970-1200 \rm \, MeV$, two lower figures: recoiling proton,
$E_{\gamma} = 1200-1450 \rm \, MeV$. }
\end{figure}
three particles in the final state. Their decay plane can be calculated from the vector product of the momenta in the three-particle rest frame of any pair of the final-state particles.  Selecting one of the three final-state particles, the angle between the reaction plane and the decay plane is called $\phi^*$. The definitions of $\phi$ and $\phi^*$ are shown in Fig.~\ref{fig:angles} for the case where the recoiling proton defines the reaction plane. With these angles, the cross section can be expressed in the form~\cite{Roberts:2004mn}:

\begin{eqnarray}
\label{eq:xsec}
\frac{\mathrm{d}\sigma}{\mathrm{d}\Omega} =
\left(\frac{\mathrm{d}\sigma} {\mathrm{d}\Omega}\right)_{0}\{1\ &+\
P_{l}&[I^{s}(\phi^{*})\sin(2\phi)\nonumber \\
&+&I^{c}(\phi^{*})\cos(2\phi)]\}.
\end{eqnarray}

\begin{figure*}[pt]
\begin{center}
\begin{tabular}{ccc}
\raisebox{5mm}{\includegraphics[width=0.48\textwidth]{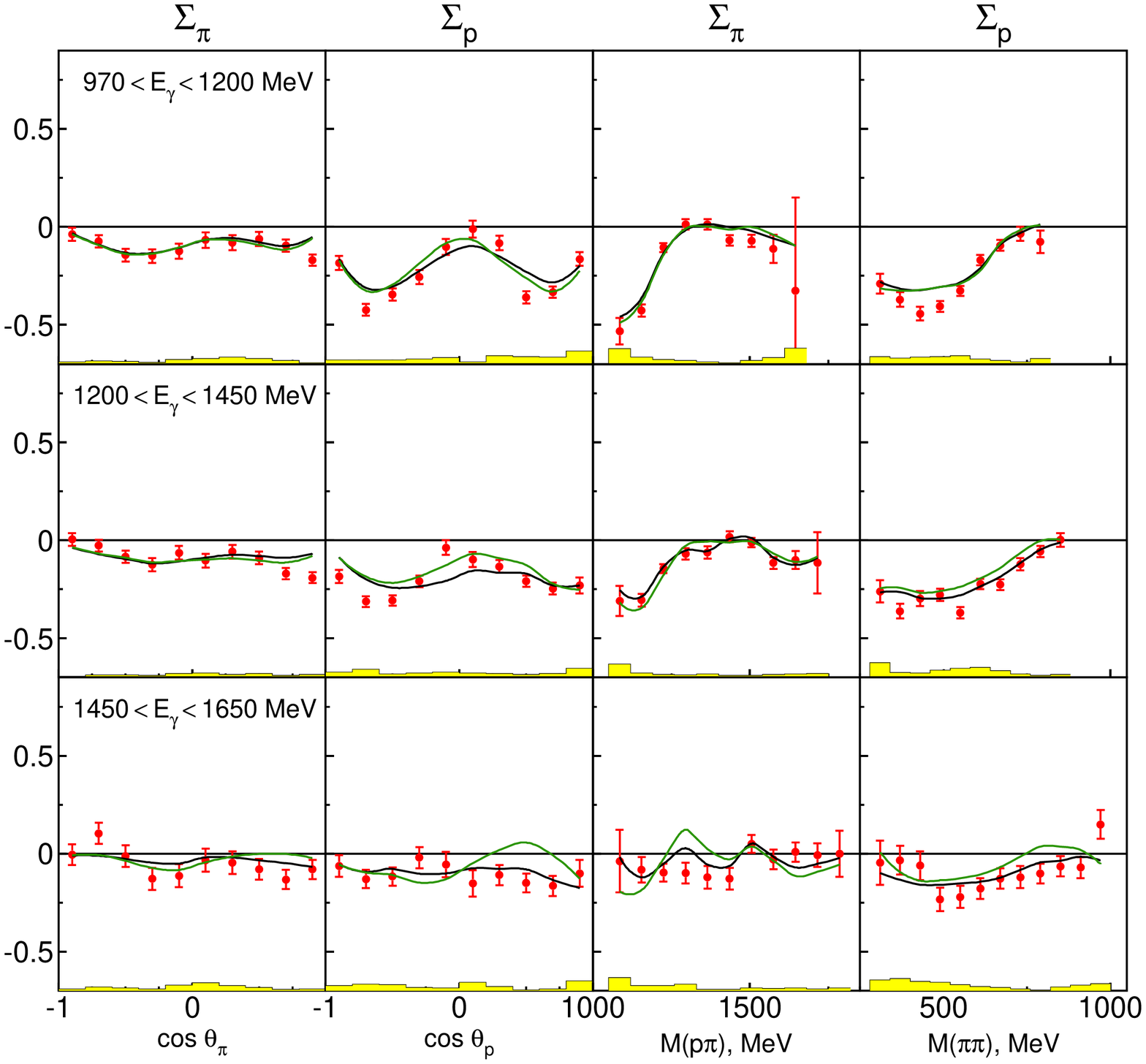}}&
\includegraphics[width=0.48\textwidth]{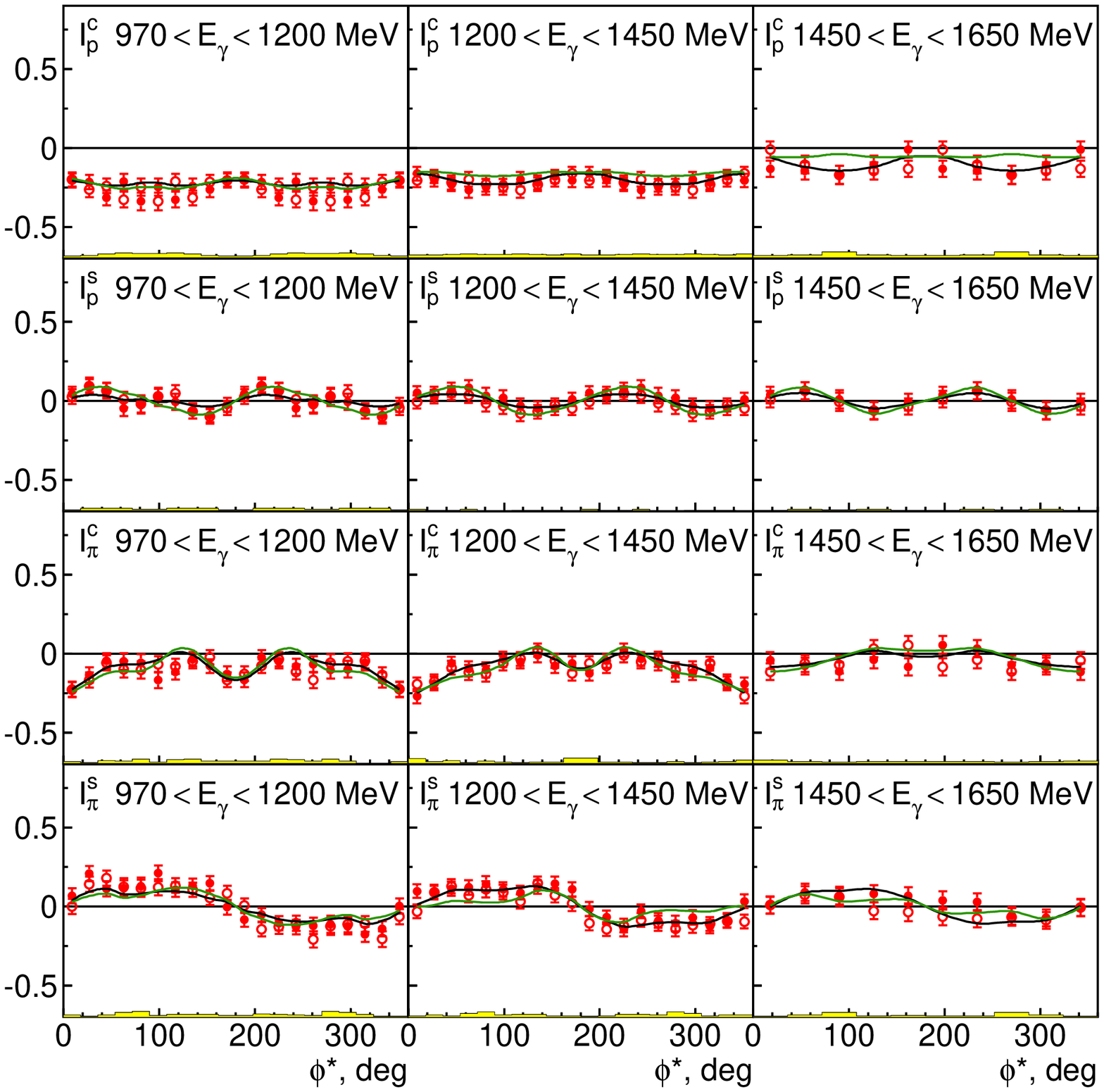}\\[-3ex]
\includegraphics[width=0.49\textwidth]{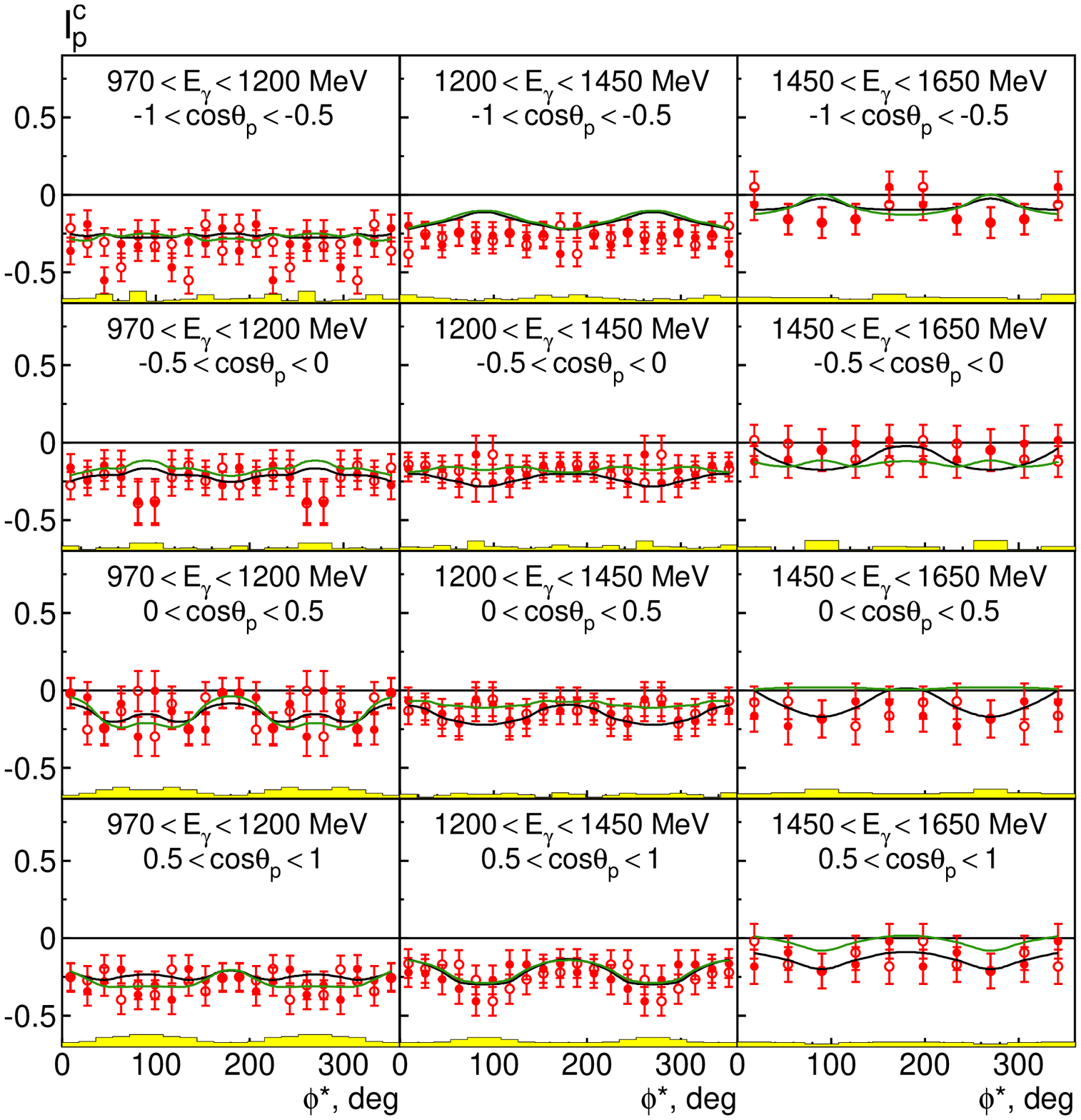}&
\includegraphics[width=0.49\textwidth]{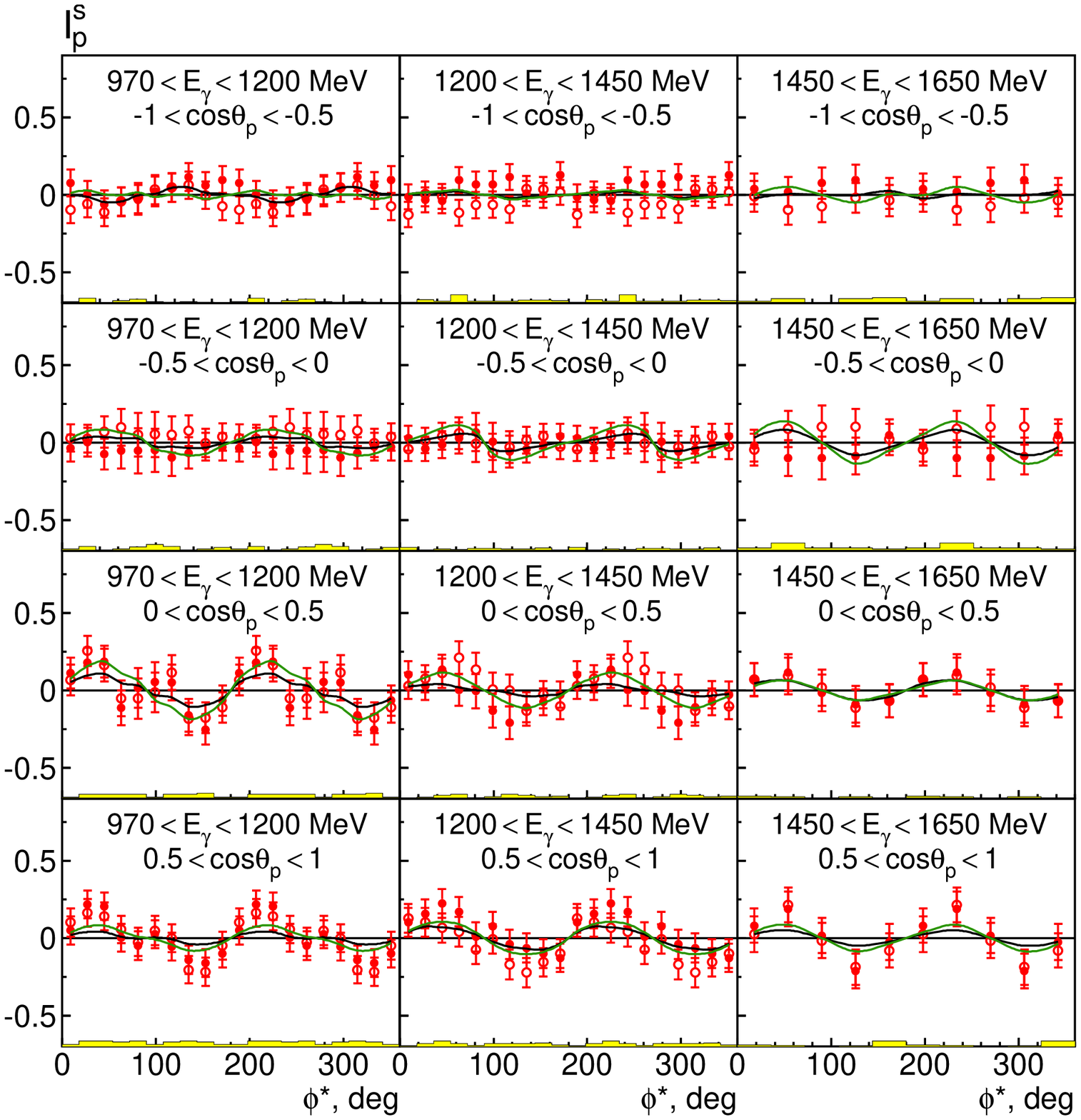}\\[-3ex]
\end{tabular}
\end{center}
\caption{\label{IsIc-phi1}Polarization observables: red points: CBELSA/TAPS data, open circles: derived from symmetry. Black curve: BnGa main PWA fit, green curve: BnGa PWA fit without $N(1900)3/2^+$. Upper left block of figures shows the beam asymmetry $\Sigma$ for three bins in photon energy (970 - 1200; 1200 - 1450; 1450 - 1650\,MeV), in the first column as a function of $\cos\theta_{\pi}$ (pion recoiling), in the second column as a function of $\cos\theta_{\pi\pi}$ (proton recoiling), in the third column of $M(p\pi)$ (pion recoiling), in the fourth column of $M(\pi\pi)$ (proton recoiling). The upper right block shows the observables  $I^{c}$ and $I^s$ for three bins in $E_\gamma$  for the proton recoiling (upper two rows) and pion recoiling (lower two rows) as function of the angle $\phi^{*}$. The figures in the lower blocks show $I^{c}_p$ (left) and $I^{s}_p$ (right) additionally binned in $\cos\theta_{p}$. Systematic errors are shown as yellow bands. }
\end{figure*}

Here, $\left(\frac{\mathrm{d}\sigma}{\mathrm{d}\Omega}\right)_{0}$
is the unpolarized cross section, $P_{l}$ the degree of linear
photon polarization. $I^s$ and $I^c$ are extracted by a fit to the
$\phi$-distributions. A few examples for the according distribution are
given in Fig.~\ref{fig:phi}.
Both, the $\cos(2\phi)$-modulation
due to  $I^{c}$ and the $\sin(2\phi)$-modulation due to $I^{s}$ are
clearly visible and are fitted with function (\ref{eq:xsec}). One can also observe opposite phase shifts in different ranges of $\phi^{*}$. The distributions are shown pairwise: a rotation of the production plane around the normal of the decay plane, with $\phi^{*}\rightarrow 2\pi - \phi^{*}$, corresponds to a sign flip. The photon polarization plane is also rotated, and the azimuthal angle is changed to $\phi\rightarrow 2\pi - \phi$. In eq.~(\ref{eq:xsec}), these changes lead to $\sin(2\cdot(2\pi-\phi)) = - \sin(2\phi)$ and
$\cos(2\cdot(2\pi-\phi)) = \cos(2\phi)$, and thus to
 \begin{eqnarray}\label{symmetries}
I^{c}(\phi^{*}) = I^{c}(2\pi - \phi^{*})\,; &\quad& I^{s}(\phi^{*}) =
-I^{s}(2\pi - \phi^{*}).
 \end{eqnarray}

The relation $\phi^{*}_{2}$ = $\phi^{*}_{1} + \pi$ holds where $\phi^{*}_{1}$ and $\phi^{*}_{2}$ refer to the two pions.  Since the two pions are indistinguishable,  $I^{c}(\phi^{*}) = I^{c}(\phi^{*} + \pi)$ and $I^{s}(\phi^{*}) = I^{s}(\phi^{*} + \pi)$ follows when the proton is taken as recoiling particle. In Fig.~\ref{fig:phi} one can see that these symmetry requirements are well  fulfilled indicating the absence of significant systematic effects in the data. The distributions can be integrated over $\phi^*$. Then, $I^s$ vanishes identically and $I^c=\Sigma$ holds. \vspace{-1mm}

\begin{figure*}[pt]
\begin{center}
\begin{tabular}{ccc}
\includegraphics[width=0.49\textwidth]{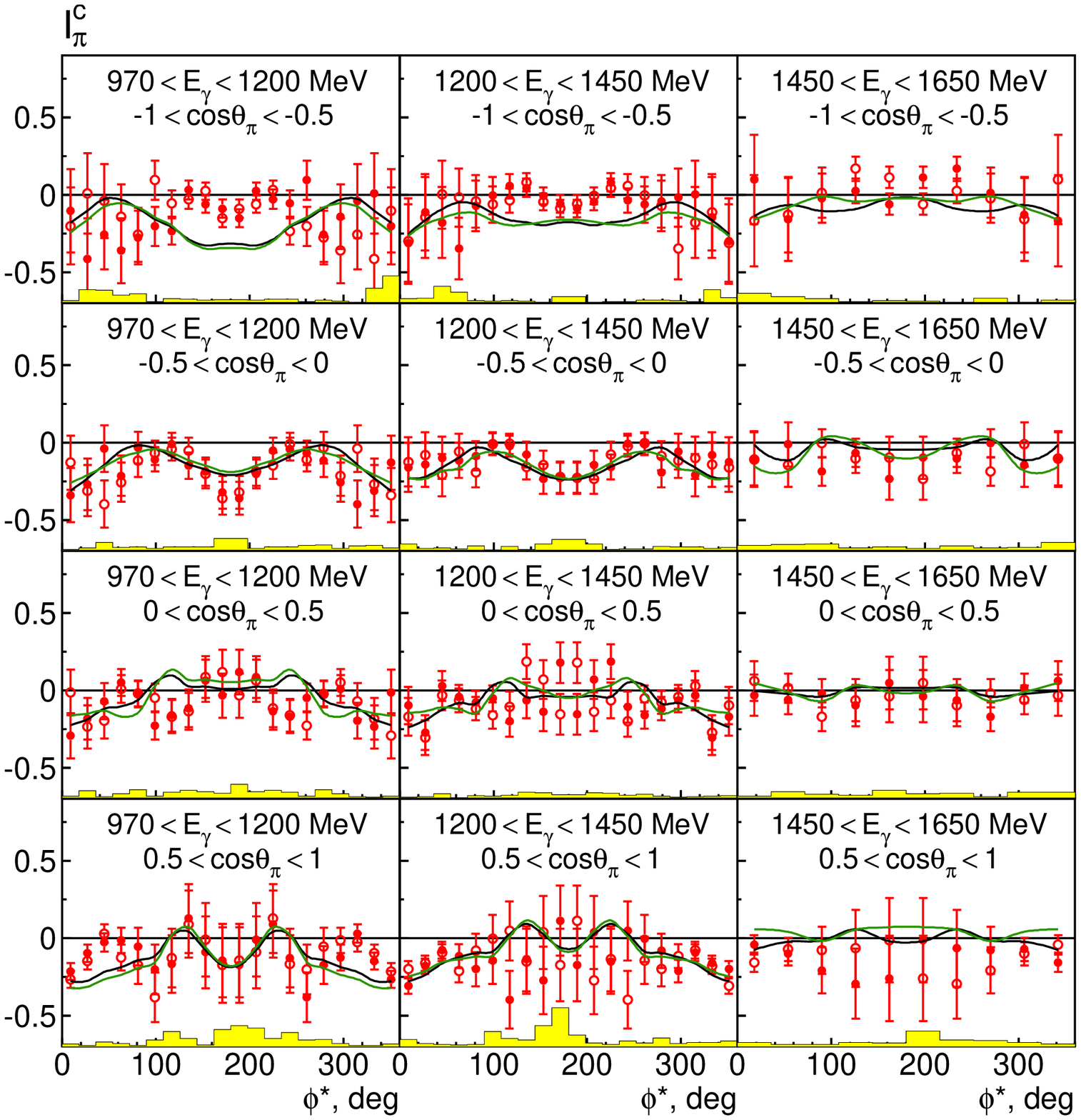}&
\includegraphics[width=0.49\textwidth]{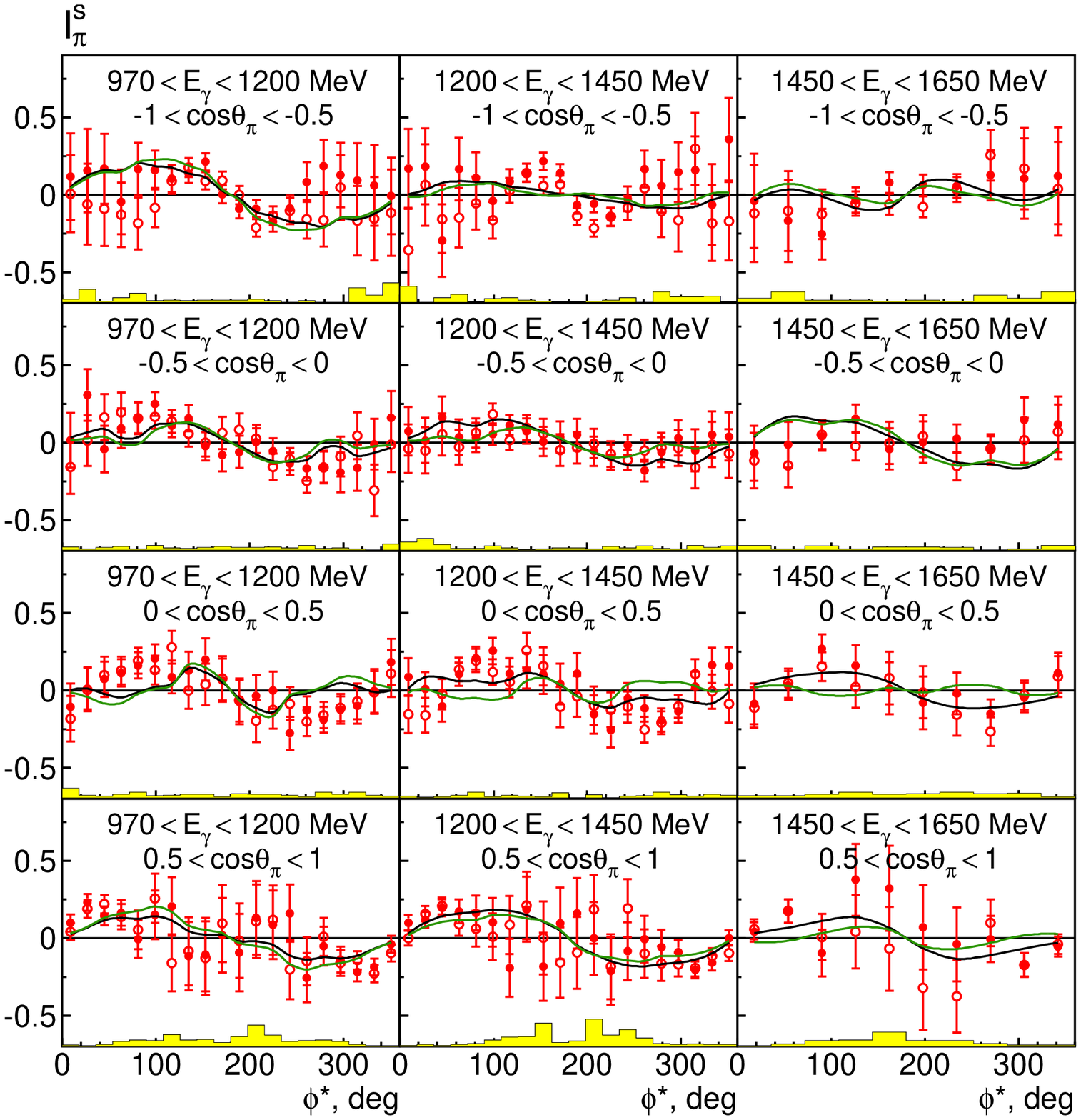}\\[-2ex]
\includegraphics[width=0.49\textwidth]{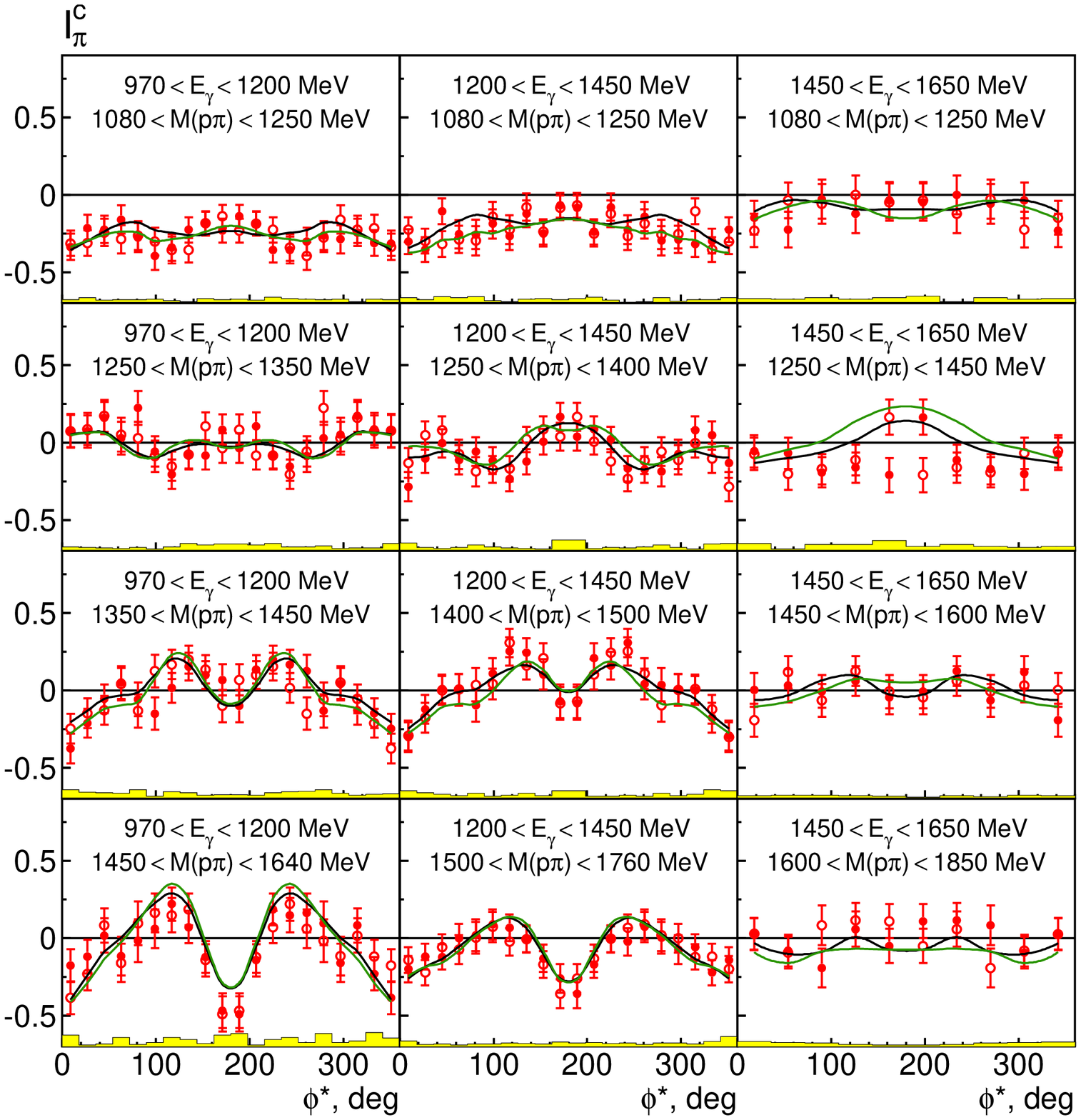}&
\includegraphics[width=0.49\textwidth]{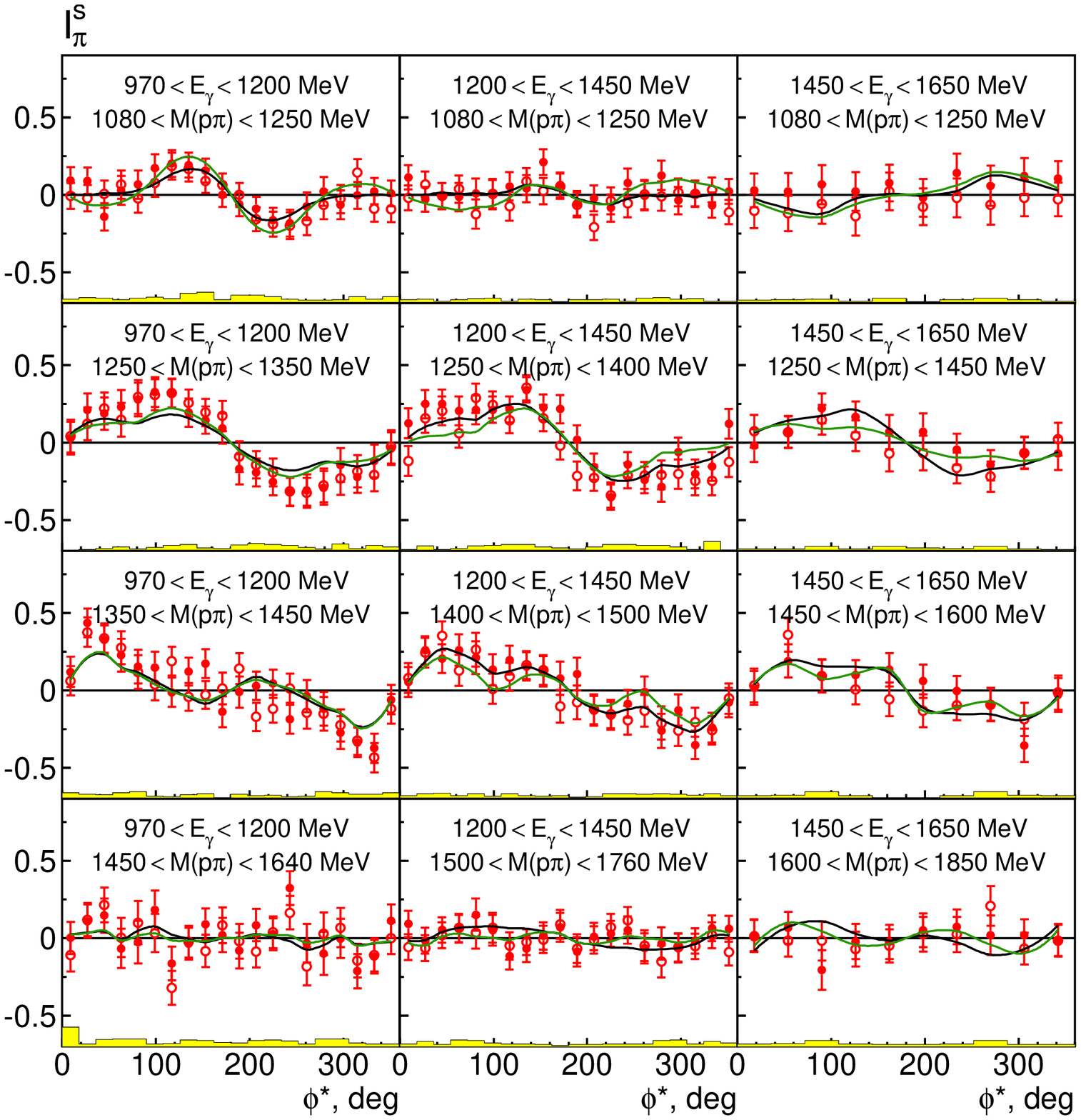}\\[-3ex]
\end{tabular}
\end{center}
\caption{\label{IsIc-phi2}Polarization observables: {red points: CBELSA/TAPS data, open circles: derived from symmetry. Black curve: BnGa main PWA fit, green curve: BnGa PWA fit without $N(1900)3/2^+$. The two upper blocks of figures show $I^{c}_{\pi}$ and $I^{s}_{\pi}$ for three bins in photon energy, additionally binned in four ranges of $\cos\theta_{\pi}$; the two lower blocks show $I^{c}_p$ and $I^{s}_p$, additionally binned in the invariant mass $M(p\pi)$.} Systematic errors are shown as yellow bands. \vspace{-3mm}}
\end{figure*}

\begin{figure*}[pt]
\begin{center}
\begin{tabular}{ccc}
\includegraphics[width=0.49\textwidth]{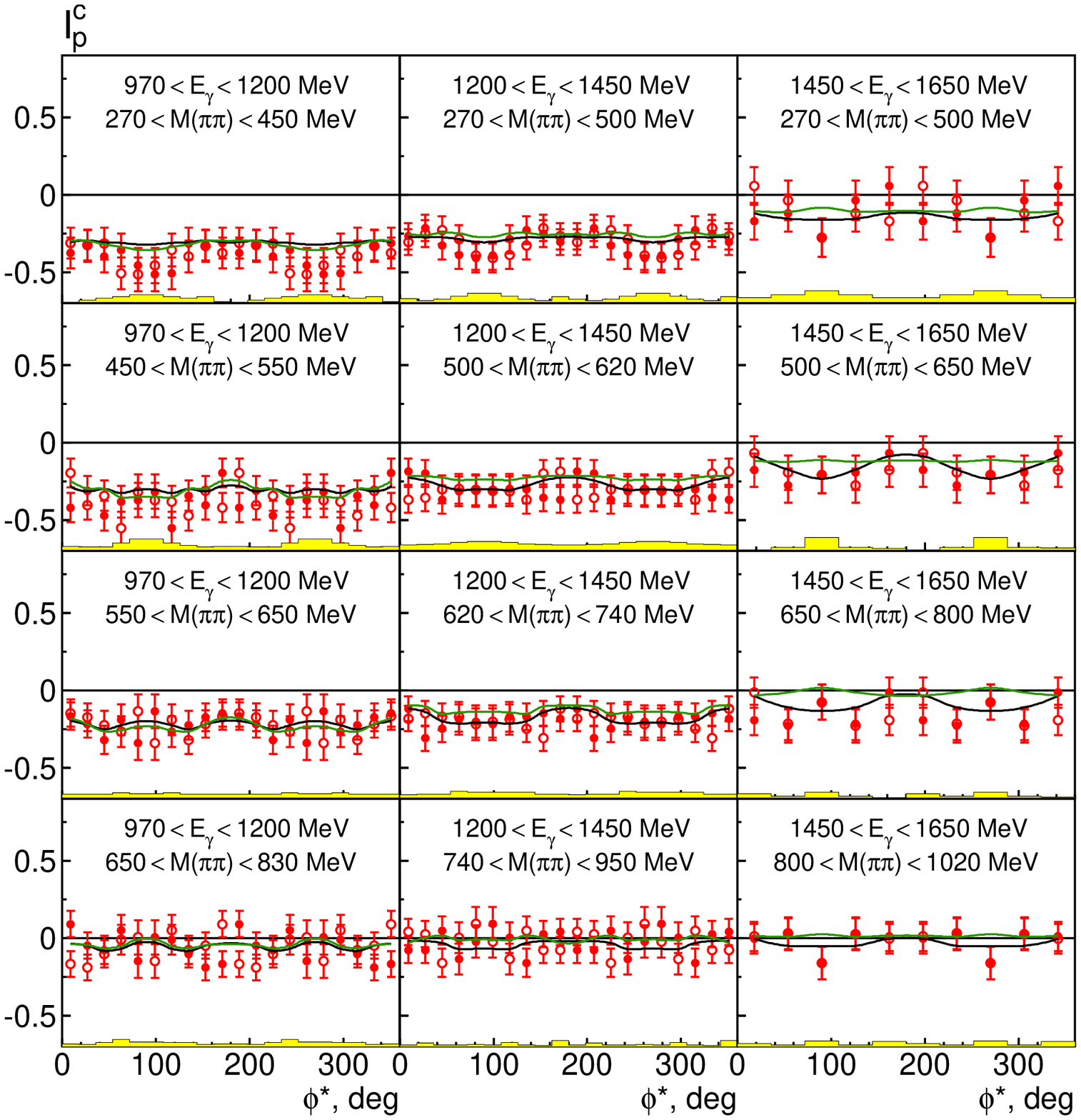}&
\includegraphics[width=0.49\textwidth]{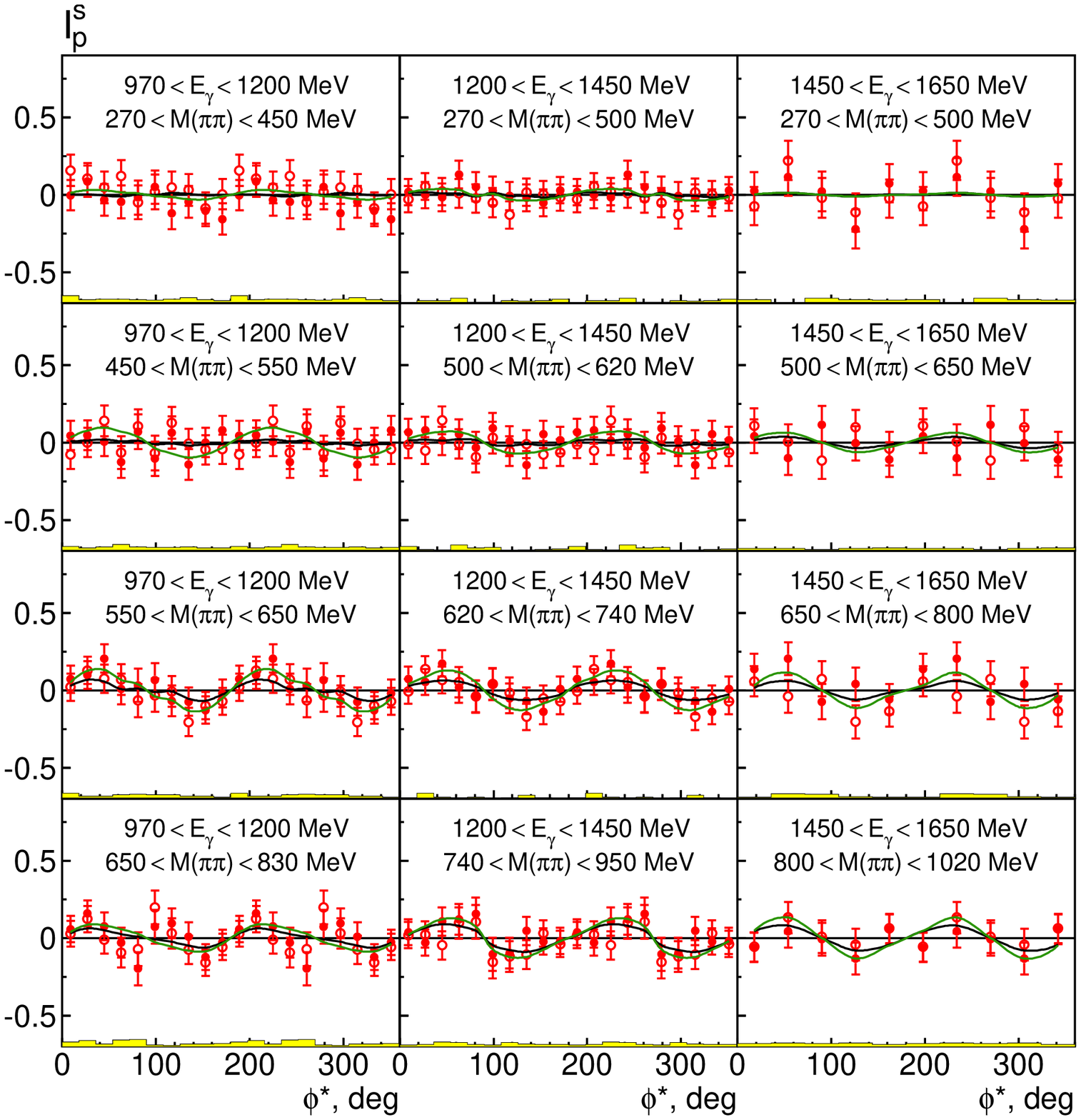}\\[-2ex]
\end{tabular}
\end{center}
\caption{\label{IsIc-m}{Red points: CBELSA/TAPS data, open circles: derived from symmetry. Black curve: full Bonn-Gatchina PWA fit, green curve: best fit without $N(1900)3/2^+$. The figures show $I^{c}_{p}$ (left) and  $I^{s}_p$ (right) for three bins in $E_\gamma$, additionally binned in the invariant mass $M(\pi\pi)$ (from 270 up to 1020\,MeV). Systematic errors are shown as yellow bands. }\vspace{-3mm}}
\end{figure*}

\paragraph{The observables:} The two upper blocks in Fig.~\ref{IsIc-phi1} show the beam asymmetry $\Sigma$ for a recoiling $\pi^0$ or proton and $I^c_p$, $I^s_p$, $I^c_\pi$, $I^s_\pi$ as functions of $\phi^*$, for different slices in photon energy. The data are shown with their statistical errors, the bold solid curve reproduces the partial-wave-analysis fit. The values can be binned in slices of $\cos\theta_{p}$ ($I^c_p$, $I^s_p$: two lower blocks of subfigures in Fig.~\ref{IsIc-phi1}), in slices of $\cos\theta_{\pi}$ ($I^c_\pi$, $I^s_\pi$: upper blocks in Fig.~\ref{IsIc-phi2}), in slices of $M(p\pi)$  ($I^c_\pi$, $I^s_\pi$: lower blocks in Fig.~\ref{IsIc-phi2}), or slices in $M(\pi\pi)$ ($I^c_p$, $I^s_p$: Fig.~\ref{IsIc-m}). The yellow bands indicate the systematic errors due to the uncovered phase space. It was obtained by (a) investigating the difference in $I^s$ and $I^c$ obtained from the BnGa PWA for the reconstructed and generated Monte Carlo events and (b) by determining the effect of a two-dimensional acceptance correction (instead of the full five-dimensional acceptance correction using the PWA) with $\phi$ and $\phi^*$ as variables on  $I^{s}$ and $I^{c}$. The larger deviation was chosen as the  systematic uncertainty. The symmetry conditions (see eq.~\ref{symmetries}) are reasonably fulfilled for the data on  $I^{s}$ and $I^{c}$ presented in
Figs.~\ref{IsIc-phi1}\,-\,\ref{IsIc-m}.
\vspace{-1mm}

\begin{figure}[pt]
\vspace{-2mm}
\begin{center}
\includegraphics[width=0.49\textwidth]{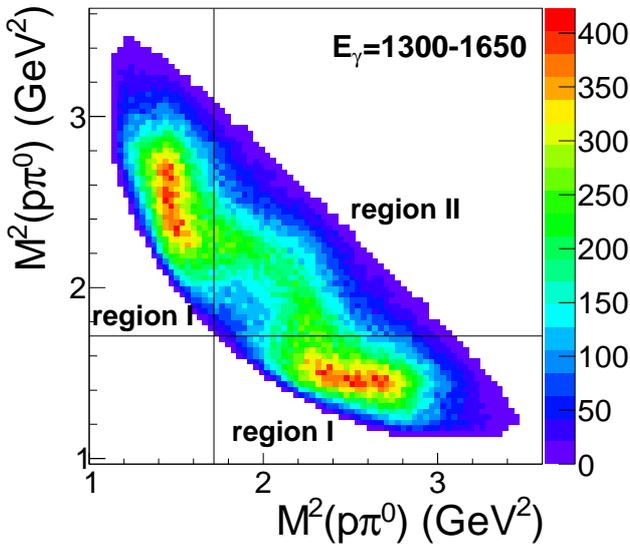}
\end{center} \caption{\label{Dalitz_IsIc}
Dalitz plot $p\pi^0\pi^0$ for $1.30 < E_\gamma < 1.65$\,GeV and polarized photons.
The phase space is divided into two regions I and II. In region II the $\Delta(1232)$
contribution is suppressed and the $N(1520)3/2^-$ contribution enhanced. \vspace{-3mm}}
\end{figure}

\begin{figure}[pt]
\begin{center}
\includegraphics[width=0.48\textwidth]{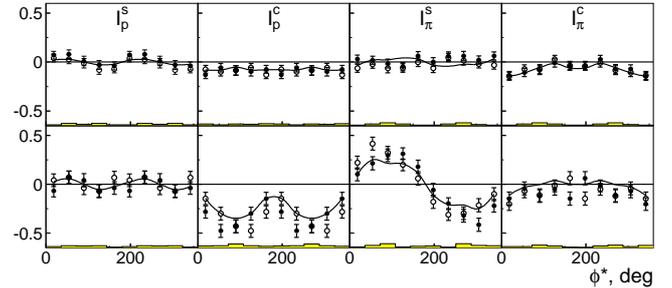}
\end{center} \caption{\label{glob:IsIc}
$I^{s}_p$, $I^{c}_p$, $I^{s}_\pi$, and $I^{c}_\pi$ as functions of
$\phi^*$ for $1300 < E_\gamma < 1650$\,MeV. The five subfigures on the
first (second) row show the distributions for events in region I (II) of the
Dalitz plot. Dots: $I^s,I^c$; open circles: derived from symmetry; grey
band: syst. uncertainties. The solid curve represents the BnGa PWA fit. \vspace{-3mm} }
\end{figure}

\paragraph{\boldmath Further evidence for $N(1900)3/2^+$:} The polarization data are sensitive to the spin-parity
of resonances decaying to $p\pi^0\pi^0$. In particular, new conclusive evidence can be deduced for $N(1900)3/2^+$.
To show this, we select a region of the Dalitz plot in which $N(1520)3/2^-$ is observed. In a first step, we show $I^c$ and $I^s$ for protons and pions integrated over $\cos\theta$ and over a fraction of the Dalitz plot as shown in Fig.~\ref{Dalitz_IsIc}. Region I contains events which are compatible with the reaction $\gamma p\to \Delta(1232)\pi$ while region II enhances the contributions from $\gamma p\to N(1520)\pi$. The corresponding $I^{s}$ and $I^{c}$ distributions are shown in Fig.~\ref{glob:IsIc} (adapted from \cite{Sokhoyan:2015eja}). The events from region I do not show any significant structure. We relate this observation to the large number of resonances with significant decay modes to $\Delta(1232)\pi$ which wash out any structure. On the contrary, events in region II show significant deviations from uniformity. This encouraged us to search for a leading contribution to the reaction chain  $\gamma p\to N^*\to N(1520)\pi$.

\begin{figure}[pt]
\begin{center}
\includegraphics[width=0.50\textwidth]{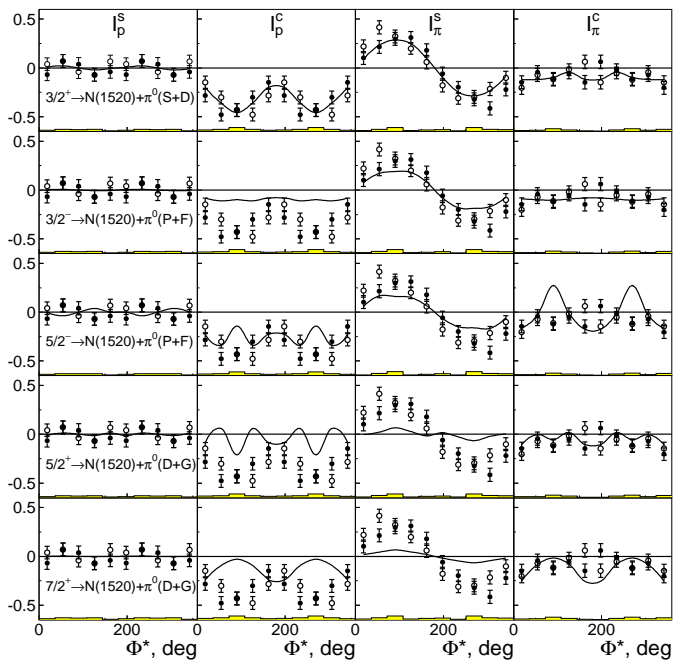}
\end{center} \caption{\label{IsIc-pred}$I^s$ and $I^c$
for events in the $N(1520)$ region. The
same data are shown in all rows. The
lines represent simulated distributions for the reaction chains
$N(J^P)$ $\to$ $N(1520)3/2^-\pi^0$, $N(1520)3/2^-$ $\to$ $N\pi^0$
for different spin-parities $(J^P)$ of the initial state. Some predictions
for $I^s_p$ vanish.
\vspace{-3mm}}
\end{figure}

The distributions in $I^{s}$ and $I^{c}$ for events in region II of Fig.~\ref{Dalitz_IsIc} are shown again in
Fig.~\ref{IsIc-pred} (adapted from \cite{Sokhoyan:2015eja}). The errors in $I^{s}$ and $I^{c}$ reflect the
statistical errors; the systematic uncertainty are shown as yellow bands. The data are shown repeatedly in four lines and compared to highly simplified predictions. It is assumed that a single
resonance at 1900\,MeV decays into $p\pi^0\pi^0$ via formation
of $N(1520)3/2^-$ as intermediate resonance. The expected
$I^{s}/I^{c}$ pattern then depends on the quantum numbers of the
primary resonance, on the ratio of the helicity amplitudes
$A_{1/2}/A_{3/2}$, and on the ratio of the decay amplitudes with
a leading and a higher-$L$ orbital angular momentum (D/S, F/P, or G/D). If there is
only one helicity amplitude, for
$J=1/2$, $I^{s}$ and $I^{c}$ vanish identically. The helicity ratio and the amplitude ratio
were used to fit the data in a two-parameter fit.

The comparison shows that an initial $J^P=3/2^+$ resonance decaying into $N(1520)3/2^-$ gives by far the best description of the data (with a $\chi^2/N_{\rm dof} =1.25$ for 40 data points). The D/S ratio is determined to 5\%.
Assuming an initial quantum number $J^P=3/2^-$ ($J^P=3/2^+$; $J^P=5/2^+$) results in a $\chi^2/N_{\rm dof}$ between 5 and 10. This is a remarkable result: the spin-parity of the initial state follows unambiguously from the observed distributions; no partial wave analysis is required to arrive at this result.

The distributions do not provide any information on the isospin. However, the helicity ratio for the $J^P=3/2^+$ hypothesis is determined from the fit to $A_{1/2}/A_{3/2}=-0.47$. PDG quotes $-0.55$ for $N(1900)3/2^+$ and $+1.7$ for $\Delta(1920)3/2^+$. Hence we might conclude that - most likely - $N(1900)3/2^+$ should be responsible for the observed pattern and should provide a substantial contribution to the reaction. As will be seen below, the partial wave analysis finds indeed a significant contribution from the decay chain $N(1900)3/2^+\to N(1520)3/2^- + \pi^0$ with $N(1520)3/2^-$ decaying into $p\pi^0$.

\section{Partial wave analysis}
\label{SectionPWA}

The data presented here are included in the large data base of the BnGa partial wave analysis. The included data are listed in \cite{Anisovich:2011fc} and \cite{Anisovich:2013vpa}. The new data allow us to extract detailed branching ratios for sequential decays of $N^*$ and $\Delta^*$ resonances. The data with three particles in the final state, like the data presented here, are fitted using an event-based likelihood method which returns the logarithmic likelihood $\ln{\cal L}_i$ of the fit for the data set $i$. However, the polarization observables enter the fit only as histograms, and the fit quality is described by a $\chi^2$ value. The fit minimizes the total log likelihood function defined by
\be
 -\ln {\cal L}_{\rm tot}= ( \frac 12\sum w_i\chi^2_i-\sum w_i\ln{\cal L}_i ) \ \frac{\sum
N_i}{\sum w_i N_i}\,. \label{likeli}
\ee
The data sets are weighted with weight factors $w_i$ to avoid that polarization data -- mostly of low statistics but very sensitive to the phases of amplitudes -- are dominated by high-statistics differential cross section. Changes of  $-\ln {\cal L}_{\rm tot}$ are converted into a changes of a pseudo-$\chi^2$ by
\be
\delta\chi^2 =  -\frac12\delta\ln {\cal L}. \label{pseudo}
\ee
The properties of the resonances used in the analysis are listed in Table~\ref{nucleon} in the Appendix.

\section{Interpretation}
\label{SectionInterpretation}

\begin{table}
\caption{\label{tab:DeltaN}Comparison of branching ratios (in \%) for decays into $\pi N$ and $\pi\Delta$. Branching ratios from the Particle Data Group~\cite{Agashe:2014kda} are given by small numbers.}
\begin{center}
\renewcommand{\arraystretch}{1.4}
{\scriptsize
\begin{tabular}{|l|cc|cc|cc|cl}
\hline
                & $N\pi$ &$L$&$\Delta(1232)\pi$ &\hspace{-3mm}$L$$<$$J$\hspace{-3mm}&$\Delta(1232)\pi$ &\hspace{-3mm}$L$$>$$J$\hspace{-3mm}\\ \hline
$N(1440)1/2^+$          &63\er 2&1&&&20\er 7&1\\[-1.5ex]
    & \tiny     55 - 75&&&&\tiny        20 - 30& \\[-0.2ex]
$N(1520)3/2^-$       &61\er 2&2& 19\er 4&0&9\er 2&2\\[-1.5ex]
    & \tiny     55 - 65&&\tiny 10 - 20&&\tiny       10 - 15& \\[-0.2ex]
$N(1535)1/2^-$       &52\er 5&0& &&2.5\er 1.5&2\\[-1.5ex]
    & \tiny     35 - 55&&&&\tiny        0 - 4& \\[-0.2ex]
$N(1650)1/2-$        &51\er 4&0&&&12\er 6&2\\[-1.5ex]
    & \tiny     50 - 90&&&&\tiny        0 - 25& \\[-0.2ex]
$N(1675)5/2^-$       &41\er 2&2&30\er 7&2&$<2$&4\\[-1.5ex]
    & \tiny     35 - 45&&\tiny 50 - 60&&\tiny        - & \\[-0.2ex]
$N(1680)5/2^+$       &62\er 4&3& 7\er 3&1&10\er 3&3\\[-1.5ex]
    & \tiny     65 - 70&&\tiny 10\er 5&&\tiny 0 - 12 & \\[-0.2ex]
$N(1700)3/2^-$       &15\er 6&2& 65\er 15&0& 9\er 5&2\\[-1.5ex]
    & \tiny     12\er 5&&\tiny 10 - 90&&\tiny 0 - 20 & \\[-0.2ex]
$N(1710)1/2^+$       & 5\er 3&1&&&25\er 10&1\\[-1.5ex]
    & \tiny     5 - 20&&\tiny &&\tiny   15 - 40& \\[-0.2ex]
$N(1720)3/2^+$       &11\er 4&1& 62\er 15&1&6\er 6&3\\[-1.5ex]
    & \tiny     11\er 3&&\tiny &&\tiny  75\er 15& \\[-0.2ex]
$N(1875)3/2^-$       & 4\er 2&2& 14\er 7&0&7\er 5&2\\[-1.5ex]
    & \tiny     12\er 10&&\tiny 40\er 10&&\tiny 17\er 10& \\[-0.2ex]
$N(1880)1/2^+$       & 6\er 3&1&&&30\er 12&1\\
$N(1895)1/2^-$       & 2.5\er 1.5&0& &&7\er 4&2\\
$N(1900)3/2^+$       & 3\er 2&1& 48\er 10&1&33\er 12&3\\[-1.5ex]
    & \tiny $\sim$10&&\tiny &&& \\[-0.2ex]
$N(1990)7/2^+$       & 1.5\er 0.5&3&16\er 6&3&&5\\
$N(2000)5/2^+$       & 8\er 4&1&22\er 10&1&34\er15&3\\
$N(2060)5/2^-$       & 11\er 2&2&&& 7\er3&2\\
$N(2120)3/2^-$       & 5\er 3&2&50\er20&0&20\er12&2\\
$N(2100)1/2^+$       &16\er 5&1&&&10\er 4&1\\[-1.5ex]
    & \tiny     11\er 3&&&&& \\[-0.2ex]
$N(2190)7/2^-$       &16\er 2&4&25\er 6&2&&4\\[-1.5ex]
    & \tiny     10 - 20&&\tiny &&& \\[-0.2ex]
$\Delta(1600)3/2^+$  &14\er 4&1&77\er 5&1&$<2$&3\\[-1.5ex]
    & \tiny     10 - 25&&\tiny &&& \\[-0.2ex]
$\Delta(1620)1/2^-$  &28\er 3&0&&&62\er10&2\\[-1.5ex]
    & \tiny     20 - 30&&\tiny &&\tiny 30 - 60& \\[-0.2ex]
$\Delta(1700)3/2^-$  &22\er 4&2&20\er 15&0& 10\er 6&2\\[-1.5ex]
    & \tiny     10 - 20&&\tiny 35 - 50&&\tiny 5 - 15& \\[-0.2ex]
$\Delta(1900)1/2^-$  & 7\er 2&0&&&50\er20&2\\[-1.5ex]
    & \tiny     10 - 30&&\tiny &&\tiny & \\[-0.2ex]
$\Delta(1905)5/2^+$  &13\er 2&3&33\er10&1&&3\\[-1.5ex]
    & \tiny     9 - 15&&\tiny &&& \\[-0.2ex]
$\Delta(1910)1/2^+$  &12\er 3&1&&&50\er16&1\\[-1.5ex]
    & \tiny     15 - 30&&\tiny &&\tiny 60\er 28& \\[-0.2ex]
$\Delta(1920)3/2^+$  &8\er 4&1&18\er 10&1&58\er14&3\\[-1.5ex]
    & \tiny     5 - 20&&\tiny &&\tiny & \\[-0.2ex]
$\Delta(1940)3/2^-$  & 2\er 1&0&46\er20&0&12\er 7&2\\
$\Delta(1950)7/2^+$  &46\er 2&3&5\er4&3&&5\\[-1.5ex]
    & \tiny     35 - 45&&\tiny 20 - 30&&\tiny & \\[-0.2ex]
\hline
\end{tabular}}
\renewcommand{\arraystretch}{1.0}\vspace{-3mm}
\end{center}
 \end{table}
Table~\ref{tab:DeltaN} lists the branching ratios of $N^*$ and $\Delta^*$ resonances for their decays into $N\pi$ or $\Delta(1232)\pi$. A detailed comparison of data and model predictions has been made in \cite{Capstick:2000qj} which we do not repeat here (even though there are now a few more entries in the data list). Instead, we discuss the branching ratios at a phenomenological level.

\begin{figure}[pt]
\begin{center}
\includegraphics[width=0.50\textwidth,height=0.40\textwidth]{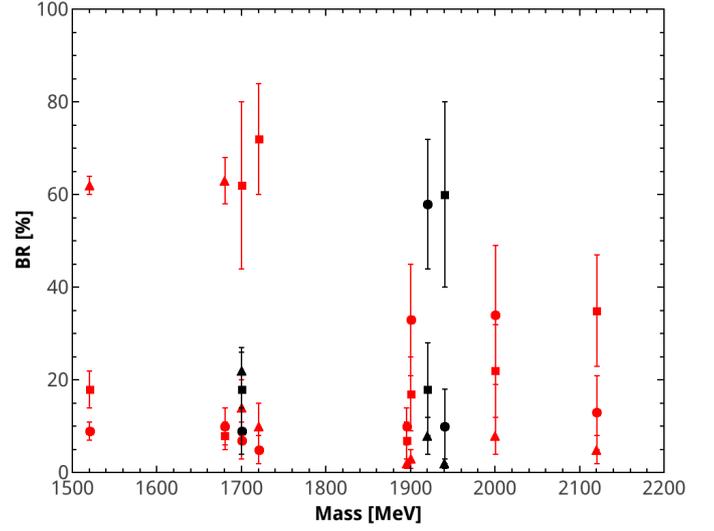}
\vspace{-2mm}
\end{center}
\caption{\label{fig:DeltaN}Branching ratios for the decays of $N^*$ (black) and $\Delta^*$ ({{red}}) resonances into $N\pi$ ($\triangle$), and into $\Delta(1232)\pi$ with $L<J$ ($\square$) and $L>J$ ($\bigcirc$).\vspace{-2mm}
}
\end{figure}

One might expect that the branching ratios should depend on the decay momentum. Figure~\ref{fig:DeltaN} shows the branching ratios as a function of the mass of the decaying resonance. No systematic dependence is observed. Likewise, in $\Delta(1232)\pi$ decays, there are often two angular momenta compatible with all selection rules but there is no obvious preference for the lower or higher orbital angular momentum. Also, we did not find any hint why for some resonances the $N\pi$ decay mode is preferred over the $\Delta(1232)\pi$ decay mode while for other resonances, the reverse preference holds.

\subsection{Wave functions of excited baryons}
\label{Cascades}

The total wave function consists of a color, flavor, spin, and spatial part. The color wave function is totally antisymmetric with respect to the exchange of any pair of quarks which implies a symmetric spin-flavor-spatial wave function. Here, we discuss $N^*$ and $\Delta^*$ resonances, hence the flavor wave function reduces to the isospin wave function.

A wave function of a three-particle system can be symmetric, antisymmetric or of mixed symmetry with respect to the exchange of a pair of quarks. The state with maximum total spin projection for $S=3/2$ can be written as $|\uparrow\uparrow\uparrow>$ and is obviously totally symmetric. For total intrinsic spin  $S=1/2$ one has, again for maximal spin projection, the states\vspace{-4mm}\\
$$-\frac{1}{\sqrt 6}|\uparrow\downarrow\uparrow> -\frac{1}{\sqrt 6}|\downarrow\uparrow\uparrow> +\sqrt{\frac{2}{3}}|\uparrow\uparrow\downarrow>
$$
$$ {\rm and}\qquad
\frac{1}{\sqrt 2}|\uparrow\downarrow\uparrow> - \frac{1}{\sqrt 2}|\downarrow\uparrow\uparrow>
$$
which are called mixed symmetric and mixed antisymmetric, respectively.

\begin{table}[pb]
\caption{%
\label{tab:Permutations} The six combinations to form totally
symmetric spatial-spin-flavor wave functions. The color wave
function guarantees the overall antisymmetry of the total wave functions.
\vspace{-3mm}}
{
  \renewcommand{\arraystretch}{1.4}
 \begin{center} \begin{tabular}{cc}
    \hline
    $S$& space spin isospin \\
    \hline
    $S_1$&$\mathcal{S}\mathcal{S}\mathcal{S}$\\
    $S_2$&$\mathcal{S} (\cal M_\mathcal{S}\cal M_\mathcal{S}
    +
    \cal M_\mathcal{A}\cal M_\mathcal{A})$\\
    $S_3$&$(\cal M_\mathcal{S}\cal M_\mathcal{S}
    +
    \cal M_\mathcal{A}\cal M_\mathcal{A}) \mathcal{S}$\\
    $S_4$&$(\cal M_\mathcal{A}\cal M_\mathcal{A}
    -
    \cal M_\mathcal{S}\cal M_\mathcal{S})
    \cal M_\mathcal{S} + (\cal M_\mathcal{S}\cal M_\mathcal{A}
    +
    \cal M_\mathcal{A}\cal M_\mathcal{S}) \cal M_\mathcal{A}$\\
    $S_5$&$(\cal M_\mathcal{S} \mathcal{S} \cal M_\mathcal{S}
    +
    \cal M_\mathcal{A} \mathcal{S} \cal M_\mathcal{A}) $\\
    $S_6$&$A (\cal M_\mathcal{A}  \cal M_\mathcal{S}
    -
    \cal M_\mathcal{S}  \cal M_\mathcal{A}) $\\
    \hline
  \end{tabular}\vspace{-3mm}\end{center}
  \renewcommand{\arraystretch}{1.0}
}
\end{table}

Likewise, isospin and spatial wave functions of defined symmetry can be constructed. In the nucleon ground state, e.g., the spin and the isospin wave functions are both of mixed symmetry, the combination (\ref{gsn}) leads to a fully symmetric spin-flavor wave function. The $\Delta(1232)$ resonance is the ground state of states with isospin $I=3/2$ and has spin $S=3/2$. The spin-flavor wave function is hence symmetric, see eq.~(\ref{gsd}). Hence the nucleon and $\Delta(1232)$ have - as expected for ground states - symmetric spatial wave functions:
\begin{subequations}\label{gs}
\begin{align}
{\cal M}_{S}(\rm spin){\cal M}_{S}(\rm isospin)+{\cal M}_{A}(\rm spin){\cal M}_{A}(\rm isospin)\label{gsn}\\
{\cal S}(\rm spin){\cal S}(\rm isospin)\,.\hspace{20mm}\label{gsd}
\end{align}
\end{subequations}

The spatial wave functions of excited baryons are more complicated. With the position vectors $\vec x_{i}
(i=1,2,3)$ of the three particles and after separation of the center-of-mass motion, the baryon excitations can be described by two oscillators, with
coordinates which are conventionally called $\vecr, \vecl$:
\begin{subequations}\label{rl}
\begin{align}
{\vecr}&= \frac{1}{\sqrt 2}\left({\vec x}_1 - {\vec x}_2\right) \label{rl1}\\
{\vecl}&= \frac{1}{\sqrt 6}\left({\vec x}_1+{\vec x}_2 - 2{\vec x}_3\right)\label{rl2}
\end{align}
\end{subequations}
where (\ref{rl1}) is antisymmetric and (\ref{rl2}) symmetric with respect to exchange of particle 1 and 2. The wave functions can be calculated for a harmonic oscillator (h.o.) potential. The h.o. wave functions for a given orbital angular momentum $L$
\be
\label{ho}
\hspace{-2mm}\Phi^L(\vec\rho, \vec\lambda)= \hspace{-3mm}\sum_{n_\rho,\ell_\rho,n_\lambda,\ell_\lambda}\hspace{-4mm}
  c^L_{n_\rho\,\ell_\rho\,n_\lambda\,\ell_\lambda}
\left[\phi_{n_\rho\,l_\rho}(\vec\rho)\times\phi_{n_\lambda\,l_\lambda}(\vec\lambda)\right]^{L}
\ee
depend on the vibrational and rotational quantum numbers of the two oscillators. By appropriate linear combinations of the functions (\ref{ho}), states of definite permutational symmetry can be
constructed. The separation of the spin and orbital-angular-momen\-tum is a non-relativistic concept; in general the spatial wave function of a baryon resonance needs to be written as a sum of h.o. wave functions. For the present discussion it is, however, sufficient to consider a non-relativistic approximation. This may look like too crude. However, the full $N^*$ and $\Delta^*$ mass spectrum can be reproduced rather well with a surprisingly simple mass formula using just three free parameters: the nucleon and $\Delta(1232)$ masses and the string tension \cite{Klempt:2002vp,Forkel:2008un}. Explicit spin-orbit interactions are not part of the formula, and that is the reason why spin and orbital angular momenta can be assigned to most $N^*$ and $\Delta^*$ resonances.

The first negative-parity excitations have components in their harmonic-oscillator wave function of the form
\begin{subequations}\label{lone}
\begin{align}
{\cal M}_S&=\left[\phi_{0s}(\vec\rho)\times\phi_{0p}(\vec\lambda)\right]^{(L=1)} \label{l1s}\\
{\cal M}_A&=\left[\phi_{0p}(\vec\rho)\times\phi_{0s}(\vec\lambda)\right]^{(L=1)}\,. \label{l1a}
\end{align}
\end{subequations}

In $\Delta$ states with $S=1/2$, the spin-spatial wave function is combined as
\be
{\cal M}_S (\rm spin){\cal M}_S (\rm space) + {\cal M}_A (\rm spin){\cal M}_A (\rm space)
\ee
to give a symmetric spin-spatial wave function; the symmetric $I=3/2$ isospin wave function then guarantees the correct overall symmetry (see $S_3$ in Table \ref{tab:Permutations}). There is no symmetric spatial wave function with $L=1$, hence $L=1,S=3/2$ is forbidden for $\Delta$ states with their symmetric isospin wave function.

For nucleons with $L=1$, states with $S=1/2$ and $S=3/2$ are both allowed. They use the same oscillator wave functions (\ref{lone}) but in different combinations ($S_4$ and $S_5$ in Table \ref{tab:Permutations}). In all cases, the orbital wave functions do not correspond to a definite excitation of the $\rho$ or $\lambda$ coordinate, and this can be viewed as oscillation
of the excitation energy from the $\rho$ oscillator to the $\lambda$ oscillator and back. It is helpful to look at the motion of the three quarks in a classical picture: either the two quarks in the $\rho$ oscillator rotate around the third quark which is frozen at the center or - in the $\lambda$ oscillator - two quarks at one end and the third quark at the other end rotate around their common center of gravity.

The five $N^*$ and two $\Delta^*$ resonances with negative parity expected in the quark model can be identified with the seven lowest-mass negative-parity nucleon and $\Delta^*$ resonances:
\renewcommand{\arraystretch}{1.2}
{\small\bc \begin{tabular}{ccc}
$N(1650)1/2^-$ & $N(1700)3/2^-$ & $N(1675)5/2^-$  \\
$N(1535)1/2^-$ & $N(1520)3/2^-$ &   \\
$\Delta(1620)1/2^-$  & $\Delta(1700)3/2^-$ & {\bf A}\\ \end{tabular}\ec}
\renewcommand{\arraystretch}{1.0}

These resonances are often assigned to the first excitation shell. Resonances with one unit in a vibrational excitation quantum number $n_\rho$ or $n_\lambda$ and resonances with $l_\rho + l_\lambda =2$ belong to the second excitation shell. Resonances with $(n_\rho, n_\lambda)=(1,0)$ or $(0,1)$ have symmetric spatial wave functions and their spin-flavor wave function is the same as that of their respective ground states, see Eq.~\ref{gs}). Obvious candidates are
\renewcommand{\arraystretch}{1.2}
{\small\bc \begin{tabular}{cccc}
$N(1440)1/2^+$ & \quad&$\Delta(1600)3/2^+$ & \quad{\bf B}\end{tabular}\ec}
\renewcommand{\arraystretch}{1.0}
These all correspond to single-mode excitations. Mixing with other configurations may occur, but their contributions to the wave function are expected (and calculated) to be small.

Resonances with wave functions of various symmetries can be formed when $l_\rho + l_\lambda =2$. The simplest case is a fully symmetric wave function:
\begin{equation}
\label{S}
\begin{footnotesize}
\mathcal{S}
    = \frac{1}{\sqrt{2}}\bigl\lbrace
    \left[\phi_{0s}(\vec\rho)\times\phi_{0d}(\vec\lambda)\right]
    +
    \left[\phi_{0d}(\vec\rho)\times\phi_{0s}(\vec\lambda)\right]\bigr\rbrace^{(L=2)}.
	\end{footnotesize}
\end{equation}
As in the first shell, both oscillators are coherently excited; the oscillation energy fluctuates from $\rho$ to $\lambda$ and back to the $\rho$ oscillator. We call these  excitations single-mode excitations. The spin-flavor wave function must be symmetric and adopts the form given in eq.~(\ref{gs}). Nucleon resonances belonging to this category must hence have spin $S=1/2$  and $\Delta$ states must have $S=3/2$ ($S_2$ and $S_1$, respectively, in Table~\ref{tab:Permutations}):
{\footnotesize
\renewcommand{\arraystretch}{1.2}
\vspace{-3mm}
\bc
\begin{tabular}{cccc}
$\Delta(1910)1/2^+$&\hspace{-2mm}$\Delta(1920)3/2^+$&
\hspace{-2mm}$\Delta(1905)5/2^+$&\hspace{-2mm}$\Delta(1950)7/2^+$\\
&$N(1720)3/2^+$&$N(1680)5/2^+$&{\bf C}
\end{tabular}
\ec
\renewcommand{\arraystretch}{1.0}
}
A total orbital angular momentum $L=2$ can also be constructed by spatial wave functions of mixed symmetry, see $S_3$ to $S_5$ in Table~\ref{tab:Permutations}. These are:
\begin{subequations}
\begin{footnotesize}
\begin{align}
\label{MS}
\mathcal{M_S}
    & = \frac{1}{\sqrt{2}}\bigl\lbrace
    \left[\phi_{0s}(\vec\rho)\times\phi_{0d}(\vec\lambda)\right]
   -\left[\phi_{0d}(\vec\rho)\times\phi_{0s}(\vec\lambda)\right]\bigr\rbrace^{(L=2)} \\[-2.5ex]
\label{MA}\mathcal{M_A} & =
    \left[\phi_{0p}(\vec\rho)\times\phi_{0p}(\vec\lambda)\right]^{(L=2)} \,.
   \end{align}
\end{footnotesize}
\end{subequations}
The wave function $\mathcal{M_S}$ represents a single excitations single-mode excitations, while the part $\mathcal{M_A}$ describes a component in which both, the $\rho$ and the $\lambda$ oscillator, are excited independently. The component represents a two-mode excitation.

$S_3$ and $S_4$ both have mixed-symmetry spin wave functions, in $S_3$ the isospin is symmetric and in $S_4$
it is of mixed symmetry. Hence we expect a doublet of $N^*$ and $\Delta^*$ resonances with $J^P=3/2^+$ and $5/2^+$. These predicted states have not been reported, they are {\it missing resonances}; the $N(1975)3/2^+$ listed in the Table~\ref{listres} of additional resonances might be one of these missing resonances. The $S_5$ configuration  corresponds to $N^*$ resonances with intrinsic spin $S=3/2$. Hence a quartet of $N^*$ resonances is expected which we assign to the four positive-parity resonances observed here:

\renewcommand{\arraystretch}{1.2}
{\footnotesize\bc \begin{tabular}{cccc}
$N(1880)1/2^+$&\hspace{-2mm}$N(1900)3/2^+$&\hspace{-2mm}$N(2000)5/2^+$&\hspace{-2mm}$N(1990)7/2^+$\\[-1ex]
&&&\hfill {\bf D}
\end{tabular}\ec}
\renewcommand{\arraystretch}{1.0}

These resonances have components in their wave functions which are single-mode and two-mode excitations.

\begin{table*}
\caption{\label{tab:Decayall}Branching ratios (in \%) for decays of nucleon and $\Delta^*$ resonances. For $N(1710)1/2^+$ and $N(1990)7/2^+$, two alternative solutions are given in separate lines. Otherwise, the spread of results from different solutions is used to estimate the uncertainties. $N(1535)\pi$ decay modes from \cite{Gutz:2014wit} - which are used in the discussion - are listed as well. $x$ signifies forbidden transitions, empty slots indicate that no second orbital angular momentum contributes. Entries marked - were fitted to small values and set to zero to avoid an excessive number of free parameters. The intermediate resonances on the right side of the double line carry orbital excitations. }
\begin{center}
\renewcommand{\arraystretch}{1.4}
{\scriptsize
\begin{tabular}{|l|cc|cc|cc||c|c|c|c|cl|}
\hline
                   &$N\pi$&$L$  &$\Delta\pi$ &\hspace{-3mm}$L$$<$$J$\hspace{-3mm}&$\Delta\pi$ &\hspace{-3mm}$L$$>$$J$\hspace{-3mm}&$N(1440)\pi$ $L$ &$N(1520)\pi$ $L$&$N(1535)\pi$ $L$&$N(1680)\pi$ $L$&$N\sigma$&$L$\\ \hline
$N(1535)1/2^-$     &52\er 5   &0&x&&2.5\er 1.5&2& \ 12\er 8\hfill 0&\quad -\hfill 1 &\quad - \hfill 1 &\quad -\hfill 2& 6\er 4&1\\
$N(1520)3/2^-$     &61\er 2   &2& 19\er 4&0&9\er 2&2&$<$1\hfill 2&\quad -\hfill 1&\quad-\hfill 1      &\quad -\hfill 2&$<2$&1\\
$N(1650)1/2-$      &51\er 4   &0&x&&12\er 6&2&16\er 10\hfill 0&\quad -\hfill 1 &\quad -\hfill 1       &\quad -\hfill 2& 10\er 8&1\\
$N(1700)3/2^-$     &15\er 6   &2& 65\er 15&0&9\er5&2& 7\er 4\hfill 2& $<$4\hfill 1 &$<$1\hfill 1      &\quad -\hfill 2& 8\er 6&1\\
$N(1675)5/2^-$     &41\er 2   &2&30\er 7&2&\quad -&4&\quad -\hfill 2 &\quad -\hfill 1 &\quad -\hfill 3&\quad -\hfill 0& 5\er 2&3\\
$\Delta(1620)1/2^-$&28\er 3   &0&x&&62\er10&2&6\er 3\hfill 0&\quad -\hfill 1 &\quad -\hfill 1         &\quad -\hfill 2& x &\\
$\Delta(1700)3/2^-$&22\er 4   &2&20\er 15&0& 10\er 6&2&$<$1  \hfill 2&3\er 2\hfill 1&$<$1    \hfill 1 &\quad -\hfill 2&x&\\
\hline
$N(1720)3/2^+$     &11\er 4   &1& 62\er 15&1&6\er 6&3&$<$2\hfill 1&3\er2 \hfill 0 &$<$2 \hfill 2      &\quad -\hfill 1& 8\er 6 &2\\
$N(1680)5/2^+$     &62\er 4   &3& 7\er 3&1&10\er 3&3&\quad -\hfill 3&$<$1\hfill 2 &\quad -\hfill 2    &\quad -\hfill 1& 14\er 5 &2 \\
$\Delta(1910)1/2^+$&12\er 3   &1&x&&50\er 16&1&6\er 3\hfill 1&\quad-\hfill 0&5\er 3\hfill 2           &\quad -\hfill 3&x&\\
$\Delta(1920)3/2^+$& 8\er 4   &1&18\er 10&1&58\er14&3&$<4$\hfill 1& $<5$\hfill 0&$<2$\hfill 2         &\quad -\hfill 1&x&\\
$\Delta(1905)5/2^+$&13\er 2   &3&33\er10&1&-&3&\quad -\hfill 3 &\quad -\hfill 2 &$<1$\hfill 2         &10\er5\hfill 1&x&\\
$\Delta(1950)7/2^+$&46\er 2   &3&5\er 4&3&\quad-&5&\quad-\hfill 3&\quad-\hfill 2&\quad-\hfill 4       & \ 6\er3\hfill 1&x&\\[0.2ex]
\hline\hline\\[-3.5ex]
$N(1880)1/2^+$     & 6\er 3   &1&x&&30\er 12&1&\quad-\hfill 1&\quad-\hfill 2&8\er 4 \hfill 0          &\quad -\hfill 3&25\er 15&0\\
$N(1900)3/2^+$     & 3\er 2   &1&17\er 8&1&33\er 12&3&$<$2 \hfill 1&15\er 8\hfill 0&7\er 3\hfill 2    &\quad -\hfill 1&4\er 3&2\\
$N(2000)5/2^+$     & 8\er 4   &3&22\er 10&1&34\er15&3&\quad-\hfill 1&21\er 10\hfill 2&\quad-\hfill 2  &16\er9 \hfill 1& 10\er 5&2\\
$N(1990)7/2^+$     & 1.5\er0.5&3&48\er 10&3&-&5&$<$2\hfill 1&$<$2\hfill 1&$<$2\hfill 4                &\quad -\hfill 1&-&4\\
$N(1990)7/2^+$     & 2\er1    &3&16\er 6&3&- &5&$<$2\hfill 1&$<$2\hfill 1&$<$2\hfill 4                &\quad -\hfill 1&-&4\\
\hline
$N(1895)1/2^-$     & 2.5\er1.5&0&x   & &7\er 4&2&8\er 8\hfill 0&\quad-\hfill 1&\quad-\hfill 1         &\quad -\hfill 2&18\er15&1\\
$N(1875)3/2^-$     & 4\er 2   &2&14\er7&0&7\er5&2&5\er3\hfill 2&$<2$\hfill 1&$<$1\hfill 1             &\quad -\hfill 2&45\er15&1\\
$\Delta(1900)1/2^-$& 7\er 2   &0&x & &50\er20&2&20\er12\hfill 0&6\er4\hfill 1&\quad-\hfill 1          &\quad -\hfill 2&x&\\
$\Delta(1940)3/2^-$& 2\er 1   &0&46\er20&0&12\er 7&2&7\er7\hfill 2&4\er3\hfill 1&8\er6 \hfill 1       &\quad -\hfill 2&x&\\
\hline
$N(2120)3/2^-$     & 5\er 3   &2&50\er20&0&20\er12&2&10\er10\hfill 2&15\er10\hfill 1&15\er8\hfill 1   &\quad -\hfill 2&11\er4&1\\
$N(2060)5/2^-$     &11\er 2   &2&7\er3&2&-&4& 9\er 5\hfill 2&15\er 6\hfill 1& \quad-\hfill 3          &15\er7 \hfill 0&6\er3&3\\
$N(2190)7/2^-$     &16\er 2   &4&25\er 6&2&-&4&\quad-\hfill 4&\quad-\hfill 3&\quad-\hfill 3           &\quad -\hfill 2&6\er3&3\\
\hline
$N(1710)1/2^+$     & 5\er 3   &1&x&&7\er4&1&30\er10\hfill 1&$<$2\hfill 2&\quad-\hfill 0               &\quad -\hfill 3&55\er15&0\\
$N(1710)1/2^+$     & 5\er 3   &1&x&&25\er10&1&$<$5\hfill 1&$<$2\hfill 2&15\er6\hfill 0                &\quad -\hfill 3&10\er5&0\\
$N(2100)1/2^+$     &16\er 5   &1&x&&10\er4&1&\quad-\hfill 1&$<$2\hfill 2&30\er4\hfill 0               &\quad -\hfill 3&20\er 6&0\\
\hline
\end{tabular}}
\renewcommand{\arraystretch}{1.0}\vspace{-3mm}
\end{center}
 \end{table*}

There is one further symmetry combination, $S_6$, which is completely antisymmetric. In this case, both oscillators are excited and carry one unit of orbital angular momentum ($l_\rho=1, l_\lambda=1$) which add to a total orbital angular momentum ($\vec l_\rho + \vec l_\lambda=\vec L$) with $L=1$.
\begin{footnotesize}
\begin{eqnarray}
\label{A}
\mathcal{A}
    & = &\frac{1}{\sqrt{2}}
    \left[\phi_{0p}(\vec\rho)\times\phi_{0p}(\vec\lambda)\right]^{(L=1)}
\end{eqnarray}
\end{footnotesize}
This orbital wave function requires a fully antisymmetric spin-flavor wave function. Hence a spin doublet of positive parity resonances with $J=1/2$ and $J=3/2$ is expected which we call $N^A_{1/2,3/2}$. Neither of the two expected states has ever been identified.

Resonances in the third shell of the harmonic oscillator all have wave functions with a component in which both oscillators are excited. We list those coupling to $L=3$.

\begin{subequations}\label{grp}
\begin{scriptsize}
\begin{align}
\cal S&={\sqrt{\frac{1}{4}}}
    \left[\phi_{0s}(\vec\rho)\times\phi_{0f}(\vec\lambda)\right]^{(L=3)}
    -{\sqrt{\frac{3}{4}}}
    \left[\phi_{0d}(\vec\rho)\times\phi_{0p}(\vec\lambda)\right]^{(L=3)}
\\
\cal A&=-{\sqrt{\frac{3}{4}}}
    \left[\phi_{0p}(\vec\rho)\times\phi_{0d}(\vec\lambda)\right]^{(L=3)}
    +{\sqrt{\frac{1}{4}}}
    \left[\phi_{0f}(\vec\rho)\times\phi_{0s}(\vec\lambda)\right]^{(L=3)}
        \\
{\cal M}_s&={\sqrt{\frac{3}{4}}}
    \left[\phi_{0s}(\vec\rho)\times\phi_{0f}(\vec\lambda)\right]^{(L=3)}
    +{\sqrt{\frac{1}{4}}}
    \left[\phi_{0d}(\vec\rho)\times\phi_{0p}(\vec\lambda)\right]^{(L=3)}
\\
{\cal M}_a&={\sqrt{\frac{3}{4}}}
    \left[\phi_{0p}(\vec\rho)\times\phi_{0d}(\vec\lambda)\right]^{(L=3)}
    +{\sqrt{\frac{1}{4}}}
    \left[\phi_{0f}(\vec\rho)\times\phi_{0s}(\vec\lambda)\right]^{(L=3)}
    \end{align}
\end{scriptsize}
\end{subequations}

From the mixed-symmetry components, a $N^*$ quartet (with $J^P=3/2^-,\cdots,9/2^-$) and a $N^*$ and a $\Delta^*$ doublet (with $J^P=5/2^-,7/2^-$) can be constructed. Three candidates,
\bc
\begin{tabular}{cccc}
$N(2120)3/2^-$ & $N(2060)5/2^-$ & $N(2190)7/2^-$\,, &{\bf E}
\end{tabular}
\ec
are observed here, two further candidates - $N(2250)9/2^-$ and $\Delta(2200)7/2^-$ - are listed in the Review of Particle Properties~\cite{Agashe:2014kda}. Three additionally expected states  $N(xxx)$ $5/2^-$, $N(xxx)7/2^-$, and $\Delta(xxx)5/2^-$ are missing.

Finally we consider the higher-mass negative-parity resonances
\bc
\begin{tabular}{ccc}
$N(1895)1/2^-$ & $N(1875)3/2^-$ & {\bf F} \\
$\Delta(1900)1/2^-$ & $\Delta(1940)3/2^-$ & $\Delta(1930)5/2^-$
\end{tabular}
\ec
(the latter resonance is not seen here). The three $\Delta^*$ resonances can be interpreted as a triplet (degenerate spin quartet), with $L=1$, $S=3/2$. Since spin and isospin wave functions are symmetric, the spatial wave function must be symmetric, too. This is possible only when the resonances are radially excited in addition. The three $\Delta^*$ resonances should be accompanied by a spin doublet of $N^*$ resonances, with a symmetric spatial wave function as well, which we identify with the two $N^*$'s listed under {\bf F}. A symmetric spatial wave function with odd parity is represented by
\begin{subequations}\label{nod}
\begin{footnotesize}
\begin{align}
\cal S&=
    \frac{1}{{2}}
    \left[\phi_{0s}(\vec\rho)\times\phi_{1p}(\vec\lambda)\right]^{(L=1)}
        -
            \frac{1}{\sqrt{3}}\nonumber
    \left[\phi_{0d}(\vec\rho)\times\phi_{0p}(\vec\lambda)\right]^{(L=1)}\\
     &-
        \sqrt{\frac{5}{12}}
    \left[\phi_{1s}(\vec\rho)\times\phi_{0p}(\vec\lambda)\right]^{(L=1)}
    \end{align}
\end{footnotesize}
\end{subequations}

The $N(1710)1/2^+$ resonance is often interpreted as the mixed-symmetry partner of the $N(1440)1/2^+$ (Roper) resonance . Then it has components in the wave function in which the $\rho$ or the $\lambda$ oscillator carries one unit of the vibrational quantum numbers but also a component in which both oscillators carry one unit of orbital angular momentum. The assignment of $N(2100)1/2^+$ is not clear. In any case, these two resonances
\bc
\begin{tabular}{ccc}
$N(1710)1/2^+$ & $N(2100)1/2^+$ & {\bf G}
\end{tabular}
\ec
have wave functions with single and with two-mode excitations.

\subsection{Evidence for two-mode excitations}
We now try to deduce the existence of the component with both oscillators excited from the measured branching ratios. The branching ratios are listed in Table~\ref{tab:Decayall}. This Table includes branching ratios of decay modes into $N(1535)\pi$ derived from the reaction $\gamma p\to p\pi^0\eta$ \cite{Gutz:2014wit} (where $N(1535)$ decayed into $N\eta$) which help in the following discussion. The uncertainties were defined from the variance of results using different fit assumptions. Forbidden decays are marked by an x; in many cases, the fits converge to a zero, then a - symbol is entered. We note that the resonances {\bf A-C} have wave functions with single-mode excitations, those listed under {\bf D-G} have wave functions with two-mode excitations. The branching ratios of resonances {\bf A} and {\bf C} are listed in the first two blocks in Table~\ref{tab:Decayall}, those of the resonances {\bf D-G} in the other four blocks.

We now ask if single-mode and two-mode excitations lead to different decay patterns. We assume (and test this idea) that the component (\ref{MA}) in resonances carrying a two-mode excitation has at most a small probability to decay into $N\pi$ or $\Delta(1232)\pi$ (nor to $N\omega$, $\Delta(1232)\eta$, $\Sigma K$, etc.). We assume instead that this component leads to an increased probability for decays in which the final-state baryon or the outgoing meson is itself excited.

Figure~\ref{fig:DeltaN} shows that one cannot expect to test this idea on individual branching ratios. There is also no model which reproduces the branching ratios for decays into $N\pi$ and $\Delta(1232)\pi$ and which could provide a guide. But we can use statistical arguments. The first two blocks in Table~\ref{tab:Decayall} lists branching ratios of resonances {\bf A} and {\bf C} which have a wave function with components in which only one oscillator is excited. Their mean branching ratio for decays into $N\pi$ and $\Delta(1232)\pi$ is about 70\% and for decays into states carrying orbital excitation 10\%. The other blocks in Table~\ref{tab:Decayall} list resonances which have additionally a type~(\ref{MA}) component. Their mean branching ratio for $N\pi$ and $\Delta(1232)\pi$ decays is less than 40\% while they decay into states carrying orbital excitation with 35\% probability. These numbers give a first hint that resonances having a type~(\ref{MA}) component have an increased chance to decay into states carrying orbital excitation as one may naively expect.

This comparison could suffer from the fact that the masses of the resonances in the first two blocks in Table~\ref{tab:Decayall} are mostly lower than the masses of the other resonances. From Fig.~\ref{fig:DeltaN} we deduce that the mass is not a very important quantity when branching ratios are discussed. Nevertheless, we may restrict the discussion to a subset of resonances with similar masses and identical quantum numbers. These are the four positive-parity nucleon and the four positive-parity $\Delta$ resonances in the 1880 to 2000\,MeV mass region. These resonances can be interpreted as a spin quartet with intrinsic orbital angular momentum $L=2$ and quark spin $S=3/2$. These four $\Delta^*$ resonances have a mean sum of $N\pi$ and $\Delta(1232)\pi$ decay branching ratios of 60\%. Branching ratios of the four $\Delta^*$ resonances into states carrying orbital excitation are small and mostly, upper limits are given only. On average, the mean branching ratio into states carrying orbital excitation of these four $\Delta^*$ resonances is 8\%. If four $N^*$ resonances are interpreted as quartet with intrinsic orbital angular momentum $L=2$ and quark spin $S=3/2$, the mixed-symmetry isospin-1/2 flavor wave function requires a mixed symmetry orbital wave function of  type~(\ref{MS}) and type~(\ref{MA}). The mean branching ratio of the four nucleon resonances for decays into $N\pi$ and $\Delta(1232)\pi$ is 47\% while the mean branching ratio into states carrying orbital excitation is 27\%. Again, resonances with a wave-function component carrying a two-mode excitation have a higher probability to decay into states carrying orbital excitation.

The individual branching ratios have large errors bars. Hence we fitted the data with two assumptions: i) we forbade decays of the four $\Delta^*$'s into states carrying orbital excitation. This has little effect on the fit and the pseudo-$\chi^2$ (see eqn.~\ref{pseudo}) deteriorated by 692 units. If these decay modes were forbidden for the four $N^*$ resonances, the change in pseudo-$\chi^2$ became 3880 units, and the fit quality deteriorated visibly.
Since the pseudo-$\chi^2$ is derived from a likelihood function, the absolute $\chi^2$ value is meaningless but it is clear that forbidding decays into states carrying orbital excitation has a much larger effect for the four positive-parity $N^*$ resonances than for the four $\Delta^*$'s.

\begin{figure}[pt]
\begin{center}
\includegraphics[width=0.49\textwidth]{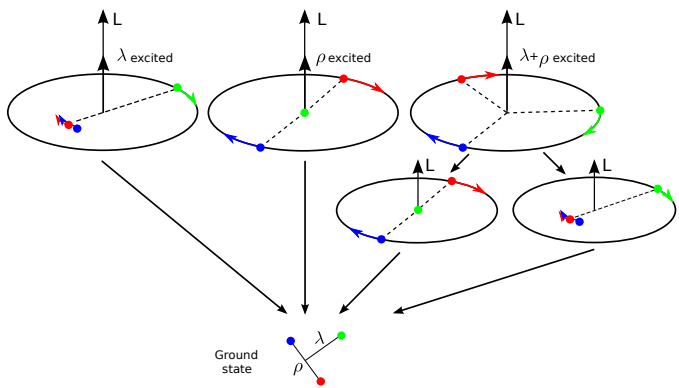}
\end{center}
\caption{\label{FigureDecay}Classical orbits of nucleon excitations with $L=2$ (upper row) and $L=1$ (lower row) (adapted from \cite{Thiel:2015xba}).
In the $\lambda$ excitation, the quark and the diquark both rotate around their common center of mass. In the $\rho$ excitation, the two quarks of the diquark carry the angular momentum. When both oscillators are excited, all three quarks rotate around their common center of mass (third picture in the upper row). In a transition of an excited nucleon in this configuration to $L=1$ (lower row), either the $\lambda$ or the $\rho$ oscillator loses its energy and the baryon  remains excited. }
\end{figure}

At the end, we visualize these observations in a semi-classical picture. Figure~\ref{FigureDecay} (adapted from \cite{Thiel:2015xba}) shows classical orbits of a three-particle system. In the first subfigure, the $\lambda$-oscillator is excited and carries two units of angular momentum while the $\rho$-oscillator diquark remains in its ground state. Of course, there is no linear oscillation, instead - due to the orbital angular momentum - the single quark and the diquark rotate around their common center of gravity. In the second subfigure, the $\rho$-oscillator is excited and the two quarks of the diquark rotate around their common center of gravity while the $\lambda$-oscillator carries no excitation. In the third subfigure, both oscillators are excited, each of them carries one unit of angular momentum, and the three quarks rotate in a Mercedes-star configuration. The latter twofold excitation cannot de-excite into the ground state; instead it de-excites by a change of angular momentum by one unit into one of the configurations shown in the lower subfigures with $\rho$ or $\lambda$ carrying one unit of angular momentum.

In this semi-classical picture we expect an intermediate state carrying one unit of internal orbital angular momentum, like $N(1520)3/2^-$ or $N(1535)1/2^-$. Decays of resonances into $N\sigma$ are also included in this discussion (the $\sigma=f_0(500)$ stands for the $\pi\pi$ S-wave). Here we suppose that in the decay a $q\bar q$ is created, the antiquark picks up a quark from one of the oscillators and the escaping $q\bar q$ pair carries away the angular momentum of the baryonic oscillator.

The list of states carrying orbital excitation (on the right side of the double line in Table~\ref{tab:Decayall}) includes further states, $N(1440)1/2^+$ and $N(1680)5/2^+$, even though they do not correspond to the picture we just have developed. In the case of $N(1440)1/2^+\pi$, the mode of both oscillators is changed in one transition; in the $N(1680)5/2^+\pi$ case, the orbital angular momentum is increased. We interpret this as a kind of Auger process. In atomic physics, an unoccupied inner energy level, e.g. a K shell, can be filled by a transition of an electron orbiting in the L shell and, in the same step, a second electron in another L shell can be ejected (KLL electron). In the type~(\ref{MA}) component, both oscillators carry one unit of orbital angular momentum. Due to the compact size and the range of the interaction, the overlap of the wave functions of the two oscillators will be large~\cite{Melde:2008yr}. We thus assume that one oscillator de-excites into its ground states and the second oscillator receives a kick to the $N(1440)1/2^+$ or $N(1680)5/2^+$ intermediate state.

\section{Summary}
\label{SectionSummary}

We have reported a study of the reaction $\gamma p\to p 2\pi^0$ with
unpolarized and with linearly polarized photons. The $p\pi^0\pi^0$
Dalitz plots reveal clear evidence for the isobars $\Delta(1232)\pi$,
$N(1520)3/2^-\pi$, and $N(1680)5/2^+\pi$. Production of $Nf_0(980)$ is clearly
seen in the $\pi\pi$ invariant mass distribution, and there is evidence
for $Nf_2(1270)$. The partial wave analysis identifies $Nf_0(500)$ and
$N(1440)1/2^+\pi$ in addition. These intermediate states are mostly produced
in the decay of high-mass resonances while $t-$ and $u-$chan\-nel exchange and direct production of the
three-body final state contribute little to the reaction. Thus a large
number of branching ratios of nucleon and $\Delta$ resonances for
decays $p\pi^0\pi^0$ via several intermediate states is reported, in
particular via $\Delta(1232)\pi$, $N(1520)3/2^-\pi$, $N(1680)5/2^+\pi$,
$N(1440)1/2^+\pi$, and $Nf_0(500)$. $Nf_0(980)$ and $Nf_2(1270)$ are also
observed to stem from the decay of high-mass resonances even though
their assignment to specific resonances suffers from ambiguities.

A large number of resonances decay into $\Delta(1232)\pi$. The masses
of the decaying resonances vary from 1440\,MeV to 2190\,MeV. Orbital angular momenta
up to $L=4$ are involved in the process, conservation of parity and of orbital angular momentum leads to selection rules. In
many cases, resonances can decay into $\Delta(1232)\pi$ by two
different angular momenta. Surprisingly, there seems to be no relation
between the observed frequency of a decay and the linear momentum or orbital
angular momentum involved. High momenta are not preferred and higher orbital
angular momenta (up to $L=3$) seem not to be much suppressed.

Particularly interesting is the observation that the symmetry
properties of the wave functions of resonances have a significant
impact on the decay modes. The data were compatible with the following conjecture:
i) Resonances having a component in their quark-model wave function
with both oscillators excited have a significant branching ratio for
decays via a decay chain -- with an intermediate resonance carrying orbital-angular-momentum or vibrational excitation -- over direct $N\pi$ or
$\Delta(1232)\pi$ decays. ii) Resonances in which only  one oscillator
is excited have,  in contrast to i),  a much reduced chance to decay into an intermediate
resonance carrying orbital angular momentum or vibrational excitation. This observation implies
that the high-mass resonances must have a three-particle component in
their wave functions. The observation is important as it excludes an
interpretation of baryon resonances as $q$-$qq$ excitations where the
diquark remains in a relative $S$-wave. The quark-diquark picture of
baryons is attractive since it offers an ``explanation'' of the small
number of observed resonances, a fact known as {\it missing-resonance
problem}.

This interpretation sheds new light on the existence and observability of resonances with only wave function components in which always both oscillators are excited (see eq.~\ref{A}). In the nucleon sector, a doublet of positive parity resonances with $J=1/2$ and $J=3/2$ is expected which in SU(6) belongs to a 20-plet. According to the discussion of branching ratios and wave functions we expect that $N^*$'s belonging to this doublet do not decay in a single-step transition. Hence we assume that the non-observation of these two expected resonances is caused by the impossibility to excite both oscillators in a single step. Possibly, these two resonances can be found in decay chains where a high-mass resonance $N^*$ is excited which decays in a triple cascade via
\begin{equation}
N^*\to N^A_{1/2,3/2}\pi\to N(1520)\pi\pi \to N\pi\pi\pi\,.
\end{equation}
The test of this idea is a challenge for future analyses.

A further important result of this analysis is that the data
strengthen the case for the existence of  $N(1900)3/2^+$, a
resonance discovered by D.~M.~Manley, R.~A.~Arndt, Y.~Goradia, and
V.~L.~Teplitz \cite{Manley:1984jz} in a study of $\pi N\to N\pi\pi$ and
not seen in elastic $\pi N$ scattering, neither by H\"ohler or Cutkosky
and their collaborators \cite{Hohler:1979yr,Cutkosky:1980rh} nor by
Arndt and collaborators using a much larger data sample
\cite{Arndt:2006bf}.
\\[2ex]

{\small We thank the technical staff at ELSA and at all the
participating institutions for their invaluable contributions to the
success of the experiment. We acknowledge support from the Deutsche
Forschungsgemeinschaft (within the SFB/TR16), the U.S. National Science
Foundation (NSF), and from the Schweizerische Nationalfonds. The
collaboration with St. Petersburg received funds from DFG and the
Russian Foundation for Basic Research (13-02-00425). Part of this work
comprises part of the PhD thesis of V. Sokhoyan. }
 \phantom{Dummy}


	\begin{appendix}
	\section{Properties of $N^*$ and $\Delta^*$ resonances}
Table~\ref{nucleon} lists the properties of the resonances used in the analysis and their PDG star rating. Additional confidence is represented by a {\raisebox{0.7mm}{$\star$}}. The uncertainties were derived from the spread of results from 12 up to 32 fits in which the number of resonances and the admitted decay modes were varied. Decay modes of resonances which led to very small contributions were set to zero to reduce the number of free parameters. A definition of the quantities listed in the Table are given in \cite{Gutz:2014wit}. Here we recall that the helicity amplitudes $\tilde A^{1/2}$, $\tilde A^{3/2}$ (photo-couplings in the helicity basis) are complex numbers. They become real and agree with the conventional helicity amplitudes $A^{1/2}$, $A^{3/2}$ if a Breit-Wigner amplitude with constant width is used.
Similarly, the elastic residues and the residues of the transition amplitudes turn into  $\Gamma_{N\pi}/2$ (where $\Gamma_{N\pi}$ is the elastic width), and to the channel coupling $\sqrt{\Gamma_{\rm i}\Gamma_{\rm f}}/\Gamma_{\rm tot}$.

In addition to the resonances quoted in Table~\ref{nucleon}, a few more resonances were introduced which are listed in Table~\ref{listres}. These resonances helped to achieve good convergence of the fit even though their masses could be changed in a large interval. We do not claim their existence and do not quote properties. Masses and widths in Table~\ref{listres} are just the values used in the final fit.
\begin{table}[pt]
\caption{\label{listres} List of additional resonances used in the fit with no entries in Table~\ref{nucleon}. The existence of these resonances is poor (one-star in PDG) or uncertain in our analysis; their properties are not well defined. }
\renewcommand{\arraystretch}{1.3}
 \begin{tabular}{cccccc}
\hline
Name& mass& width &Name& mass& width \\
\hline\hline \hspace{-2mm}$N(1975)3/2^+$\hspace{-2mm} & 1880&  550&
\hspace{-2mm}$N(2450)3/2^+$\hspace{-2mm} & 2450 & 400\\
$N(1860)5/2^+$ & 1830 & 280&
\hspace{-2mm}$N(2500)7/2^+$\hspace{-2mm} & 2550&  450\\
\hspace{-2mm}$\Delta(2100)1/2^+$\hspace{-2mm} & 2100 & 400\\
\hspace{-2mm}$\Delta(2100)3/2^+$\hspace{-2mm} & 2050 & 560 &
\hspace{-2mm}$\Delta(2100)3/2^-$\hspace{-2mm} & 2070 & 700\\
\hline
\hline
\renewcommand{\arraystretch}{1.}
\end{tabular}\vspace{-3mm}
\end{table}

In our previous analysis \cite{Sarantsev:2007aa}, two alternative
classes of solutions were found. In the first class of solution the resonant part of the
amplitude at 1.87\,GeV was dominated by the
$I(J^P)=3/2\,(3/2^-)$ partial wave and particularly by production
of $\Delta(1700)3/2^-$. In the second solution the contribution
from the $I(J^P)=3/2\,(3/2^-)$  partial wave was found to be
much smaller and showed a rather flat energy dependence. In
both classes of solutions there is a destructive interference of the resonant
(K-matrix) part and $t$- and $u$-channel exchange amplitudes in the
$I(J^P)=3/2\,(3/2^-)$ partial wave. Resonances decay into different charge states
according to isotopic relations; this is not true for for $t$- and $u$-channel exchange amplitudes. In \cite{Kashevarov:2012wy}, isotopic relation were applied to
the resonant and non-resonant part of the amplitudes derived in \cite{Sarantsev:2007aa}, and it was claimed that the amplitudes in \cite{Sarantsev:2007aa} are inconsistent with
their data. However, this critique was based on the (wrong) assumption that isotopic relations can be applied to the full amplitude.

The new polarization data provide important information on the partial
wave contributions also in the low-mass region. A small adjustment of
the fit parameters was sufficient to get a good description of the
data when starting from a solution of the second class. In contrast,
solutions from the first class started with a bad $\chi^2$, and the fit showed no proper
convergence. Thus we now have only one class of solutions with a relatively small
and flat contribution from the resonant part of the
$I(J^P)=3/2\,(3/2^-)$ partial wave to the total cross section
below $W=1.7$\,GeV (see Fig.~\ref{FigureTerms}, right).

		\begin{table*}[pt]
\caption{\label{nucleon} Properties of $N^*$ and $\Delta^*$ resonances from the BnGa
multichannel partial wave analysis. Along with the name of the resonance, the star rating
of the Particle Data Group \cite{Agashe:2014kda} is given. Suggested upgrades
are marked by a $\star$. The helicity couplings $A_{1/2,3/2}$ are given in GeV$^{-\frac 12}$.
The electromagnetic transition amplitudes $\gamma p$,  $\gamma p^{1/2}$, $\gamma p^{3/2}$
are defined as elements of the transition matrix and dimensionless.
$\pi N\to \pi N$
stands for the elastic pole residue, $2\,(\pi N\to X)/\Gamma$ for
inelastic pole residues. They are normalized by a factor $2/\Gamma$
with $\Gamma=\Gamma_{\rm pole}$. The phase of small complex numbers is often undefined (not def.).
}
\renewcommand{\arraystretch}{1.2}
\begin{scriptsize}

\end{tabular}
\end{scriptsize}
\end{table*}
\end{appendix}
		

\begin{thebibliography}{99}

  \bibitem{Manley:1984jz}
  D.~M.~Manley, R.~A.~Arndt, Y.~Goradia and V.~L.~Teplitz,
  Phys.\ Rev.\  D {\bf 30},  904 (1984).

 \bibitem{Vandermeulen:1992eh}
  J.~Vandermeulen,
  Z.\ Phys.\ A {\bf 342}, 329 (1992).

 \bibitem{Capstick:1986bm}
  S.~Capstick and N.~Isgur,
  Phys.\ Rev.\  D {\bf 34}, 2809 (1986).

 \bibitem{Loring:2001kx}
  U.~L\"oring, B.~C.~Metsch and H.~R.~Petry,
  Eur.\ Phys.\ J.\  A {\bf 10}, 395 (2001).

 \bibitem{Edwards:2011jj}
  R.~G.~Edwards {\it et al.}, 
  Phys.\ Rev.\ D {\bf 84}, 074508 (2011).

  \bibitem{Capstick:2000qj}
  S.~Capstick and W.~Roberts,
  Prog.\ Part.\ Nucl.\ Phys.\  {\bf 45}, S241 (2000).

\bibitem{Capstick:1992uc}
  S.~Capstick,
  Phys.\ Rev.\  D {\bf 46}, 2864 (1992).

\bibitem{Hillert:2006yb}
  W.~Hillert,
  Eur.\ Phys.\ J.\  A {\bf 28S1}, 139 (2006).

 \bibitem{Gutz:2014wit}
  E.~Gutz {\it et al.} [CBELSA/TAPS Collaboration],
  Eur.\ Phys.\ J.\ A {\bf 50}, 74 (2014).

\bibitem{Anisovich:2011fc}
  A.~V.~Anisovich, R.~Beck, E.~Klempt, V.~A.~Nikonov, A.~V.~Sarantsev and U.~Thoma,
  Eur.\ Phys.\ J.\ A {\bf 48}, 15 (2012).

\bibitem{Anisovich:2013vpa}
  A.~V.~Anisovich, E.~Klempt, V.~A.~Nikonov, A.~V.~Sarantsev and U.~Thoma,
  Eur.\ Phys.\ J.\ A {\bf 49}, 158 (2013).

\bibitem{Thiel:2015xba}
  A.~Thiel {\it et al.}  [CBELSA/TAPS Collaboration],
  Phys.\ Rev.\ Lett.\  {\bf 114}, 091803 (2015).

\bibitem{Sokhoyan:2015eja}
  V.~Sokhoyan {\it et al.}  [CBELSA/TAPS Collaboration],
  Phys.\ Lett.\ B {\bf 746}, 127 (2015).

\bibitem{CambridgeBubbleChamberGroup:1968zz}
  Cambridge Bubble Chamber Group,
  Phys.\ Rev.\  {\bf 169}, 1081 (1968).

\bibitem{Erbe:1970cq}
  R.~Erbe {\it et al.},
  Phys.\ Rev.\  {\bf 188}, 2060 (1969).

\bibitem{Ballam:1971wq}
  J.~Ballam {\it et al.},
  Phys.\ Rev.\  D {\bf 5}, 15 (1972).

\bibitem{Ballam:1971yd}
  J.~Ballam {\it et al.},
  Phys.\ Rev.\  D {\bf 5}, 545 (1972).

\bibitem{Davier:1973fy}
  M.~Davier {\it et al.},
  Nucl.\ Phys.\  B {\bf 58}, 31 (1973).

\bibitem{Gialanella:1969ng}
  G.~Gialanella {\it et al.},
  Nuovo Cim.\  A {\bf 63}, 892 (1969).

\bibitem{Carbonara:1976tg}
  F.~Carbonara {\it et al.},
  Nuovo Cim.\  A {\bf 36}, 219 (1976).

\bibitem{Braghieri:1994rf}
  A.~Braghieri {\it et al.},
  Phys.\ Lett.\  B {\bf 363}, 46 (1995).

\bibitem{Kotulla:2003cx}
  M.~Kotulla {\it et al.},
  Phys.\ Lett.\  B {\bf 578}, 63 (2004).

\bibitem{Harter:1997jq}
  F.~H\"arter {\it et al.},
  Phys.\ Lett.\ B {\bf 401}, 229 (1997).

\bibitem{Wolf:2000qt}
  M.~Wolf {\it et al.},
  Eur.\ Phys.\ J.\ A {\bf 9}, 5 (2000).

\bibitem{Langgartner:2001sg}
  W.~Langgartner {\it et al.},
  Phys.\ Rev.\ Lett.\  {\bf 87}, 052001 (2001).

\bibitem{Ahrens:2005ia}
  J.~Ahrens {\it et al.},
  Phys.\ Lett.\ B {\bf 624}, 173 (2005).

\bibitem{Ahrens:2007zzj}
  J.~Ahrens {\it et al.},
  Eur.\ Phys.\ J.\  A {\bf 34}, 11 (2007).

\bibitem{Krambrich:2009te}
  D.~Krambrich {\it et al.}  [Crystal Ball at MAMI and TAPS and A2 Collaborations],
  Phys.\ Rev.\ Lett.\  {\bf 103}, 052002 (2009).  

\bibitem{Kashevarov:2012wy}
  V.~L.~Kashevarov {\it et al.}
    [Crystal Ball at MAMI, TAPS and A2 Collaborations],
  Phys.\ Rev.\ C {\bf 85}, 064610 (2012).

\bibitem{Zehr:2012tj}
  F.~Zehr
    {\it et al.},
  Eur.\ Phys.\ J.\ A {\bf 48}, 98 (2012).  

\bibitem{Oberle:2013kvb}
  M.~Oberle
    {\it et al.},
  Phys.\ Lett.\ B {\bf 721}, 237 (2013).

    \bibitem{Assafiri:2003mv}
  Y.~Assafiri {\it et al.},
  Phys.\ Rev.\ Lett.\  {\bf 90}, 222001 (2003).

\bibitem{Ajaka:2007zz}
  J.~Ajaka  {\it et al.},
  Phys.\ Lett.\ B {\bf 651}, 108 (2007).

\bibitem{Wu:2005wf}
  C.~Wu {\it et al.},
  Eur.\ Phys.\ J.\  A {\bf 23}, 317 (2005).

\bibitem{Thoma:2007bm}
  U.~Thoma {\it et al.}  [CBELSA Collaboration],
  Phys.\ Lett.\  B {\bf 659,}  87 (2008).

\bibitem{Sarantsev:2007aa}
  A.~V.~Sarantsev {\it et al.},
  Phys.\ Lett.\  B {\bf 659}, 94 (2008).

\bibitem{Hirose:2009zz}
  K.~Hirose
    {\it et al.},
  Phys.\ Lett.\ B {\bf 674}, 17 (2009).

\bibitem{Mokeev:2012vsa}
  V.~I.~Mokeev {\it et al.}  [CLAS Collaboration],
  Phys.\ Rev.\ C {\bf 86}, 035203 (2012).  


  \bibitem{Tiator:2011pw}
  L.~Tiator, D.~Drechsel, S.~S.~Kamalov and M.~Vanderhaeghen,
  Eur.\ Phys.\ J.\ ST {\bf 198}, 141 (2011).  

\bibitem{Aznauryan:2011qj}
  I.~G.~Aznauryan and V.~D.~Burkert,
  Prog.\ Part.\ Nucl.\ Phys.\  {\bf 67}, 1 (2012).  

\bibitem{Strauch:2005cs}
  S.~Strauch {\it et al.},
  Phys.\ Rev.\ Lett.\  {\bf 95}, 162003 (2005).  

\bibitem{Luke:book}
D. L\"uke, P. S\"oding, Springer Tracts in Modern Physics, Vol. {\bf
59} (Springer, Berlin, Heidelberg, 1971) p. 39.

\bibitem{GomezTejedor:1995kj}
  J.~A.~Gomez Tejedor, F.~Cano and E.~Oset,
  Phys.\ Lett.\  B {\bf 379}, 39 (1996).

\bibitem{GomezTejedor:1995pe}
  J.~A.~Gomez Tejedor and E.~Oset,
  Nucl.\ Phys.\  A {\bf 600}, 413 (1996).

\bibitem{Hirata:1997xva}
  M.~Hirata, K.~Ochi and T.~Takaki,
  ``Effect of $\rho N$ channel in the $\gamma N \to\pi\pi N$ reactions,''
HUPD-9722, arXiv:nucl-th/9711031.

\bibitem{Nacher:2000eq}
  J.~C.~Nacher, E.~Oset, M.~J.~Vicente and L.~Roca,
  Nucl.\ Phys.\  A {\bf 695}, 295 (2001).

\bibitem{Penner:2002md}
  G.~Penner and U.~Mosel,
  Phys.\ Rev.\  C {\bf 66}, 055212 (2002).

\bibitem{Hirata:2002tp}
  M.~Hirata, N.~Katagiri and T.~Takaki,
  Phys.\ Rev.\  C {\bf 67}, 034601 (2003).

  \bibitem{Fix:2005if}
  A.~Fix and H.~Arenh\"ovel,
  Eur.\ Phys.\ J.\  A {\bf 25}, 115 (2005).

\bibitem{SAID}
R. A. Arndt {\it et al.}, {\ttfamily http://gwdac.phys.gwu.edu}.

  \bibitem{Anisovich:2004zz}
  A.~Anisovich, E.~Klempt, A.~Sarantsev and U.~Thoma,
  Eur.\ Phys.\ J.\  A {\bf 24}, 111 (2005).

\bibitem{Fix:2012ds}
  A.~Fix and H.~Arenh\"ovel,
  Phys.\ Rev.\ C {\bf 85}, 035502 (2012).  

\bibitem{Aker:1992ny}
  E.~Aker {\it et al.}  [Crystal Barrel Collaboration],
  Nucl.\ Instrum.\ Meth.\ A {\bf 321}, 69 (1992).  

\bibitem{Novotny:1991ht}
  R.~Novotny [TAPS Collaboration],
  IEEE Trans.\ Nucl.\ Sci.\  {\bf 38}, 379-385 (1991).

\bibitem{Gabler:1994ay}
  A.~R.~Gabler
    {\it et al.},
  Nucl.\ Instrum.\ Meth.\  A {\bf 346}, 168-176 (1994).


\bibitem{Suft:2005cq}
  G.~Suft {\it et al.},
  Nucl.\ Instrum.\ Meth.\  A {\bf 538}, 416 (2005).

\bibitem{Elsner:2008sn}
  D.~Elsner {\it et al.}  [CBELSA/TAPS Collaboration],
  Eur.\ Phys.\ J.\  A {\bf 39}, 373-381 (2009).

\bibitem{Natter:2003}
F.A.~Natter, P.~Grabmayr, T.~Hehla, R.O.~Owens and S.~Wunderlich,
  Nucl.\ Instrum.\ Meth.\ B {\bf 211}, 465 (2003).

\bibitem{vanPee:2007tw}
  H.~van Pee {\it et al.}  [CBELSA Collaboration],
  Eur.\ Phys.\ J.\ A {\bf 31}, 61 (2007).

\bibitem{Agashe:2014kda}
  K.~A.~Olive {\it et al.}  [Particle Data Group Collaboration],
  Chin.\ Phys.\ C {\bf 38}, 090001 (2014).

\bibitem{Abashian:1960zz}
  A.~Abashian, N.~E.~Booth and K.~M.~Crowe,
  Phys.\ Rev.\ Lett.\  {\bf 5}, 258 (1960).  

\bibitem{Adlarson:2011bh}
  P.~Adlarson {\it et al.}  [WASA-at-COSY Collaboration],
  Phys.\ Rev.\ Lett.\  {\bf 106}, 242302 (2011).

    \bibitem{Adlarson:2012fe}
  P.~Adlarson {\it et al.}  [WASA-AT-COSY Collaboration],
  Phys.\ Lett.\ B {\bf 721}, 229 (2013).

\bibitem{Schilling:1969um}
  K.~Schilling, P.~Seyboth and G.~E.~Wolf,
  Nucl.\ Phys.\ B {\bf 15}, 397 (1970)
  [Erratum-ibid.\ B {\bf 18}, 332 (1970)].

	\bibitem{Roberts:2004mn}
  W.~Roberts and T.~Oed,
  Phys.\ Rev.\ C {\bf 71}, 055201 (2005).

\bibitem{Klempt:2012fy}
  E.~Klempt and B.~C.~Metsch,
  Eur.\ Phys.\ J.\ A {\bf 48}, 127 (2012).

\bibitem{Klempt:2002vp}
  E.~Klempt,
  Phys.\ Rev.\ C {\bf 66}, 058201 (2002).

\bibitem{Forkel:2008un}
  H.~Forkel and E.~Klempt,
  Phys.\ Lett.\ B {\bf 679}, 77 (2009).

\bibitem{Melde:2008yr}
  T.~Melde, W.~Plessas and B.~Sengl,
  Phys.\ Rev.\ D {\bf 77}, 114002 (2008).

    \bibitem{Hohler:1979yr}
  G.~H\"ohler {\it et al.}, F.~Kaiser, R.~Koch and E.~Pietarinen,
  ``Handbook Of Pion Nucleon Scattering,''
Published by Fachinform. Zentr. Karlsruhe 1979, 440 P.
Physics Data, No.12-1 (1979).

\bibitem{Cutkosky:1980rh}
  R.E.~Cutkosky {\it et al.}, C.~P.~Forsyth, J.~B.~Babcock, R.~L.~Kelly and R.~E.~Hendrick, ``Pion - Nucleon Partial Wave Analysis,''
    4th Int. Conf. on Baryon Resonances, Toronto, Canada, Jul 14-16, 1980. Publ.
    in: Baryon 1980:19 (QCD161:C45:1980).

\bibitem{Arndt:2006bf}
  R.A.~Arndt {\it et al.}, 
  Phys.\ Rev.\  C {\bf 74}, 045205 (2006).

    \end{thebibliography}
    \end{document}